\documentclass[
 reprint,
superscriptaddress,
 amsmath,amssymb,
 aps,
pra,
nofootinbib,longbibliography]{revtex4-1}
\usepackage{hyperref}
\newcommand{\bi}{\mathbf}
\usepackage{graphicx}
\usepackage{dcolumn}
\usepackage{bm}
\usepackage{verbatim}
\usepackage{float}

\usepackage{balance}

\begin{document}


\title{The Free Electron Gas in Cavity Quantum Electrodynamics} 

\author{Vasil~Rokaj}
\email{vasil.rokaj@mpsd.mpg.de}
\affiliation{Max Planck Institute for the Structure and Dynamics of Matter,
Center for Free Electron Laser Science, 22761 Hamburg, Germany}

\author{Michael~Ruggenthaler}
\email[]{michael.ruggenthaler@mpsd.mpg.de}

\affiliation{Max Planck Institute for the Structure and Dynamics of Matter,
Center for Free Electron Laser Science, 22761 Hamburg, Germany}

\author{Florian~G.~Eich}
\affiliation{Max Planck Institute for the Structure and Dynamics of Matter,
Center for Free Electron Laser Science, 22761 Hamburg, Germany}

\author{Angel~Rubio}
\email{angel.rubio@mpsd.mpg.de}
\affiliation{Max Planck Institute for the Structure and Dynamics of Matter,
Center for Free Electron Laser Science, 22761 Hamburg, Germany}
\affiliation{Center for Computational Quantum Physics (CCQ), Flatiron Institute, 162 Fifth Avenue, New York NY 10010}


\date{\today}

\begin{abstract}
 Cavity modification of material properties and phenomena is a novel research field largely motivated by the advances in strong light-matter interactions. Despite this progress, exact solutions for extended systems strongly coupled to the photon field are not available, and both theory and experiments rely mainly on finite-system models. Therefore a paradigmatic example of an exactly solvable extended system in a cavity becomes highly desireable. To fill this gap we revisit Sommerfeld's theory of the free electron gas in cavity quantum electrodynamics (QED). We solve this system analytically in the long-wavelength limit for an arbitrary number of non-interacting electrons, and we demonstrate that the electron-photon ground state is a Fermi liquid which contains virtual photons. In contrast to models of finite systems, no ground state exists if the diamagentic $\bi{A}^2$ term is omitted. Further, by performing linear response we show that the cavity field induces plasmon-polariton excitations and modifies the optical and the DC conductivity of the electron gas. Our exact solution allows us to consider the thermodynamic limit for both electrons and photons by constructing an effective quantum field theory. The continuum of modes leads to a many-body renormalization of the electron mass, which modifies the fermionic quasiparticle excitations of the Fermi liquid and the Wigner-Seitz radius of the interacting electron gas. Lastly, we show how the matter-modified photon field leads to a repulsive Casimir force and how the continuum of modes introduces dissipation into the light-matter system. Several of the presented findings should be experimentally accessible.

\end{abstract}

\pacs{Valid PACS appear here}
\maketitle


\section{Introduction}

The free electron gas introduced by Sommerfeld in 1928~\cite{Sommerfeld1928} is a paradigmatic model for solid state and condensed matter physics. It was originally developed for the description of thermal and conduction properties of metals, and has served since then as one of the fundamental models for understanding and describing materials. The free electron gas with the inclusion of the electron-electron interactions, was transformed into the homogeneous electron gas~\cite{Mermin, Vignale}, known also as the jellium model, and with the advent of density functional theory (DFT) and the local density approximation (LDA)~\cite{HohenbergKohn} has become one of the most useful computational tools and methods in physics, chemistry and materials science~\cite{RubioReview}. Also within the Fermi liquid theory, developed by Landau~\cite{LandauFermiLiquid}, the free electron gas model was used as the fundamental building block~\cite{Nozieres}. In addition, the free electron gas in the presence of strong magnetic fields has also been proven extremely important for the description of the quantum Hall effect~\cite{Klitzing, LaughlinPRB}.

On the other hand, the cornerstone of the modern description of the interaction between light and matter, in which both constituents are treated on equal quantum mechanical footing, and both enter as dynamical entities, is quantum electrodynamics~\cite{greiner1996, spohn2004, ruggenthaler2017b, Weinberg, cohen1997photons}. This description of light and matter has led to a number of great fundamental discoveries, like the laser cooling~\cite{PhillipsNobel, TannoudjiNobel, ChuNobel}, the first realization of Bose-Einsten condensation in dilute gases and the atom laser~\cite{KetterleNobel, CornellNobel}, the theory of optical coherence~\cite{GlauberNobel} and laser-based precision spectroscopy~\cite{HanschNobel, HallNobel}, and the manipulation of individual quantum systems with photons~\cite{HarocheNobel, WinelandNobel}.

In most cases simplifications of QED are employed for the practical use of the theory (due to its complexity) in which matter is described by a few states. This leads to the well-known models of quantum optics, like the Rabi, Jaynes-Cummings or Dicke models~\cite{shore1993,dicke1954, Kirton}. Although, these models have served well and have been proven very succesful~\cite{kockum2019ultrastrong}, recently they are being challenged by novel developments in the field of cavity QED materials~\cite{ruggenthaler2017b}. For this, first-principle approaches have already been put forward using Green's functions methods~\cite{Melo2016}, the exact density-functional reformulation of QED, known as QEDFT~\cite{TokatlyPRL, ruggenthaler2014, CamillaPRL}, hybrid-orbital approaches~\cite{Buchholz2020, Nielsen2019}, or generalized coupled cluster theory for electron-photon systems~\cite{KochCC, mordovinaCC}.

Cavity QED materials~\cite{flick2017, flick2015, ruggenthaler2017b} is an emerging field, combining many different platforms for manipulating and engineering quantum materials with electromagnetic fields, ranging from quantum optics~\cite{cohen1997photons}, polaritonic chemistry~\cite{ebbesen2016, george2016, hutchison2013, hutchison2012, orgiu2015, ruggenthaler2017b, feist2017polaritonic, galego2016, schafer2019modification, TaoLi}, and light-induced states of matter using either classical fields~\cite{Basov2017, Cavalleri} or quantum fields originating from a cavity~\cite{SentefRonca, Kiffner, Eckstein2020, AshidaPRL}. A plethora of pathways have been explored recently from both theorists and experimenters. Quantum Hall systems under cavity confinement, in both the integer~\cite{Hagenmuller2010cyclotron, rokaj2019, Keller2020, ScalariScience, li2018} and the fractional~\cite{AtacPRL, SmolkaAtac} regime, have demonstrated ultrastrong coupling to the light field and modifications of transport~\cite{paravacini2019}. Light-matter interactions have been suggested to modify electron-phonon coupling and superconductivity~\cite{SchlawinSuperconductivity, AtacSuperconductivity, sentef2018, Galitski} with the first experimental evidence already having appeared~\cite{A.Thomas2019}. Cavity control of excitons has been investigated~\cite{LatiniRonca, ExcitonControl, LevinsenPRR} and exciton-polariton condensation has been achieved~\cite{Littlewood, KeelingKenaCohen}. Further, the implications of coupling to chiral electromagnetic fields has also attracted interest and is currently investigated~\cite{ChiralCavities, ChiralPetersen67, PRAChiral, ChiralQuantumOptics}. 
 
Much of our understanding and theoretical description of light-matter interactions and of these novel experiments, is based on finite-system models from quantum optics. However, extended systems like solids behave very much differently than finite systems and it is questionable whether the finite-system models can be straightforwardly extended to describe macroscopic systems, like materials, strongly coupled to a cavity. It is therefore highly desirable, in analogy to the Rabi and the Dicke model~\cite{shore1993,dicke1954, Kirton}, to have a paradigmatic example of an extended system strongly coupled to the quantized cavity field. 

The aim of this work is to fill this gap, by revisiting Sommerfeld's theory~\cite{Sommerfeld1928} of the free electron gas in the framework of QED and providing a new paradigm for many-body physics in the emerging field of cavity QED materials. 

In this article we introduce and study in full generality the 2D free electron gas (2DEG) coupled to a cavity. We show that this system in the long-wavelength limit and for a finite amount of cavity modes is analytically solvable and we find the full set of eigenstates and the eigenspectrum of the system. Specializing to the paradigmatic case of just one effective mode (with both polarizations included) we highlight that in the large $N$ or thermodynamic limit the ground state of the electrons is a Slater determinant of plane waves with the momenta of the electrons distributed on the 2D Fermi sphere, thus it is a Fermi liquid. On the other hand, the photon field gets strongly renormalized by the full electron density and the combined light-matter ground state exhibits quantum fluctuation effects and contains virtual photons. Moreover, we study the full phase diagram of the system (see Fig.~\ref{Phase Diagram}) and we find that when the coupling approaches its maximum value (critical coupling) a critical situation appears with the ground state being infinitely degenerate. Above the critical coupling (which in principle is forbidden) the system is unstable and has no ground state. The lack of a ground state shows up also when the diamagnetic $\bi{A}^2$ term is neglected in the Hamiltonian. This is in stark contrast to the standard quantum optics models, like the Rabi or the Dicke model, which have a ground state even without the diamagnetic $\bi{A}^2$ term. This highlights that the $\bi{A}^2$ term is necessary for the stability of extended systems like the 2DEG. This result we believe sheds light on the ongoing discussion about whether the $\bi{A}^2$ term can be eliminated or not~\cite{schaeferquadratic,vukics2014, bernadrdis2018breakdown, DiStefano2019} which is related to the existence of the superradiant phase transition~\cite{Lieb, Wang, Birula, CiutiSuperradiance, MarquardtNoGO, MazzaSuperradiance, AndolinaNoGo, Jaako2016, Andolina2020, Guerci2020, Stokes2020}. 

Performing linear response~\cite{kubo, flick2018light, Vignale} for the interacting electron-photon system in the cavity, we compute the optical conductivity $\sigma(w)$ in which we identify diamagnetic modifications to the standard conductivity of the free electrons gas, coming from the cavity field. Further, in the static limit we find that the cavity field suppresses the DC conductivity and the Drude peak of the 2DEG. This shows that a cavity can alter the conduction properties of 2D materials as suggested also experimentally~\cite{paravacini2019, A.Thomas2019}. Our linear response formalism demonstrates that \textit{plasmon-polariton} resonances exist for this interacting electron-photon system~\cite{TodorovHEG, Todorov2015} and provides a microscopic quantum electrodynamical description of plasmon-polaritons. 

To overcome the discrepancy between the electronic sector, in which the energy density of the electrons is finite, and the photonic sector, whose energy density in the thermodynamic limit vanishes, we promote the single mode theory into an \textit{effective quantum field theory} in the 2D continuum by integrating over the in-plane modes of the photon field. The area of integration in the photonic momentum space is directly connected to the \textit{effective cavity volume} and the upper cutoff in the photon momenta defines the \textit{effective coupling} of the theory. Moreover, in the effective field theory the energy density of the photon field becomes finite and renormalizes the electron mass~\cite{CHEN20082555, Frohlich2010, HiroshimaSpohn}. The renormalized mass depends on the full electron density in the cavity which means that we have a many-body contribution to the renormalized mass due to the collective coupling of the electrons to the cavity field. In addition, the renormalized electron mass shows up in the expression for the chemical potential and modifies the fermionic quasiparticle excitations of the Fermi liquid. Upon the inclusion of the Coulomb interaction, the mass renormalization leads also to a shrinking of the Wigner-Seitz radius, which implies a localization effect for the electrons. From the energy density of the photon field in the cavity we compute the corresponding Casimir force~\cite{Casimir:1948dh, casimir1948influence} (pressure) and we find that due to the interaction of the cavity field with the 2DEG, the Casimir force is \textit{repulsive}~\cite{Munday2009}. Furthermore, we are able to describe consistently and from first principles dissipation and absorption processes without the need of any artificial damping parameter~\cite{flick2018light, Vignale}.

\textit{Outline of the Paper}---In section~\ref{EG in cavity QED} we introduce the 2DEG in cavity QED and we solve the system exactly. In section~\ref{Ground State} we find the ground state of the system in the large $N$ (or thermodynamic) limit. In section~\ref{Instability and A2} we provide the phase diagram of the system for any value of the coupling constant and we discuss under which conditions the system is stable and has a ground state. In section~\ref{Cavity Responses} we perform linear response, we introduce the four fundamental responses (matter-matter, photon-photon, photon-matter and matter-photon) and we compute the optical and the DC conductivity of the 2DEG in the cavity. In section~\ref{Effective Field Theory} out of the single mode theory we construct an effective quantum field theory in the continuum. Finally, in section~\ref{Summary} we conclude and highlight the experimental implications of this work and give an overview of the future perspectives.

\section{Electron Gas in Cavity QED}\label{EG in cavity QED}
Our starting point is the Pauli-Fierz Hamiltonian which describes slowly moving electrons in the non-relativistic limit, minimally coupled to the photon field~\cite{rokaj2017, spohn2004, cohen1997photons}
\begin{eqnarray}\label{Pauli-Fierz} 
\hat{H}&=&\frac{1}{2m_{\textrm{e}}}\sum\limits^{N}_{j=1}\left(\textrm{i}\hbar \mathbf{\nabla}_{j}+e \hat{\mathbf{A}}(\bi{r}_j)\right)^2 +\frac{1}{4\pi\epsilon_0}\sum\limits^{N}_{j< k}\frac{e^2}{|\mathbf{r}_j-\mathbf{r}_k|}\nonumber\\
&+&\sum\limits^{N}_{j=1}v_{ext}(\mathbf{r}_{j})+\sum\limits_{\bm{\kappa},\lambda}\hbar\omega(\bm{\kappa})\left[\hat{a}^{\dagger}_{\bm{\kappa},\lambda}\hat{a}_{\bm{\kappa},\lambda}+\frac{1}{2}\right] .
\end{eqnarray}
where we neglected the Pauli (Stern-Gerlach) term, i.e., $\hat{\bm{\sigma}}\cdot\hat{ \bi{B}}(\bi{r})$. The quantized vector potential $\hat{\mathbf{A}}(\bi{r})$ of the electromagnetic field in Coulomb gauge is~\cite{spohn2004, greiner1996} \begin{equation}\label{AinCoulomb}
\hat{\bi{A}}(\bi{r})=\sum_{\bm{\kappa},\lambda}\sqrt{\frac{\hbar}{\epsilon_0V2\omega(\bm{\kappa})}}\left[ \hat{a}_{\bm{\kappa},\lambda}\bi{S}_{\bm{\kappa},\lambda}(\bi{r})+\hat{a}^{\dagger}_{\bm{\kappa},\lambda}\bi{S}^*_{\bm{\kappa},\lambda}(\bi{r})\right].
\end{equation}
Further, $\bm{\kappa}=(\kappa_x,\kappa_y,\kappa_z)$ are the wave vectors of the photon field, $\omega(\bm{\kappa})=c|\bm{\kappa}|$ are the allowed frequencies in the quantization volume $V=L^2L_z$, $\lambda=1,2$ the two transversal polarization directions and $\bi{S}_{\bm{\kappa},\lambda}(\bi{r})$ are the vector valued mode functions, chosen such that the Coulomb gauge is satisfied~\cite{spohn2004, greiner1996}. The operators $\hat{a}_{\bm{\kappa},\lambda}$ and $\hat{a}^{\dagger}_{\bm{\kappa},\lambda}$ are the annihilation and creation operators of the photon field and obey bosonic commutation relations $[\hat{a}_{\bm{\kappa},\lambda},\hat{a}^{\dagger}_{\bm{\kappa}^{\prime},\lambda^{\prime}}]=\delta_{\bm{\kappa}\bm{\kappa}^{\prime}}\delta_{\lambda\lambda^{\prime}}$. 
\begin{figure}[h]
\includegraphics[width=\columnwidth]{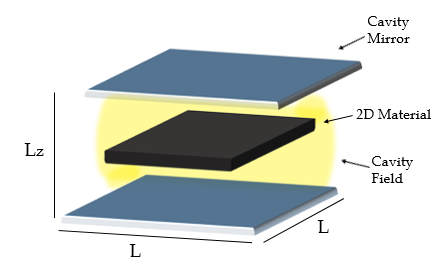}
\caption{\label{HEG_Cavity} Schematic depiction of a 2D material inside a cavity with mirrors of length $L$ and area $S=L^2$. The area of the material is also $S$, and the distance between the mirrors is $L_z$. We note that the Pauli-Fierz Hamiltonian of Eq.~(\ref{Pauli-Fierz}) is in 3D, which is highlighted by the 3D Coulomb potential, while the electrons are restricted in the 2D plane. Further, we mention that in experimental setups, to increase the light-matter coupling, the space between the 2D material and the cavity is filled with a highly polarizable medium.} 
\end{figure}

Here we are interested in the 2D free electron gas confined in a cavity as depicted in Fig.~\ref{HEG_Cavity}. Thus, we take $v_{\textrm{ext}}(\bi{r})=0$ and we neglect the Coulomb interaction as in the original free electron model introduced by Sommerfeld~\cite{Sommerfeld1928}. Since we restrict our considerations in two dimensions\footnote{We would like to mention that all the derivations we present here do not depend on the choice of dimensions and can be performed also in the 3D case.}, the momentum operator has only two components $\nabla=(\partial_x,\partial_y)$. We thus assume the 2DEG restricted on the $(x,y)$ plane, in which the system is considered macroscopic. Then the electrons can be described with the use of periodic boundary conditions, as in the original Sommerfeld model~\cite{Sommerfeld1928}. We would like to mention that for macroscopic systems the choice of the boundary conditions does not affect the bulk properties~\cite{Lieb_Boundary_Conditions}.

For the mode functions $\bi{S}_{\bm{\kappa},\lambda}(\bi{r})$ to satisfy the boundary conditions of the cavity, the momenta of the photon field take the values $\bm{\kappa}=(\kappa_x,\kappa_y,\kappa_z)=(2\pi n_x/L,2\pi n_y/L,\pi n_z/L_z)$ with $\bi{n}=(n_x,n_y,n_z)\in\mathbb{Z}^3$. In the long-wavelength limit~\cite{rokaj2017, faisal1987}, which has been proven adequate for cavity QED systems\cite{schafer2019modification,ruggenthaler2017b}, the mode functions $\bi{S}_{\bm{\kappa},\lambda}(\bi{r})$ become spatially independent vectors $\bi{S}_{\bm{\kappa},\lambda}(\bi{r})=\bm{\varepsilon}_{\lambda}(\bm{\kappa})$, which satisfy the condition $\bi{\varepsilon}_{\lambda}(\bm{\kappa})\cdot \bi{\varepsilon}_{\lambda^{\prime}}(\bm{\kappa})=\delta_{\lambda\lambda^{\prime}}$. The long-wavelength limit or dipole approximation is justified in cases where the size of the matter system is much smaller than the wavelength of the electromagnetic field. This means that the spatial extension of the material in the direction confined by the cavity has to be much smaller than the wavelength of the mode. In our case the long-wavelength limit is respected and justified, because we are considering a 2D material confined in the cavity, as depicted in Fig.~\ref{HEG_Cavity}. In addition, since our aim is to revisit the Sommerfeld model in QED, the assumption of spatially non-varying fields it is necessary because otherwise homogeneity and translational invariance would not be respected. These assumptions are fundamental for the electrons in the Sommerfeld model~\cite{Sommerfeld1928, Mermin} and it is necessary to enforce them to the photon field as well. We note that the description of solids in QED beyond the dipole approximation remains an open research question.

As a starting point, we consider the case where the electromagnetic field consists of a single mode of frequency $\omega$ but with both polarization vectors $\bm{\varepsilon}_{\lambda}$ kept. Although, as shown in appendix~\ref{Mode-Mode Interactions}, we can solve this problem even for arbitrarily many discrete modes analytically, the one-mode case serves as a stepping stone to construct an effective quantum field theory that takes into account the continuum of modes.  In this way the fact that the cavity is open is also taken into account. In section~\ref{Effective Field Theory} we then show, with the help of the exact analytic solution for the many-mode case, that the presented effective field theory is a good approximation for most current experimental situations. 

The polarization vectors are chosen to be in the $(x,y)$ plane such that the mode to interact with the 2DEG. The polarization vectors have to be orthogonal and we choose $\bm{\varepsilon}_1=\mathbf{e}_x$ and $\bm{\varepsilon}_2=\mathbf{e}_y$. Under these assumptions the Pauli-Fierz Hamiltonian of Eq.~(\ref{Pauli-Fierz}), after expanding the covariant kinetic energy, is
\begin{eqnarray}\label{single mode Hamiltonian}
\hat{H}&=&\sum\limits^{N}_{j=1}\left[-\frac{\hbar^2}{2m_{\textrm{e}}}\nabla^2_j +\frac{\textrm{i}e\hbar}{m_{\textrm{e}}} \hat{\mathbf{A}}\cdot\nabla_j\right]\nonumber\\ &+&\underbrace{ \frac{Ne^2}{2m_{\textrm{e}}} \hat{\mathbf{A}}^2+\sum\limits^{2}_{\lambda=1}\hbar\omega\left(\hat{a}^{\dagger}_{\lambda}\hat{a}_{\lambda}+\frac{1}{2}\right)}_{\hat{H}_p},
\end{eqnarray}
and the quantized vector potential of Eq.~(\ref{AinCoulomb}) is
\begin{eqnarray}\label{AinDipole}
\hat{\bi{A}}=\left(\frac{\hbar}{\epsilon_0V}\right)^{\frac{1}{2}}\sum^{2}_{\lambda=1}\frac{\bm{\varepsilon}_{\lambda}}{\sqrt{2\omega}}\left( \hat{a}_{\lambda}+\hat{a}^{\dagger}_{\lambda}\right).
\end{eqnarray}
In the Hamiltonian of Eq.~(\ref{single mode Hamiltonian}) we have a purely photonic part $\hat{H}_p$ which depends only on the annihilation and creation operators of the photon field $\{\hat{a}^{\dagger}_{\lambda},\hat{a}_{\lambda}\}$. Substituting the expression for the vector potential $\hat{\mathbf{A}}$ given by Eq.~(\ref{AinDipole}) and introducing the diamagnetic shift $\omega_p$
\begin{eqnarray}\label{plasma frequency}
    \omega_p=\sqrt{\frac{e^2 n_{\textrm{e}}}{m_{\textrm{e}}\epsilon_0}}=\sqrt{\frac{e^2n_{\textrm{2D}}}{m_{\textrm{e}}\epsilon_0L_z}},
\end{eqnarray}
the photonic part $\hat{H}_p$ takes the form
\begin{eqnarray}\label{photonicpart}
\hat{H}_p=\sum^{2}_{\lambda=1}\left[\hbar\omega\left(\hat{a}^{\dagger}_{\lambda}\hat{a}_{\lambda}+\frac{1}{2}\right)+\frac{\hbar \omega^2_p}{4\omega}\left(\hat{a}_{\lambda}+\hat{a}^{\dagger}_{\lambda}\right)^2\right].
\end{eqnarray}
The diamagnetic shift $\omega_p$ is induced due to the collective coupling of the full electron density $n_{\textrm{e}}=N/V$ to the transversal quantized field~\cite{rokaj2017, rokaj2019, todorov2010, todorov2012, Todorov2015, faisal1987}. This means that $\omega_p=\sqrt{e^2 n_{\textrm{e}}/m_{\textrm{e}}\epsilon_0}$ is the plasma frequency in the cavity. We note that the electron density $n_{\textrm{e}}=N/V$ is defined via the 2D electron density of the material inside the cavity $n_{2\textrm{D}}=N/S$ and the distance between the mirrors of the cavity $L_z$ as $n_{\textrm{e}}=n_{2\textrm{D}}/L_z$.

The photonic part $\hat{H}_p$ can be brought into diagonal form by introducing a new set of bosonic operators $\{\hat{b}^{\dagger}_{\lambda},\hat{b}_{\lambda}\}$
\begin{eqnarray}\label{boperators}
\hat{b}_{\lambda}&=&\frac{1}{2\sqrt{\omega\widetilde{\omega}}}\left[\hat{a}_{\lambda}\left(\widetilde{\omega}+\omega\right)+\hat{a}^{\dagger}_{\lambda}\left(\widetilde{\omega}-\omega\right)\right]\\
\hat{b}^{\dagger}_{\lambda}&=&\frac{1}{2\sqrt{\omega\widetilde{\omega}}} \left[\hat{a}_{\lambda}\left(\widetilde{\omega}-\omega\right)+\hat{a}^{\dagger}_{\lambda}\left(\widetilde{\omega}+\omega\right)\right].\nonumber
\end{eqnarray}
where the frequency
\begin{eqnarray}\label{plasmon polariton}
    \widetilde{\omega}=\sqrt{\omega^2+\omega^2_p}
\end{eqnarray}
is a dressed frequency which depends on the cavity frequency $\omega$ and the diamagnetic shift (or plasma frequency) $\omega_p$. Thus, the dressed frequency $\widetilde{\omega}$ should be interpreted as a $\textit{plasmon-polariton}$ frequency, and as we will show in section~\ref{Cavity Responses} it corresponds to a plasmon-polariton excitation (or resonance) of the system.  The operators $\{\hat{b}_{\lambda},\hat{b}^{\dagger}_{\lambda}\}$ satisfy bosonic commutation relations $[\hat{b}_{\lambda},\hat{b}^{\dagger}_{\lambda^{\prime}}]=\delta_{\lambda,\lambda^{\prime}}$ for $\lambda,\lambda^{\prime}=1,2$. In terms of this new set of operators the photonic part $\hat{H}_p$ of our Hamiltonian, is equal to the sum of two non-interacting harmonic oscillators
\begin{equation}\label{Hpinb}
    \hat{H}_{p}=\sum^{2}_{\lambda=1}\hbar\widetilde{\omega}\left(\hat{b}^{\dagger}_{\lambda}\hat{b}_{\lambda}+\frac{1}{2}\right)
\end{equation}
and the quantized vector potential $\hat{\bi{A}}$ is
\begin{eqnarray}\label{Ainb}
\hat{\bi{A}}=\left(\frac{\hbar}{\epsilon_0V}\right)^{\frac{1}{2}}\sum^{2}_{\lambda=1}\frac{\bm{\varepsilon}_{\lambda}}{\sqrt{2\widetilde{\omega}}}\left( \hat{b}_{\lambda}+\hat{b}^{\dagger}_{\lambda}\right).
\end{eqnarray}
From this expression we see that the vector potential $\hat{\bi{A}}$ got renormalized and depends on the dressed frequency $\widetilde{\omega}$~\cite{rokaj2019}. Substituting back into the Hamiltonian of Eq.~(\ref{single mode Hamiltonian}) the expressions for the photonic part $\hat{H}_p$ and the vector potential $\hat{\bi{A}}$ given by Eqs.~(\ref{Hpinb}) and~(\ref{Ainb}) respectively, and introducing the parameter $g$
\begin{eqnarray}\label{g coupling}
g=\frac{e\hbar}{m_{\textrm{e}}}\left(\frac{\hbar}{\epsilon_0V2\widetilde{\omega}}\right)^{\frac{1}{2}},
\end{eqnarray}
the Hamiltonian of Eq.~(\ref{single mode Hamiltonian}) looks as
\begin{eqnarray}\label{H in bs}
\hat{H}&=&-\frac{\hbar^2}{2m_{\textrm{e}}}\sum^N_{j=1}\nabla^2_j+\sum^2_{\lambda=1}\hbar\widetilde{\omega}\left(\hat{b}^{\dagger}_{\lambda}\hat{b}_{\lambda}+\frac{1}{2}\right) .\nonumber\\
&+&\textrm{i}g\sum^2_{\lambda=1}\left( \hat{b}_{\lambda}+\hat{b}^{\dagger}_{\lambda}\right)\bm{\varepsilon}_{\lambda}\cdot \sum^N_{j=1}\nabla_j.
\end{eqnarray}
The parameter $g$ in Eq.~(\ref{g coupling}) can be interpreted as the single-particle light-matter coupling constant. The Hamiltonian is invariant under translations in the electronic configuration space, since it only includes the momentum operator of the electrons. This implies that $\hat{H}$ commutes with the momentum operator $\nabla$, $[\hat{H},\nabla]$=0, and they share eigenfunctions. As we already stated, for the electrons we employ periodic boundary conditions~\cite{Sommerfeld1928, Mermin}. Thus, the eigenfunctions of the momentum operator $\nabla$ and the Hamiltonian are plane waves of the form
\begin{eqnarray}\label{single electron}
    \phi_{\mathbf{k}_j}(\mathbf{r}_j)=\frac{e^{\textrm{i}\mathbf{k}_j\cdot \mathbf{r}_j} }{\sqrt{S}} \;\; \textrm{with} \;\; 1\leq j \leq N . 
\end{eqnarray}
where $\mathbf{k}_j=2\pi (n^x_{j}/L,n^y_{j}/L)$ are the momenta of the electrons, with $\mathbf{n}_j=\left(n^x_j,n^y_j\right)\in \mathbb{Z}^2$, and $S=L^2$ is the areas of the material inside the cavity depicted in Fig.~\ref{HEG_Cavity}. The wavefunctions of Eq.~(\ref{single electron}) are the single-particle eigenfunctions. But the electrons are fermions and the many-body wavefunction must be antisymmetric under exchange of any two electrons. To satisfy the fermionic statistics we use a Slater determinant $\Phi(\bi{r}_1\sigma_1,..,\bi{r}_N\sigma_N)$  built out of the single-particle eigenfunctions of Eq.~(\ref{single electron}). For convenience we denote this Slater determinant as 
\begin{eqnarray}\label{Slater determinant}
\Phi_{\bi{K}}\equiv \Phi(\bi{r}_1\sigma_1,..,\bi{r}_N\sigma_N),    
\end{eqnarray}
where $\bi{K}=\sum_j\bi{k}_j$ is the collective momentum of the electrons. This makes the notation shorter but also indicates the fact that the ground state and the excited states of the system depend on the distribution of the electrons in $\bi{k}$-space and particularly on the collective momentum $\bi{K}$. Applying $\hat{H}$ of Eq.~(\ref{H in bs}) on the wavefunction $\Phi_{\bi{K}}$ we obtain
\begin{eqnarray}\label{H on Psik}
    \hat{H}\Phi_{\bi{K}}=&\Bigg\{&\sum^2_{\lambda=1}\left[\hbar\widetilde{\omega}\left(\hat{b}^{\dagger}_{\lambda}\hat{b}_{\lambda}+\frac{1}{2}\right)-g\left( \hat{b}_{\lambda}+\hat{b}^{\dagger}_{\lambda}\right)\bm{\varepsilon}_{\lambda}\cdot\bi{K}\right]\nonumber\\
    &+& \frac{\hbar^2}{2m_{\textrm{e}}}\sum^N_{j=1}\bi{k}^2_j\Bigg\} \Phi_{\bi{K}} \;\; \textrm{where}\;\; \bi{K}=\sum^N_{j=1}\bi{k}_j.
\end{eqnarray}
Defining now another set of annihilation and creation operators $\{\hat{c}^{\dagger}_{\lambda},\hat{c}_{\lambda}\}$
\begin{eqnarray}\label{c operators}
    \hat{c}_{\lambda}=\hat{b}_{\lambda}-\frac{g\bm{\varepsilon}_{\lambda}\cdot \bi{K}}{\hbar\widetilde{\omega}} \;\;\; \textrm{and}\;\;\; \hat{c}^{\dagger}_{\lambda}=\hat{b}^{\dagger}_{\lambda}-\frac{g\bm{\varepsilon}_{\lambda}\cdot \bi{K}}{\hbar\widetilde{\omega}},
\end{eqnarray}
the operator $\hat{H}\Phi_{\bi{K}}$ given by Eq.~(\ref{H on Psik}) simplifies as follows 
\begin{eqnarray}\label{Projection H}
    \hat{H}\Phi_{\bi{K}}=&\Bigg\{& \sum^2_{\lambda=1}\left[\hbar\widetilde{\omega}\left(\hat{c}^{\dagger}_{\lambda}\hat{c}_{\lambda}+\frac{1}{2}\right)-\frac{g^2}{\hbar\widetilde{\omega}}\left(\bm{\varepsilon}_{\lambda}\cdot \mathbf{K}\right)^2\right]+\nonumber\\
    &+&\frac{\hbar^2}{2m_{\textrm{e}}}\sum\limits^{N}_{j=1}\mathbf{k}^2_j\Bigg\} \Phi_{\bi{K}}.
\end{eqnarray}
 The operators defined in Eq.~(\ref{c operators}) also satisfy bosonic commutation relations $[\hat{c},\hat{c}^{\dagger}_{\lambda^{\prime}}]=\delta_{\lambda\lambda^{\prime}}$ for $\lambda, \lambda^{\prime}=1,2$. For the quadratic operator $\hat{H}_{\lambda}=\hbar\widetilde{\omega}\left(\hat{c}^{\dagger}_{\lambda}\hat{c}_{\lambda}+1/2\right)$ which is of the form of a harmonic oscillator we know that the full set of eigenstates is given by the expression\cite{GriffithsQM}
\begin{eqnarray}\label{c eigenstates}
    |n_{\lambda},\bm{\varepsilon}_{\lambda}\cdot\bi{K}\rangle_{\lambda}=\frac{(\hat{c}^{\dagger}_{\lambda})^{n_{\lambda}}}{\sqrt{n_{\lambda}!}}|0,\bm{\varepsilon}_{\lambda}\cdot\mathbf{K}\rangle_{\lambda} \;\; \textrm{with}\;\; n_{\lambda}\in\mathbb{Z}, \lambda=1,2\nonumber\\
\end{eqnarray}
where $|0,\bm{\varepsilon}_{\lambda}\cdot\mathbf{K}\rangle_{\lambda}$ is the ground state of $\hat{H}_{\lambda}$, which gets annihilated by $\hat{c}_{\lambda}$~\cite{GriffithsQM}, and the eigenergies of $\hat{H}_{\lambda}$ are $\hbar\widetilde{\omega}(n_{\lambda}+1/2)$. The $\hat{H}\Phi_{\bi{K}}$ given by Eq.~(\ref{Projection H}) in terms of the operators $\{\hat{c}^{\dagger}_{\lambda},\hat{c}_{\lambda}\}$ contains only the sum over $\hat{H}_{\lambda}$ and consequently applying $\hat{H}\Phi_{\bi{K}}$ on the states $\prod_{\lambda}|n_{\lambda},\bm{\varepsilon}_{\lambda}\cdot\bi{K}\rangle_{\lambda}$ we obtain 
\begin{widetext}
\begin{eqnarray}\label{eigenvalue eigenstate Equation}
    \hat{H}\left[\Phi_{\bi{K}}   \prod^2_{\lambda=1}|n_{\lambda},\bm{\varepsilon}_{\lambda}\cdot\bi{K}\rangle_{\lambda}
    \right]=\left( \sum^2_{\lambda=1}\left[\hbar\widetilde{\omega}\left(n_{\lambda}+\frac{1}{2}\right)-\frac{g^2\left(\bm{\varepsilon}_{\lambda}\cdot \mathbf{K}\right)^2}{\hbar\widetilde{\omega}}\right]+\sum\limits^{N}_{j=1}\frac{\hbar^2\mathbf{k}^2_j}{2m_{\textrm{e}}}\right) \left[\Phi_{\bi{K}}   \prod^2_{\lambda=1}|n_{\lambda},\bm{\varepsilon}_{\lambda}\cdot\bi{K}\rangle_{\lambda}\right].
    \end{eqnarray}
\end{widetext}
From the above equation we conclude that the full set of eigenstates of the electron-photon hybrid system described by the Hamiltonian of Eq.~(\ref{single mode Hamiltonian}) is
\begin{eqnarray}\label{eigenstates}
   \Phi_{\bi{K}}  \prod^2_{\lambda=1}|n_{\lambda},\bm{\varepsilon}_{\lambda}\cdot\bi{K}\rangle_{\lambda} \;\; \textrm{with}\;\; \lambda=1,2\;\; \textrm{and}\;\;  \bi{K}=\sum^N_{j=1}\bi{k}_j\nonumber\\
\end{eqnarray}
and its eigenspectrum is. 
\begin{eqnarray}\label{eigenspectrum}
   E_{n_{\lambda},\bi{k}}=\sum^2_{\lambda=1}\left[\hbar\widetilde{\omega}\left(n_{\lambda}+\frac{1}{2}\right)-\frac{\gamma}{N}\frac{\left(\bm{\varepsilon}_{\lambda}\cdot \hbar\bi{K}\right)^2}{2m_{\textrm{e}}}\right]+\sum\limits^{N}_{j=1}\frac{\hbar^2\mathbf{k}^2_j}{2m_{\textrm{e}}}\nonumber\\
\end{eqnarray}
It is important to mention that the electron-photon eigenstates constitute a correlated eigenbasis, because the bosonic eigenstates $|n_{\lambda},\bm{\varepsilon}_{\lambda}\cdot \bi{K}\rangle_{\lambda}$ depend on the collective momentum of the electrons $\bi{K}$. Moreover, from the expression of the eigenspectrum we see that there is a negative term which is proportional to the square of the collective momentum of the electrons $\sim\left(\bm{\varepsilon}_{\lambda}\cdot \hbar\bi{K}\right)^2 $. This is an all-to-all photon-mediated interaction between the electrons in which the momentum of each electron couples to the momenta of all the others. This photon-mediated interaction as we will see in section~\ref{Effective Field Theory} has implications for the effective electron mass and the quasiparticle excitations of this Fermi liquid.  

To obtain the expression of Eq.~(\ref{eigenspectrum}) we substituted in Eq.~(\ref{eigenvalue eigenstate Equation}), the definition of the the single-particle coupling $g$ given by Eq.~(\ref{g coupling}), and we introduced the parameter $\gamma$
\begin{eqnarray}\label{collective coupling}
   \gamma=\frac{2m_{\textrm{e}}N}{\hbar^2}\frac{g^2}{\hbar\widetilde{\omega}}=\frac{\omega^2_p}{\widetilde{\omega}^2}=\frac{\omega^2_p}{\omega^2+\omega^2_p}\leq 1.
\end{eqnarray}
The parameter $\gamma$ can be viewed as the collective coupling of the electron gas to the cavity mode and depends on the cavity frequency and the full electron density $n_{\textit{e}}$ via the frequency $\omega_p$ defined in Eq.~(\ref{plasma frequency}). This implies that the more charges in the system the stronger the coupling between light and matter in the cavity. Further, we note that the collective coupling parameter $\gamma$ is dimensionless and most importantly $\gamma$ has an upper bound and cannot be larger than one. As we will see in section~\ref{Instability and A2} this upper bound guarantees the stability of the system. Lastly, we highlight that also in the case of a multi-mode quantized field, with the mode-mode interactions included, the structure of the many-body spectrum stays the same with the one in Eq.~(\ref{eigenspectrum}), but with a different coupling constant, frequencies and polarizations (due to the mode-mode interactions) and a sum over all the modes~\cite{faisal1987}. This is shown in detail in appendix~\ref{Mode-Mode Interactions}. 

\section{Ground State in the Large $N$ Limit}\label{Ground State}

\begin{figure}
\includegraphics[height=5cm,width=6cm]{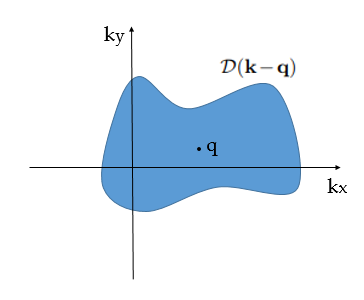}
\caption{\label{Distribution Kspace}Schematic depiction of a generic distribution $\mathcal{D}(\bi{k}-\bi{q})$ in $\bi{k}$-space. The shape $\mathcal{D}$ of the distribution as well as its origin $\bi{q}$ are arbitrary. To find the ground state distribution in $\bi{k}$-space one needs to minimize the energy density of the system with respect to both the shape $\mathcal{D}$ and the origin $\bi{q}$. }
\end{figure}

Having diagonalized the Hamiltonian of Eq.~(\ref{single mode Hamiltonian}) we want now to find the ground state of this many-body system in the large $N$ limit. For this we need to minimize the energy of the many-body spectrum given by Eq.~(\ref{eigenspectrum}) in the limit where the number of electrons $N$ and the area $S$ become arbitrarily large and approach the thermodynamic limit, but in such a way that the 2D electron density $n_{2\textrm{D}}=N/S$ stays fixed. The electron density can be defined by the number of allowed states in a region of $\bi{k}$-space, of volume $\Omega_{\mathcal{D}}$ with respect to a distribution $\mathcal{D}$ in $\bi{k}$-space~\cite{Mermin}. The number of states in the volume $\Omega_{\mathcal{D}}$ is: $\#\textrm{states}=\Omega_{\mathcal{D}}S/(2\pi)^2$. The volume $\Omega_{\mathcal{D}}$ with respect to an arbitrary distribution $\mathcal{D}(\mathbf{k}-\mathbf{q})$  whose origin $\mathbf{q}$ is also arbitrary (see Fig.~\ref{Distribution Kspace}) is 
\begin{eqnarray}
\Omega_{\mathcal{D}}=\iint \limits_{-\infty}^{+\infty} \mathcal{D}(\mathbf{k}-\mathbf{q}) d^2k=\iiint \limits_{-\infty}^{+\infty} \mathcal{D}(\mathbf{k}^{\prime}) d^2k^{\prime},
\end{eqnarray}
where we performed the shift $\bi{k}^{\prime}=\bi{k}-\bi{q}$. The number of electrons $N$ we can accommodate in the volume $\Omega_{\mathcal{D}}$ is 2 times (due to spin degeneracy) the number of allowed states. Thus, the 2D electron density is $n_{2\textrm{D}}=2\Omega_{\mathcal{D}}/(2\pi)^2$.

The energy of Eq.~(\ref{eigenspectrum}) minimizes for $n_{\lambda}=0$ for both $\lambda=1,2$. Thus, the photonic contribution to the ground state energy is constant $E_p=\hbar\widetilde{\omega}$ and does not influence the electrons in $\bi{k}$-space. Then, the ground state energy $E_{gs}(\equiv E_{0,\bi{k}})=E_p+E_{\bi{k}}$ is the sum of the photonic contribution $E_p$ and the part which depends on the electronic momenta $E_{\bi{k}}$, which includes two terms: a positive one, which is the sum over the kinetic energies of all the electrons and we denote by $T$, and a negative one which is minus the square of the collective momentum $\bi{K}=\sum_j\bi{k}_j$. To find the ground state we need to minimize the energy density $E_{\bi{k}}/S$ with respect to the distribution $\mathcal{D}(\bi{k}-\bi{q})$. In the large $N,S$ limit the sums in the expression for the energy density $E_{\bi{k}}/S$ turn into integrals. Thus, the kinetic energy density $T/S$ (with doubly occupied momenta) is~\cite{Mermin}
\begin{eqnarray}\label{Kinetic energy}
\frac{T}{S}&=&\frac{\hbar^2}{2m_{\textrm{e}}}\lim_{S\to\infty}\frac{2}{S}\sum_{\mathbf{k}}\bi{k}^2=\frac{\hbar^2}{2m_{\textrm{e}}}\frac{2}{(2\pi)^2}\iint \limits^{+\infty}_{-\infty} d^2k \mathcal{D}(\mathbf{k}-\mathbf{q})\mathbf{k}^2\nonumber\\
\end{eqnarray}
and after performing the transformation $\bi{k}^{\prime}=\bi{k}-\bi{q}$ we obtain 
\begin{widetext}
\begin{eqnarray}\label{TperV}
\frac{T}{S}&=&\frac{\hbar^2}{2m_{\textrm{e}}}\Bigg[\underbrace{\frac{2}{(2\pi)^2}\iint \limits^{+\infty}_{-\infty} d^2k^{\prime} \mathcal{D}(\mathbf{k}^{\prime})\left(\mathbf{k}^{\prime}\right)^2}_{t_{\mathcal{D}}}+2\mathbf{q}\underbrace{\frac{2}{(2\pi)^2}\iint \limits^{+\infty}_{-\infty} d^2k^{\prime} \mathcal{D}(\mathbf{k}^{\prime})\mathbf{k}^{\prime}}_{\mathbf{K}_{\mathcal{D}}}+\mathbf{q}^2\underbrace{\frac{2}{(2\pi)^2}\iint \limits^{+\infty}_{-\infty} d^2k^{\prime} \mathcal{D}(\mathbf{k}^{\prime})}_{n_{2\textrm{D}}}\Bigg]=\nonumber\\
&=&\frac{\hbar^2}{2m_{\textrm{e}}}\left(t_{\mathcal{D}}+2\mathbf{q}\cdot\mathbf{K}_{\mathcal{D}}+\mathbf{q}^2n_{2\textrm{D}}\right).
\end{eqnarray}
\end{widetext}
The term $t_{\mathcal{D}}$ is the kinetic energy of free electrons with respect to a distribution centered at zero $\mathcal{D}(\mathbf{k}^{\prime})$~\cite{Mermin}. The term $\mathbf{K}_{\mathcal{D}}$ is the collective momentum of the electrons with respect to $\mathcal{D}(\mathbf{k}^{\prime})$, and $\mathbf{q}^2 n_{2\textrm{D}}$ is the kinetic energy due to the arbitrary origin of the distribution (see Fig.~\ref{Distribution Kspace}). This last term depends on the 2D density $n_{2\textrm{D}}$ and the origin $\bi{q}$, but not on the shape of the distribution $\mathcal{D}$. 
Let us compute now the negative term appearing in Eq.~(\ref{eigenspectrum}). The square of the collective momentum per area $\left(\bm{\varepsilon}_{\lambda}\cdot\bi{K}/S\right)^2$ (for doubly occupied momenta) in the large $N$ limit is
\begin{eqnarray}
\left(\frac{\bm{\varepsilon}_{\lambda}\cdot \bi{K}}{S}\right)^2&=&\left(\frac{2}{(2\pi)^2}\iint \limits^{+\infty}_{-\infty} d^2k \mathcal{D}(\mathbf{k}-\mathbf{q})\bm{\varepsilon}_{\lambda}\cdot\mathbf{k}\right)^2.
\end{eqnarray}
Performing the transformation $\bi{k}^{\prime}=\bi{k}-\bi{q}$ and multiplying by the area $S$ we find 
\begin{eqnarray}\label{negative term}
 \frac{\left(\bm{\varepsilon}_{\lambda}\cdot\mathbf{K}\right)^2}{S}=S\left(\bm{\varepsilon}_{\lambda}\cdot\mathbf{K}_{\mathcal{D}}+\bm{\varepsilon}_{\lambda}\cdot\mathbf{q}n_{2\textrm{D}}\right)^2.
  \end{eqnarray}
 Summing the two contributions which we computed in Eqs.~(\ref{TperV}) and~(\ref{negative term}) we find the energy density as function of the shape of the distribution $\mathcal{D}$ and the origin $\bi{q}$
 \begin{widetext}
 \begin{eqnarray}\label{energy density}
  \mathcal{E}[\mathcal{D}]\equiv\frac{E_{\bi{k}}}{S}=\frac{\hbar^2}{2m_{\textrm{e}}}\left[t_{\mathcal{D}}+2\mathbf{q}\cdot\mathbf{K}_{\mathcal{D}}+\mathbf{q}^2n_{2\textrm{D}}-\frac{\gamma}{n_{2\textrm{D}}}\sum^2_{\lambda=1}\left(\bm{\varepsilon}_{\lambda}\cdot\mathbf{K}_{\mathcal{D}}+\bm{\varepsilon}_{\lambda}\cdot\mathbf{q}n_{2\textrm{D}}\right)^2\right].
 \end{eqnarray}
 \end{widetext}
 The energy density has to be minimized with respect to the origin of the distribution $\bi{q}=(q_x,q_y)$. For that we compute the derivative of the energy density $ \mathcal{E}[\mathcal{D}]$ with respect to $\bi{q}$ \begin{eqnarray}\label{optimal q}
  \frac{\partial \mathcal{E}[\mathcal{D}]}{\partial \bi{q}}=(1-\gamma)\left(\bi{K}_{\mathcal{D}}+\bi{q}n_{2\textrm{D}}\right)=0 \Longrightarrow  \bi{q}_0=-\frac{\bi{K_{\mathcal{D}}}}{n_{2\textrm{D}}}.\nonumber\\
 \end{eqnarray}
 The optimal origin $\bi{q}_0$ is independent of the coupling $\gamma$, and substituting $\bi{q}_0$ into Eq.~(\ref{energy density}) we find
 \begin{eqnarray}\label{optimal energy density}
  \mathcal{E}[\mathcal{D}]|_{\bi{q_0}}=\frac{\hbar^2}{2m_{\textrm{e}}}\left[t_{\mathcal{D}}-\frac{\bi{K}^2_{\mathcal{D}}}{n_{2\textrm{D}}}\right].
 \end{eqnarray}
 The remaining task now is to optimize the energy density with respect to the shape $\mathcal{D}$ of the  distribution. In general to perform such a minimization it is not an easy task. Thus, to find the optimal $\bi{k}$-space distribution we will use some physical intuition.
 
The energy density $\mathcal{E}[\mathcal{D}]$ (as well as $\bi{q}_0$) given by Eq.~(\ref{optimal energy density}) is independent of the coupling constant $\gamma$. This indicates that the ground state and the ground state energy in the thermodynamic limit are independent of the coupling to the cavity. Driven by this observation let us compare the energy density in Eq.~(\ref{optimal energy density}) with the energy density of the original free electrons gas~\cite{Mermin} without any coupling to a cavity mode. 
 
\begin{figure}
\includegraphics[height=5cm,width=6cm]{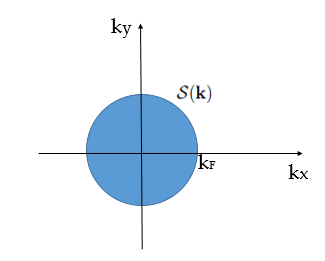}
\caption{\label{Fermi Sphere} Graphic representation of the ground state distribution of the 2DEG not coupled to a cavity. The ground state distribution is the 2D Fermi sphere $\mathcal{S}(\bi{k})$ (circle) with radius $|\bi{k}_{\textrm{F}}|$ (Fermi wavevector). For the 2DEG coupled to the cavity we find that the ground state distribution in $\bi{k}$-space is also the 2D Fermi sphere $\mathcal{S}(\bi{k})$ with radius $|\bi{k}_{\textrm{F}}|$.}
\end{figure}

In the original free electron model the energy of the system is the sum over the kinetic energies of all the electrons $E^{nc}_{\bi{k}}=\sum_{j}\hbar^2\bi{k}_j/2m_{\textrm{e}}$~\cite{Sommerfeld1928,Mermin}, and due to rotational symmetry the ground state momentum distribution is the standard Fermi sphere $\mathcal{S}(\bi{k})$~\cite{Mermin}, which in our case is a 2D sphere (circle) as shown in Fig.~\ref{Fermi Sphere}. 
 But let us forget for a moment the fact that we know the ground state distribution of the electrons, and
let us consider again a generic distribution in $\bi{k}$-space $\mathcal{D}(\bi{k}-\bi{q})$ as the one shown in Fig.~\ref{Distribution Kspace}. For such a distribution the ground state energy density, as we found in Eq.~(\ref{TperV}), is
\begin{eqnarray}\label{eg no coupling}
 \mathcal{E}^{nc}[\mathcal{D}]=\frac{\hbar^2}{2m_{\textrm{e}}}\left(t_{\mathcal{D}}+2\mathbf{q}\cdot\mathbf{K}_{\mathcal{D}}+\mathbf{q}^2n_{2\textrm{D}}\right).
\end{eqnarray}
Minimizing $\mathcal{E}^{nc}[\mathcal{D}]$ with respect to the origin $\bi{q}$ we find that the optimal origin of the distribution is $\bi{q}_0=-\bi{K}_{\mathcal{D}}/n_{2\textrm{D}}$. This is the same with the one we found in Eq.~(\ref{optimal q}) for the 2DEG coupled to the cavity mode. Substituting $\bi{q}_0=-\bi{K}_{\mathcal{D}}/n_{2\textrm{D}}$ into the expression for the energy density of the uncoupled electron gas in Eq.~(\ref{eg no coupling}) we find $\mathcal{E}^{nc}[\mathcal{D}]|_{\bi{q_0}}$ to be equal to the energy density $\mathcal{E}[\mathcal{D}]|_{\bi{q_0}}$ of the coupled system
\begin{eqnarray}\label{optimal energy no coupling}
 \mathcal{E}^{nc}[\mathcal{D}]|_{\bi{q_0}}= \mathcal{E}[\mathcal{D}]|_{\bi{q_0}}.
\end{eqnarray}
This means that both energy functionals, the coupled and the uncoupled, get minimized by the same $\bi{k}$-space distribution $\mathcal{D}$. For the uncoupled 2DEG, the shape of the distribution in $\bi{k}$-space is the 2D Fermi sphere $\mathcal{S}(\bi{k}-\bi{q}_0)$. For a sphere the collective momentum is zero, $\bi{K}_{\mathcal{S}}=0$, and consequently the optimal origin is also zero $\bi{q}_0=0$. Thus, for the coupled system the ground state momentum distribution is the 2D Fermi sphere $\mathcal{S}(\bi{k})$ centered at zero, as depicted in Fig.~\ref{Fermi Sphere}. Most importantly since the collective momentum is zero the ground state of the 2DEG coupled to the cavity is 
\begin{eqnarray}\label{Thermodynamic GS}
  |\Psi_{gs}\rangle=|\Phi_{0}\rangle \otimes  \prod^2_{\lambda=1}|0,0\rangle_{\lambda},
\end{eqnarray}
where $\Phi_0$ is the Slater determinant given by Eq.~(\ref{Slater determinant}) with zero collective momentum $\bi{K}=0$. It is important to mention that since in the ground state the collective momentum is zero, the ground state is a tensor-product state between the electrons and the photons. The fact that the ground state distribution of the electrons in $\bi{k}$-space is the Fermi sphere implies that the electronic system is a \textit{Fermi liquid}~\cite{LandauFermiLiquid}. Further, having found the ground state of the electrons, we can compute also the ground state energy density of the electrons as a function of the Fermi wavevector and we find
\begin{eqnarray}
 \mathcal{E}[\mathcal{S}]=\frac{\hbar^2}{2m_{\textrm{e}}}\frac{2}{(2\pi)^2}\iint \limits^{+\infty}_{-\infty} d^2k \mathcal{S}(\mathbf{k})\mathbf{k}^2=\frac{\hbar^2k^4_{\textrm{F}}}{16\pi m_{\textrm{e}}}.
\end{eqnarray}

\textit{Mismatch of Energies.}---Moreover, we would like to point out a fundamental discrepancy which appears between the electronic and photonic sector, with respect to their contributions in the ground state energy density. The contribution of the (single-mode) photon field, to the ground state energy, as we can deduce from Eq.~(\ref{eigenspectrum}) is $E_p/S=\hbar\widetilde{\omega}/S$. In the large $N,S$ (or thermodynamic) limit this contribution is miniscule and strictly speaking goes to zero. On the other hand the electrons have a finite energy density $ \mathcal{E}[\mathcal{S}]$. This implies that only the 2DEG contributes to the ground state energy density of the interacting electron-photon hybrid system in the cavity. This energy mismatch shows up because in the electronic sector we have $N$ electrons in the thermodynamic limit, while in the photonic sector we have only one mode. This discrepancy between the two sectors hints towards the fact that for both sectors to contribute on the same level, a continuum of modes of the photon field have to be taken into account such that the photon field to acquire a finite energy density in its ground state. We explore this direction further in section~\ref{Effective Field Theory}. Before we continue we note that the photon field in its highly excited states can still contain arbitrarily large amounts of energy. Yet for the considerations of the ground state these highly-excited photon-states do not play a role. 

From the fact that the ground state of the electrons is the standard Fermi sphere and that the energy density of the photon field in the thermodynamic limit is negligible, one might conclude that the electron-photon ground state of the system is trivial and there are no quantum fluctuation effects due to the electron-photon coupling. However, this is not the case. To classify completely the electron-photon ground state one needs to look also at the ground state photon occupation.

\subsection{Ground State Photon Occupation}

The photon number operator is
\begin{eqnarray}
\hat{N}_{\textrm{ph}}=\sum^2_{\lambda=1}\hat{a}^{\dagger}_{\lambda}\hat{a}_{\lambda}.    
\end{eqnarray}
To compute the ground state photon occupation we need to write the number operator in terms of the bosonic operators $\{\hat{c}^{\dagger}_{\lambda}, \hat{c}_{\lambda}\}$ defined in Eq.~(\ref{c operators}). Using Eqs.~(\ref{boperators}) and~(\ref{c operators}) we find that the number operator in terms of $\hat{c}^{\dagger}_{\lambda}$ and $ \hat{c}_{\lambda}$  is 
\begin{widetext}
\begin{eqnarray}
   \hat{N}_{\textrm{ph}}&=&\sum^2_{\lambda=1} \frac{1}{4\omega\widetilde{\omega}}\Bigg [ \left(\omega^2-\widetilde{\omega}^2\right)\left(\hat{c}_{\lambda}+\frac{g\bm{\varepsilon}_{\lambda}\cdot \bi{K}}{\hbar\widetilde{\omega}}\right)^2 +\left(\omega-\widetilde{\omega}\right)^2\left(\hat{c}_{\lambda}+\frac{g\bm{\varepsilon}_{\lambda}\cdot \bi{K}}{\hbar\widetilde{\omega}}\right)\left(\hat{c}^{\dagger}_{\lambda}+\frac{g\bm{\varepsilon}_{\lambda}\cdot \bi{K}}{\hbar\widetilde{\omega}}\right)\nonumber\\
   &+&\left(\omega+\widetilde{\omega}\right)^2\left(\hat{c}_{\lambda}+\frac{g\bm{\varepsilon}_{\lambda}\cdot \bi{K}}{\hbar\widetilde{\omega}}\right)\left(\hat{c}^{\dagger}_{\lambda}+\frac{g\bm{\varepsilon}_{\lambda}\cdot \bi{K}}{\hbar\widetilde{\omega}}\right)+\left(\omega^2-\widetilde{\omega}^2\right)\left(\hat{c}^{\dagger}_{\lambda}+\frac{g\bm{\varepsilon}_{\lambda}\cdot \bi{K}}{\hbar\widetilde{\omega}}\right)^2\Bigg].
\end{eqnarray}
\end{widetext}
In the ground state the collective momentum is zero, $\bi{K}=0$, and out of all the terms appearing above only the term that first creates and then destroys a bosonic excitation $\hat{c}_{\lambda}\hat{c}^{\dagger}_{\lambda}$ gives a non-zero contribution. Thus, we find for the ground state photon occupation
\begin{eqnarray}
    \langle \hat{N}_{\textrm{ph}} \rangle_{gs}\equiv \langle \Psi_{gs}|\hat{N}_{\textrm{ph}}|\Psi_{gs}\rangle=\frac{\left(\widetilde{\omega}-\omega\right)^2}{2\omega\widetilde{\omega}}.
\end{eqnarray}
From the result above we see that the ground state photon occupation is non-zero. This means that there are virtual photons in the ground state of the interacting electron-photon system. This phenomenon has also been reported for dissipative systems~\cite{DeLiberato2017}. From the fact that the ground state of the 2DEG in the cavity contains photons we conclude that there are quantum fluctuations of the photon field in the ground state due to the electron-photon coupling. Thus, our system is not a trivial Fermi liquid, but rather it is a Fermi liquid dressed with photons.

Further, the ground state photon occupation shows an interesting dependence on the electron density. For electron densities small enough for the plasma frequency $\omega_p=\sqrt{e^2 n_{\textrm{e}}/m_{\textrm{e}}\epsilon_0}$ to be much smaller than the cavity frequency, $\omega_p\ll\omega$, the dressed frequency $\widetilde{\omega}=\sqrt{\omega^2_p+\omega^2}$ is approximately equal to the cavity frequency, $\widetilde{\omega} \approx \omega$. In this case the ground state photon occupation is zero, $ \langle \hat{N}_{\textrm{ph}} \rangle_{gs}=0$. However, for large electronic densities such that $\omega_p \gg \omega$, the dressed frequency is $\widetilde{\omega} \approx \omega_p$ and the numerator in the expression for $\langle \hat{N}_{\textrm{ph}} \rangle_{gs}$ is approximately $\omega^2_p$. Thus, we find that for large electron densities the ground state photon occupation has a square root dependence on the electron density
\begin{eqnarray}
\langle \hat{N}_{\textrm{ph}} \rangle_{gs} \sim \sqrt{n_{\textrm{e}}}.
\end{eqnarray}
This implies that the amount of photons in the ground state increases with the number of electrons. This behavior might be related to the superradiant phase transition~\cite{Lieb} and could potentially provide some insights on how to achieve this phase transition, which remains still elusive.

\section{Critical Coupling, Instability \& the Diamagnetic $\bi{A}^2$ term}\label{Instability and A2}
So far we have examined rigorously and in full generality the behavior of the 2DEG coupled to the cavity, in the regime where the cavity mode $\omega$ is finite and the collective coupling parameter $\gamma$, defined in Eq.~(\ref{collective coupling}), is less than one. But now the following questions arises: what happens in the limit where the frequency of the quantized field goes to zero, $\omega\rightarrow 0$, and the collective coupling parameter takes its maximum value $\gamma\rightarrow 1$? 

\begin{figure}[ht]
\includegraphics[width=\columnwidth]{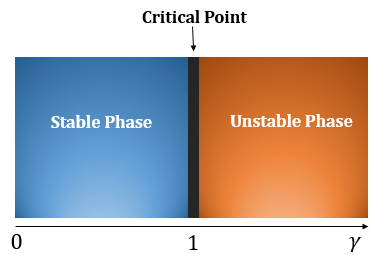}
\caption{\label{Phase Diagram} Phase diagram for the free electron model in cavity QED. The system has a stable ground state for coupling constant $\gamma<1$. At the critical coupling $\gamma_c=1$ the ground state is infinitely degenerate. Beyond the critical coupling $\gamma_c=1$ the system is unstable and the system has no ground state.}
\end{figure}

We will refer to the maximum value of the coupling constant $\gamma$ as \textit{critical coupling}, $\gamma_c=1$, because as we will see at this point an interesting transition happens for the system, from a stable phase to an unstable phase, as it is also summarized by the phase diagram in Fig.~\ref{Phase Diagram}.  

\subsection{Critical Coupling and Infinite Degeneracy}

At the critical coupling $\gamma_c=1$ the energy density $\mathcal{E}[\mathcal{D}]$ given by Eq.~(\ref{energy density}) becomes independent of the origin $\bi{q}$
\begin{eqnarray}\label{critical energy density}
  \mathcal{E}[\mathcal{D}]|_{\gamma_c}=\frac{\hbar^2}{2m_{\textrm{e}}}\left[t_{\mathcal{D}}-\frac{\bi{K}^2_{\mathcal{D}}}{n_{2\textrm{D}}}\right].
\end{eqnarray}
The fact that the energy density becomes degenerate with respect to the origin $\bi{q}$ means that the ground state of the system is not unique. Moreover, Eq.~(\ref{optimal q}) from which we determined the optimal value for the vector $\bi{q}$, gets trivially zero. 

The energy density of Eq.~(\ref{critical energy density}), as we explained in the previous section, minimizes for a sphere $\mathcal{S}(\bi{k}-\bi{q})$. But since the energy density $\mathcal{E}[\mathcal{D}]|_{\gamma_c}$ is degenerate with respect to the origin $\bi{q}$ and the optimal $\bi{q}$ cannot be determined from Eq.~(\ref{optimal q}), \textit{all spheres} of the form $\mathcal{S}(\bi{k}-\bi{q})$ are degenerate and have exactly the same ground state energy. This means that the optimal ground state $\bi{k}$-space distribution it is \textit{not unique} but rather the ground state of the system at the critical coupling $\gamma_c=1$ is \textit{infinitely degenerate} with respect to origin of the $\bi{k}$-space distribution of the electrons. 

Such an infinite degeneracy appears also for a 2D electron gas in the presence of perpendicular, homogeneous magnetic field where we have the Landau levels demonstrating exactly this behavior~\cite{Landau}. The infinite degeneracy is also directly connected to the quantum Hall effect~\cite{Klitzing}. The connection between quantum electrodynamics and the quantum Hall effect has also been explored recently in the context of quantum electrodynamical Bloch theory~\cite{rokaj2019}.

Lastly, we note that the fact that all spheres $\mathcal{S}(\bi{k}-\bi{q})$ of arbitrary origin $\bi{q}$ are degenerate means that the ground state energy of our system, at the critical coupling $\gamma_c$, is invariant under shifts in $\bi{k}$-space, which implies that is invariant under Galilean boosts.

\subsection{No Ground State Beyond the Critical Coupling}\label{No Ground State}

For completeness we would also like to consider the case where the coupling constant goes beyond the critical coupling $\gamma_c$ and becomes larger than one, $\gamma > 1$. In principle from its definition in Eq.~(\ref{collective coupling}) the coupling constant $\gamma$ is not allowed to take such values, but investigating this scenario will provide further physical insight why this should not happen. 

For simplicity and without loss of generality, we simplify our consideration to the case where the cavity field has only one polarization vector $\bm{\varepsilon}_1=\bi{e}_x$ and $\bm{\varepsilon}_2=0$. In this case the energy density $\mathcal{E}[\mathcal{D}]$ given by Eq.~(\ref{energy density}) as a function of the $x$-component of the vector $\bi{q}=(q_x,q_y)$ is
\begin{eqnarray}
  \mathcal{E}(q_x)=\frac{\hbar^2}{2m_{\textrm{e}}}(1-\gamma)\left(2q_xK^x_{\mathcal{D}}+q^2_xn_{2\textrm{D}}\right),
\end{eqnarray}
 where we neglected all terms in Eq.~(\ref{energy density}) independent of $q_x$. For $\gamma>1$ the energy density above has no minimum and it is unbounded from below because $1-\gamma<0$ is negative and taking the limit for $q_x$ to infinity the energy density goes to minus infinity
 \begin{eqnarray}\label{negative divergence}
  \lim_{q_x \to \infty}\mathcal{E}(q_x)=\frac{\hbar^2}{2m_{\textrm{e}}} \lim_{q_x \to \infty}(1-\gamma)\left(2q_xK^x_{\mathcal{D}}+q^2_xn_{\textrm{e}}\right)\rightarrow -\infty.\nonumber\\
\end{eqnarray}
 This proves that the free electron gas coupled to the cavity mode for $\gamma>1$ has no ground state and the system in this case is unstable, because shifting further and further the distribution in $\bi{k}$-space, by moving its center $\bi{q}$, we can lower indefinitely the energy density\footnote{We would like to point out that this argument is similar to the one for the lack of ground state in the length gauge when the dipole self-energy is omitted. In the length gauge the energy can be lowered indefinitely by shifting further and further in real space the charge distribution~\cite{rokaj2017}. }. Thus, we conclude that the upper bound for the collective coupling $\gamma$ given by Eq.~(\ref{collective coupling}) guarantees the stability of the coupled electron-photon system. Lastly, we would like to mention that due to the lack of ground state, equilibrium is not well-defined in the unstable phase and equilibrium phenomena cannot be described properly.
 
\subsection{No-Go Theorem and the $\bi{A}^2$ Term}
In what follows we are interested in the importance of the often neglected~\cite{vukics2014} diamagnetic $\bi{A}^2$ term for the 2DEG coupled to the single mode quantized field. The influence of this quadratic term has been studied theoretically in multiple publications~\cite{schaeferquadratic,vukics2014, bernadrdis2018breakdown, DiStefano2019} and its influence has also been experimentally measured~\cite{li2018}. Moreover, the elimination of the $\bi{A}^2$ is responsible for the notorious superradiant phase transition of the Dicke model~\cite{dicke1954}. The superradiant phase transition was firstly predicted by Hepp and Lieb~\cite{Lieb} for the Dicke model in the thermodynamic limit and soon after derived in an alternative way by Wang and Hioe~\cite{Wang}. The existence though of the superradiant phase was challenged by a no-go theorem~\cite{Birula} which showed that the superradiant phase transition in atomic systems appeared completely due to the fact that the $\bi{A}^2$ term was not taken into account. More recently, another demonstration of a superradiant phase transition was predicted in the framework of circuit QED~\cite{CiutiSuperradiance}, which again was challenged by another no-go theorem which applied also to circuit QED systems~\cite{MarquardtNoGO}. Also the inclusion of qubit-qubit interactions was shown to be important for such circuit QED systems~\cite{Jaako2016}. Further, the application of these no-go theorems it is argued that it depends on the gauge choice~\cite{Stokes2020, DeBernardis2018, bernadrdis2018breakdown}. Nevertheless, the debate over the existence of the superradiant phase transition is still ongoing, with new demonstrations coming from the field of cavity QED materials~\cite{HagenmullerSPT,MazzaSuperradiance} accompanied though by the respective no-go theorems~\cite{ChirolliNoGO, AndolinaNoGo}. Lastly, the possibility of a superradiant phase transition beyond the dipole approximation has also been investigated~\cite{Guerci2020, Andolina2020}.

For our system to examine the importance of the diamagnetic $\bi{A}^2$ term here, we study the free electron gas coupled to the cavity \textit{in the absence} of the $\bi{A}^2$ term. From the Hamiltonian $\hat{H}$ in Eq.~(\ref{single mode Hamiltonian}) it is straightforward to derive the Hamiltonian $\hat{H}^{\prime}$ for the electron gas coupled to the cavity mode when the $\bi{A}^2$ term is neglected  $\hat{H}^{\prime}=\hat{H}-Ne^2\hat{\bi{A}}^2/2m_{\textrm{e}}$ 
 \begin{eqnarray}\label{noA2}
	\hat{H}^{\prime}&=&\sum\limits^{N}_{j=1}\left[-\frac{\hbar^2\nabla^2_j}{2m_{\textrm{e}}} +\frac{\textrm{i}e\hbar}{m_{\textrm{e}}} \hat{\mathbf{A}}\cdot\nabla_j\right]+\sum^2_{\lambda=1}\hbar\omega \left[\hat{a}^{\dagger}_{\lambda}\hat{a}_{\lambda}+\frac{1}{2}\right].\nonumber\\
\end{eqnarray}
As we explained in section~\ref{EG in cavity QED} in the electronic configuration space we have translational symmetry, and the electronic eigenfunction is the Slater determinant given by Eq.~(\ref{Slater determinant}). Introducing now the parameter
\begin{eqnarray}\label{g no a2}
g^{\prime}=\frac{e\hbar}{m_{\textrm{e}}}\left(\frac{\hbar}{2\epsilon_0 \omega V}\right)^{1/2},
\end{eqnarray}
applying the Hamiltonian $\hat{H}^{\prime}$ on the Slater determinant $\Phi_{\bi{K}}$ and substituting the definition for quantized field $\hat{\bi{A}}$ given by Eq.~(\ref{AinDipole}) we obtain
 \begin{eqnarray}\label{H no A2 on Psik}
    \hat{H}^{\prime}\Phi_{\bi{K}}=&\Bigg\{&\sum^2_{\lambda=1}\left[\hbar\omega\left(\hat{a}^{\dagger}_{\lambda}\hat{a}_{\lambda}+\frac{1}{2}\right)-g^{\prime}\left( \hat{a}_{\lambda}+\hat{a}^{\dagger}_{\lambda}\right)\bm{\varepsilon}_{\lambda}\cdot\bi{K}\right]\nonumber\\
    &+& \frac{\hbar^2}{2m_{\textrm{e}}}\sum^N_{j=1}\bi{k}^2_j\Bigg\} \Phi_{\bi{K}} \;\; \textrm{where}\;\; \bi{K}=\sum^N_{j=1}\bi{k}_j.
\end{eqnarray}
The Hamiltonian $\hat{H}^{\prime}$ is of exactly the same form as $\hat{H}$ of Eq.~(\ref{H on Psik}). Following exactly the same procedure for diagonalizing $\hat{H}$, which we showed in section~\ref{EG in cavity QED}, we can diagonalize also $\hat{H}^{\prime}$ and we find that its eigenspectrum is
\begin{eqnarray}\label{eigenspectrum no A2}
   E_{n_{\lambda},\bi{k}}=\sum^2_{\lambda=1}\left[\hbar\omega\left(n_{\lambda}+\frac{1}{2}\right)-\frac{\gamma^{\prime}}{N}\frac{\hbar^2\left(\bm{\varepsilon}_{\lambda}\cdot \mathbf{K}\right)^2}{2m_{\textrm{e}}}\right]+\sum\limits^{N}_{j=1}\frac{\hbar^2\mathbf{k}^2_j}{2m_{\textrm{e}}}\nonumber\\
\end{eqnarray}
where we substituted the parameter $g^{\prime}$ of Eq.~(\ref{g no a2}) and we introduced the parameter $\gamma^{\prime}$
\begin{eqnarray}
   \gamma^{\prime}=\frac{2m_{\textrm{e}}N}{\hbar^2}\frac{(g^{\prime})^2}{\hbar\omega}=\frac{\omega^2_p}{\omega^2}
\end{eqnarray}
in complete analogy to the coupling constant $\gamma$ given by Eq.~(\ref{collective coupling}). The dressed frequency $\widetilde{\omega}$ does not show up anymore neither in the coupling $\gamma^{\prime}$ nor in the energy spectrum~(\ref{eigenspectrum no A2}), because the quantized field and the energy of the cavity mode do not get renormalized by the $\bi{A}^2$, since it is absent. Comparing now the spectrum of Eq.~(\ref{eigenspectrum no A2}) for the Hamiltonian $\hat{H}^{\prime}$,  with the spectrum given by Eq.~(\ref{eigenspectrum}) derived for the Hamiltonian $\hat{H}$ of Eq.~(\ref{single mode Hamiltonian}) which included the $\bi{A}^2$ term, we see that they are exactly the same, up to replacing $\widetilde{\omega}$ with $\omega$ and $\gamma$ with $\gamma^{\prime}$. The last one is a very important difference, because the coupling constant $\gamma^{\prime}$ has no upper bound and can be arbitrarily large, as $\omega_p$ can be larger than $\omega$. In section~\ref{No Ground State} we proved that the spectrum of the form given by Eq.~(\ref{eigenspectrum no A2}), has no ground state if the coupling constant gets larger than one. For large densities $\omega_p$ can become larger than $\omega$ and $\gamma^{\prime}$ will be larger than one, $\gamma^{\prime}>1$. Consequently, the Hamiltonian $\hat{H}^{\prime}$ will be unstable and will not have a ground state. 

This highlights that eliminating the diamagnetic $\bi{A}^2$ term, is a \textit{no-go} situation for the free electron gas coupled to the cavity, and that for a sound description of such a macroscopic solid state system the diamagnetic $\bi{A}^2$ term is absolutely necessary. For finite-system models like the Rabi or the Dicke model the $\bi{A}^2$ term is of course important, but these models have a stable ground state even without the $\bi{A}^2$ term. This is in stark contrast to the 2DEG (which is macroscopic) coupled to the cavity mode which has no ground state without the diamagnetic term. This demonstrates explicitly that finite-system models should be applied to extended systems with extra care. Our demonstration strongly suggests that the quadratic term should be included for the description of extended systems, like 2D materials, coupled to a cavity and we believe contributes substantially to the ongoing discussion about the proper description of light-matter interactions~\cite{rokaj2017, schaeferquadratic, GalegoCasimir, bernadrdis2018breakdown, DiStefano2019, vukics2014} and particularly for the emerging field of cavity QED materials. 

Finally, we emphasize that our proof can be extended also for interacting electrons. This is because the Coulomb interaction involves only the relative distances of the electrons and preserves translational symmetry. Thus, one can go to the center of mass and relative distances frame in which the relative distances decouple from the quantized vector potential $\bi{A}$ and from the center of mass. The center of mass though stays still coupled to $\bi{A}$. Then one can follow the proof we presented here and show that without the $\bi{A}^2$ term the coupling constant has no upper bound and the center of mass can obtain an arbitrarily large momentum which subsequently leads to an arbitrarily negative energy. This implies that energy of the system is unbounded from below and the system has no ground state. 

\section{Cavity Modified Responses}\label{Cavity Responses}

So far we have considered the electron gas inside the cavity to be in equilibrium without any external perturbations, like fields, potentials, forces or other kind of sources being applied to it. The aim of this section is exactly to go in this direction and apply external perturbations to our interacting electron-photon system, and compute how particular observables of the system respond to the external perturbations. 

In standard quantum mechanics and solid state physics usually one applies to the system an external field, force or potential and then focuses on how the electrons respond to the perturbation by computing matter-matter response functions, like the current-current response function $\chi^J_j$, which is related to the conductive properties of the electrons~\cite{Mermin, Vignale}.  On the other side in quantum optics one focuses on the responses of the electromagnetic field by computing photon-photon response functions, like the $\bi{A}$-field response function $\chi^A_A$. 

Quantum electrodynamics combines both perspectives under a common unified framework and except of perturbing by external fields, forces and potentials offers the possibility of coupling to \textit{external currents}. This implies that QED gives us the opportunity to access novel observables and response functions which might provide new insights in the emerging field of cavity QED~\cite{ruggenthaler2017b}. In addition to the matter-matter and photon-photon responses QED allows to access also cross-correlated response functions, like matter-photon and photon-matter. As we will see in what follows, all four sectors (matter-matter, photon-photon, matter-photon and photon-matter) have the same pole structure \textit{but} with different strengths. More specifically we will show that all sectors exhibit \textit{plasmon-polariton} excitations or resonances, which modify the radiation and conductive properties of the electron gas in the cavity. 

\subsection{Linear Response Formalism}\label{Response Formalism}

Our considerations throughout this section will remain within the framework of \textit{linear response}, in which a system originally assumed to be at rest and described by a Hamiltonian $\hat{H}$ , is perturbed by a time-dependent external perturbation of the form $\hat{H}_{\textrm{ext}}(t)=f_{\textrm{ext}}(t)\hat{\mathcal{P}}$. The external perturbation $f_{\textrm{ext}}(t)$ couples to some observable of the system represented by an operator $\hat{\mathcal{P}}$. The strength of the perturbation is considered to be \textit{small}, such that the response of the system to be of first order in perturbation theory. This is how linear response formalism (also known as Kubo formalism) is usually formulated~\cite{kubo,Vignale, flick2018light}. Then, by going into the interaction picture the response of any observable $\hat{\mathcal{O}}$ to the external perturbation is defined as~\cite{Vignale, kubo, flick2018light}
\begin{eqnarray}\label{response Observable}
\delta\langle \hat{\mathcal{O}}(t)\rangle=-\frac{\textrm{i}}{\hbar}\int^t_{t_0}dt^{\prime}\langle[\hat{\mathcal{O}}_I(t),\hat{\mathcal{P}}_{I}(t^{\prime})]\rangle f_{\textrm{ext}}(t^{\prime}).
\end{eqnarray}
The correlator above is computed with respect to the ground state  $|\Psi_{gs}\rangle$ of the unperturbed Hamiltonian $\hat{H}$, $\langle[\hat{\mathcal{O}}_I(t),\hat{\mathcal{P}}_{I}(t^{\prime})]\rangle\equiv\langle \Psi_{gs}|[\hat{\mathcal{O}}_I(t),\hat{\mathcal{P}}_{I}(t^{\prime})]|\Psi_{gs}\rangle$, and $\hat{\mathcal{O}}_I(t)=e^{\textrm{i}\hat{H}t/\hbar}\hat{\mathcal{O}}e^{-\textrm{i}\hat{H}t/\hbar}$ is the operator $\hat{\mathcal{O}}$ in the interaction picture. From Eq.~(\ref{response Observable}) by introducing the theta function $\Theta(t-t^{\prime})$ we can re-write the response  $\delta\langle \hat{\mathcal{O}}(t)\rangle$ with the help of a function $\chi^{\mathcal{O}}_{\mathcal{P}}(t-t^{\prime})$ as
\begin{eqnarray}\label{reponse O in chi}
    \delta\langle \hat{\mathcal{O}}(t)\rangle=\int^{\infty}_{t_0}dt^{\prime}\chi^{\mathcal{O}}_{\mathcal{P}}(t-t^{\prime})f_{\textrm{ext}}(t),
\end{eqnarray}
with the function $\chi^{\mathcal{O}}_{\mathcal{P}}(t-t^{\prime})$ defined as
\begin{eqnarray}\label{def chi}
    \chi^{\mathcal{O}}_{\mathcal{P}}(t-t^{\prime})=-\frac{\textrm{i}\Theta(t-t^{\prime})}{\hbar}\langle[\hat{\mathcal{O}}_I(t),\hat{\mathcal{P}}_{I}(t^{\prime})]\rangle.
\end{eqnarray}
Functions of this form are known as \textit{response functions} and are of great importance because they give us information about how observables of the system respond to an external perturbation~\cite{kubo, Vignale, flick2018light}. From Eq.~(\ref{reponse O in chi}) by performing a Laplace transform we can also obtain the response of the observable $\hat{\mathcal{O}}$ in the frequency domain
\begin{eqnarray}\label{reponse frequency O}
    \delta\langle \hat{\mathcal{O}}(w)\rangle=\chi^{\mathcal{O}}_{\mathcal{P}}(w)f_{\textrm{ext}}(w),
\end{eqnarray}
where $\chi^{\mathcal{O}}_{\mathcal{P}}(w)$ and $f_{\textrm{ext}}(w)$ are the response function and the external perturbation respectively in the frequency domain~\cite{kubo, Vignale, flick2018light}.

\subsection{Radiation \& Absorption Properties in Linear Response}\label{Photonic Response}

\begin{figure}[H]
\includegraphics[width=\columnwidth]{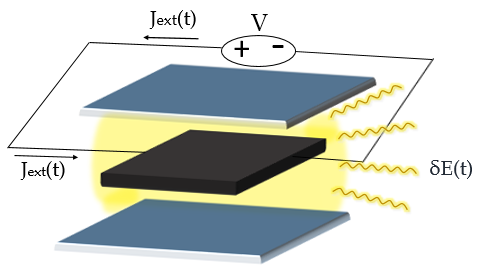}
\caption{\label{Cavity_Induction} Material confined inside a cavity, perturbed by external time dependent current $\bi{J}_{\textrm{ext}}(t)$. The external current perturbs the interacting light-matter system, a time-dependent electric field is induced, and the cavity radiates. We note that in an experiment the emitted radiation can be accessed through the openness of the cavity.}
\end{figure}

Let us start by applying linear response to the photonic sector by computing response functions related to the electromagnetic field. From such responses we obtain information about the radiation and absorption properties of the electron gas coupled to the cavity. To compute these properties, we apply an external time dependent current $\bi{J}_{\textrm{ext}}(t)$ as shown in Fig.~\ref{Cavity_Induction}. We would like to emphasize that in standard quantum mechanics the possibility of perturbing with an external current does not exist and only QED makes this available.

To couple the external current to our system we need to add to the Hamiltonian $\hat{H}$ of Eq.~(\ref{single mode Hamiltonian}) an external time dependent term $\hat{H}_{\textrm{ext}}(t)=-\mathbf{J}_{\textrm{ext}}(t)\cdot \hat{\mathbf{A}}$ as it is done in quantum electrodynamics~\cite{flick2018light, spohn2004, greiner1996}. The external current is chosen to be in only in the $x$-direction $\bi{J}_{\textrm{ext}}(t)=\bi{e}_x|\bi{J}_{\textrm{ext}}(t)|$. Adding the external perturbation the full time-dependent Hamiltonian is 
\begin{eqnarray}\label{Current Perturbation}
    \hat{H}(t)=\hat{H}-\mathbf{J}_{\textrm{ext}}(t)\cdot \hat{\mathbf{A}}.
\end{eqnarray}
 The external current influences the hybrid system in the cavity, and induces electromagnetic fields, as depicted in Fig.~\ref{Cavity_Induction}. The influence of the external current on the photonic observables is exactly what we are interested in here.

\subsubsection{$\bi{A}$-Field Response \& Absorption}\label{A field propagator}

\begin{figure}[ht]
\includegraphics[height=6cm, width=\columnwidth]{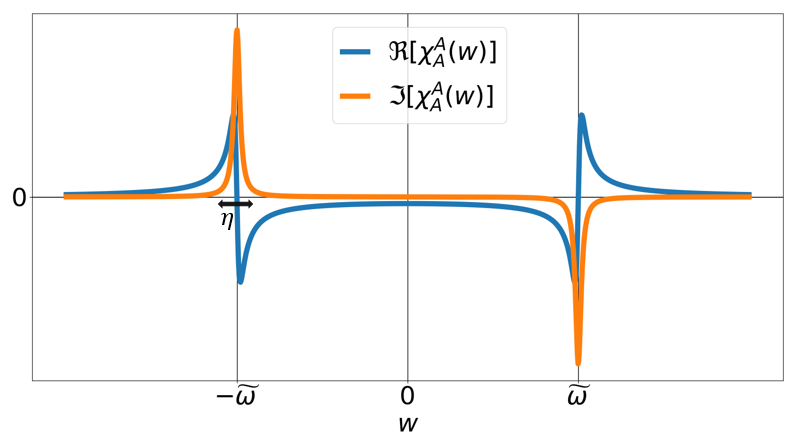}
\caption{\label{A-field Response} Real $\Re[\chi^A_A(w)]$ and imaginary $\Im[\chi^A_A(w)]$ parts of the $\bi{A}$-field response function $\chi^A_A(w)$ in the frequency domain, plotted with a finite $\eta$. The resonances for both parts appear at the plasmon-polariton frequency $w=\pm\widetilde{\omega}$.}
\end{figure}

The first thing we would like to compute is the response of the $\bi{A}$-field $\delta\langle \hat{\bi{A}}(t)\rangle$ due to the external time-dependent current $\bi{J}_{\textrm{ext}}(t)$. The response of the vector potential $\delta\langle \hat{\bi{A}}(t)\rangle$ is defined via Eq.~(\ref{response Observable}) and is given by the $\bi{A}$-field response function $\chi^A_A(t-t^{\prime})$. From Eq.~(\ref{def chi}) we can define the response function $\chi^A_A(t-t^{\prime})$, and performing the computation for the $\bi{A}$-field response function, which we show in detail in appendix~\ref{appendix B}, and we find 
\begin{eqnarray}\label{A field response in time}
    \chi^A_A(t-t^{\prime})=-\frac{\Theta(t-t^{\prime})\sin(\widetilde{\omega}(t-t^{\prime}))}{\epsilon_0\widetilde{\omega}V}.
\end{eqnarray}
Performing a Laplace transform on the response function $\chi^A_A(t-t^{\prime})$ we can find the response function $\chi^A_A(w)$ in the frequency domain, which is given in appendix~\ref{appendix B}, and we deduce the real $\Re[\chi^A_A(w)]$ and the imaginary $\Im[\chi^A_A(w)]$ parts of $\chi^A_A(w)$ 
\begin{eqnarray}\label{Re Im A-field}
    \Re[\chi^A_A(w)]&=&\frac{1}{2\epsilon_0 \widetilde{\omega}V}\left[\frac{w-\widetilde{\omega}}{(w-\widetilde{\omega})^2+\eta^2}-\frac{w+\widetilde{\omega}}{(w+\widetilde{\omega})^2+\eta^2}\right],\nonumber\\
    \Im[\chi^A_A(w)]&=&\frac{\eta}{2\epsilon_0\widetilde{\omega}V}\left[\frac{1}{(w+\widetilde{\omega})^2+\eta^2}-\frac{1}{(w-\widetilde{\omega})^2+\eta^2}\right]\nonumber\\
\end{eqnarray}
which are depicted in Fig.~\ref{A-field Response}. From this expression we see that the pole of the response function is at frequency $w=\pm\widetilde{\omega}$. The frequency $\widetilde{\omega}$ defined in Eq.~(\ref{plasmon polariton}) depends on the cavity frequency $\omega$ and the plasma frequency $\omega_p$ in the cavity. This means that the electron gas in the cavity has a \textit{plasmon-polariton} resonance.

For a self-adjoint operator the real and the imaginary part of any response function have to be respectively even and odd~\cite{Vignale}. In our case the $\bi{A}$-field is self-adjoint and we see that the real and imaginary parts of $\chi^A_A(w)$ shown in Fig.~\ref{A-field Response} satisfy these properties. 

Before we continue let us comment on how  of these parts of the response function should be interpreted. The real part $\Re[\chi^A_A(w)]$ is the component of the response function which is \textit{in-phase} with the external current that drives the system. The real part describes a polarization process in which the wavefunction is modified periodically without any energy being absorbed or released on average by the external driving~\cite{Vignale}. On the other hand, the imaginary part $\Im[\chi^A_A(w)]$ is the \textit{out-of-phase} component of $\chi^A_A(w)$, with respect to the external driving current. The imaginary part is responsible for the appearance of \textit{energy absorption} in the system, with the absorption rate $W$ given by the expression~\cite{Vignale}
\begin{eqnarray}\label{Absorption Rate}
    W=-w\Im[\chi^A_A(w)]|\bi{J}_{\textrm{ext}}(w)|^2.
\end{eqnarray}

\subsubsection{Electric Field Response \& Current Induced Radiation}

\begin{figure}[H]
\includegraphics[height=6cm, width=\columnwidth]{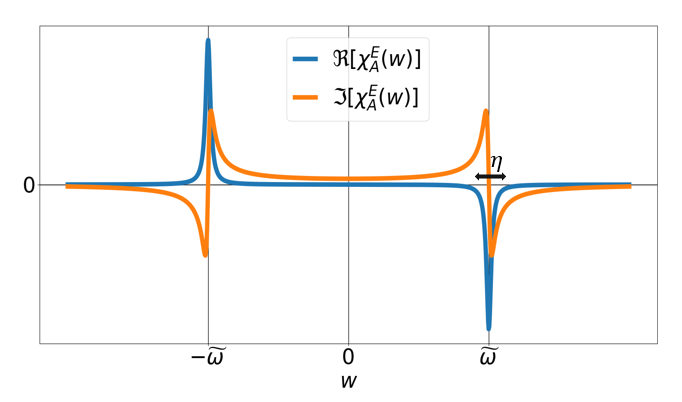}
\caption{\label{E field Responses} Real $\Re[\chi^E_A(w)]$ and imaginary $\Im[\chi^E_A(w)]$ parts of the $\bi{E}$-field response function $\chi^E_A(w)$ in the frequency domain with a finite broadening parameter $\eta$. The poles of $\Re[\chi^E_A(w)]$ and $\Im[\chi^E_A(w)]$ both appear at the frequency $w=\pm\widetilde{\omega}$ and signify the frequency at which an time-dependent electric field is oscillating. Radiation should come out of the cavity at this frequency. }
\end{figure}

Having computed the response of the $\bi{A}$-field we would also like to compute the response of the electric field $\bi{E}$ due to the external current. The electric field operator in dipole approximation and polarized in the $x$-direction is~\cite{rokaj2017}
\begin{equation}\label{Electric field}
	\hat{\mathbf{E}}=\textrm{i}\left(\frac{\hbar\omega }{2\epsilon_0 V}\right)^{\frac{1}{2}}\left(\hat{a}_{1}-\hat{a}^{\dagger}_{1}\right) \bi{e}_x.
\end{equation}
With the definition of the electric field we can compute the electric field response function $\chi^E_A(t-t^{\prime})$ using the definition of Eq.~(\ref{def chi}). The computation of  $\chi^E_A(t-t^{\prime})$ is presented in appendix~\ref{appendix C} and we find
\begin{eqnarray}\label{E field response function}
    \chi^E_{A}(t-t^{\prime})=\frac{\Theta(t-t^{\prime})\cos(\widetilde{\omega}(t-t^{\prime}))}{\epsilon_0 V}.
\end{eqnarray}
The response function above describes the generation of a time dependent electric field due to the external time dependent current $\bi{J}_{\textrm{ext}}(t)$. This means that the external current makes the coupled light-matter system radiate. From Eq.~(\ref{E field response function}) we see the radiation is at the plasmon-polariton frequency $\widetilde{\omega}$ since the response function in time is a cosine of $\widetilde{\omega}$. This fact can also be  understood from the response function in the frequency domain $\chi^{E}_{A}(w)$, whose real $\Re[\chi^E_A(w)]$ and imaginary parts $\Im[\chi^E_A(w)]$ are
\begin{eqnarray}
    \Re[\chi^E_A(w)]=\frac{\eta}{2\epsilon_0V}\left[\frac{1}{(w+\widetilde{\omega})^2+\eta^2}-\frac{1}{(w-\widetilde{\omega})^2+\eta^2}\right],\nonumber\\
    \Im[\chi^E_A(w)]=\frac{1}{2\epsilon_0V}\left[\frac{w+\widetilde{\omega}}{(w+\widetilde{\omega})^2+\eta^2}-\frac{w-\widetilde{\omega}}{(w-\widetilde{\omega})^2+\eta^2}\right]\nonumber\\
\end{eqnarray}
from which we see that the poles are at the plasmon-polariton resonance $w=\pm\widetilde{\omega}$ as shown also in Fig.~\ref{E field Responses}.

Lastly, we would like to mention that the response function of the electric field in time $\chi^E_A(t-t^{\prime})$ of Eq.~(\ref{E field response function}), and the response function of the $\bi{A}$-field $\chi^A_A(t-t^{\prime})$ of Eq.~(\ref{A field response in time}) satisfy Maxwell's equation $\chi^{E}_{A}(t-t^{\prime})=-\partial_t \chi^{A}_{A}(t-t^{\prime})$~\cite{JacksonEM}. This is a beautiful consistency check for our computations and of the whole linear response formalism in QED~\cite{flick2018light}, because it shows that linear response theory even for coupled electron-photon systems respects the classical Maxwell equations.  

\subsection{Cavity Modified Conductivity \& Drude Peak Suppression}\label{Electronic Response}

In what follows we are interested in the conduction properties of the 2DEG inside the cavity and more specifically on whether the cavity field modifies the conductivity of the 2DEG. This is a question of current theoretical and experimental interest, because recently cavity modifications of transport and conduction properties have been observed for 2D systems of Landau polaritons~\cite{paravacini2019}, as well as modifications of the critical temperature of superconductors due to cavity confinement~\cite{sentef2018, A.Thomas2019}. 

\begin{figure}[h]
\includegraphics[width=\columnwidth]{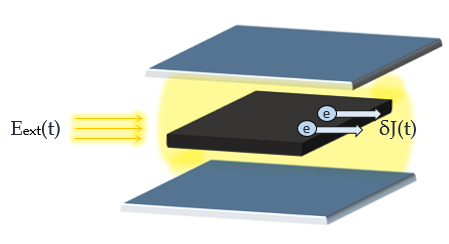}
\caption{\label{Cavity_Conductivity} An external time dependent electric field $\bi{E}_{\textrm{ext}}(t)$ perturbs the combined light-matter system, electrons start to flow, and a current is generated in the material. }
\end{figure}

To describe such processes we will follow what is usually done in condensed matter physics, namely perturb the system with an external, uniform, time-dependent electric field $\mathbf{E}_{\textrm{ext}}(t)$, as depicted in Fig.~\ref{Cavity_Conductivity}, and then compute how much current flows due to the perturbation. Here, the electric field is chosen to be polarized in the $x$-direction $\mathbf{E}_{\textrm{ext}}(t)=|\mathbf{E}_{\textrm{ext}}(t)|\bi{e}_x$ and can be represented as the time derivative of a vector potential $\mathbf{E}_{\textrm{ext}}(t)=-\partial_t\bi{A}_{\textrm{ext}}(t)$. We note that to have a causal external perturbation the electric field needs to be zero for all times prior to an instant of time $t_0$. This implies that in the frequency domain the electric field and vector potential are related via $\bi{E}_{\textrm{ext}}(w)=\textrm{i}(w+\textrm{i}\eta)\bi{A}(w)$ with $\eta \rightarrow 0^{+}$.

To couple the external field we need to add the external vector potential $\bi{A}_{\textrm{ext}}(t)$ in the covariant kinetic energy of the Pauli-Fierz Hamiltonian of Eq.~(\ref{Pauli-Fierz}), which becomes then $(\textrm{i}\hbar \mathbf{\nabla}_{j}+e \hat{\bi{A}}+e\bi{A}_{\textrm{ext}}(t))^2$~\cite{Landau, TokatlyPRL, spohn2004}. In linear response the current is computed to first order in perturbation theory and the conductivity is defined as the function relating the induced current to the external electric field~\cite{kubo, flick2018light, Vignale}. The Pauli-Fierz Hamiltonian with the electrons coupled to a single mode, in dipole approximation, to first order in the external field $\bi{A}_{\textrm{ext}}(t)$ is
\begin{eqnarray}
    \hat{H}(t)&=&\hat{H}+\sum^{N}_{j=1}\left(\frac{\textrm{i}e\hbar}{m_{\textrm{e}}}\nabla_j+\frac{e^2}{m_{\textrm{e}}}\hat{\mathbf{A}}\right)\cdot \bi{A}_{\textrm{ext}}(t)\nonumber\\
    &=&\hat{H}-\left(\hat{\bi{J}}_\textrm{p}+\hat{\bi{J}}_{\textrm{d}}\right)\cdot\bi{A}_{\textrm{ext}}(t)
\end{eqnarray}
where $\hat{H}$ is the Hamiltonian of Eq.~(\ref{single mode Hamiltonian}). The external field couples to the \textit{internal} parts of the current operator, which are the paramagnetic part $\hat{\bi{J}}_{\textrm{p}}=(-\textrm{i}e\hbar/m_{\textrm{e}})\sum_j\nabla_j$, and the diamagnetic part $\hat{\bi{J}}_{\textrm{d}}=-e^2N\hat{\bi{A}}/m_{\textrm{e}}$. The full physical current includes also the contribution due to the external vector potential $\bi{A}_{\textrm{ext}}(t)$ ~\cite{Landau, TokatlyPRL}
\begin{eqnarray}\label{Current Operator}
    \hat{\bi{J}}=-\underbrace{\frac{\textrm{i}e\hbar}{m_{\textrm{e}}}\sum^N_{j=1}\nabla_j}_{\hat{\bi{J}}_{\textrm{p}}}\underbrace{-\frac{e^2N}{m_{\textrm{e}}}\hat{\bi{A}}}_{\hat{\bi{J}}_{\textrm{d}}}-\frac{e^2N}{m_{\textrm{e}}}\bi{A}_{\textrm{ext}}(t).
\end{eqnarray}
Following the standard linear response formalism the expectation value for the full physical $\hat{\bi{J}}$ current is~\cite{Vignale, kubo}
\begin{eqnarray}\label{current Expectation Value}
    \langle \hat{\bi{J}}(t)\rangle=\langle \hat{\bi{J}}\rangle +\delta\langle \hat{\bi{J}}(t)\rangle= \langle \hat{\bi{J}}\rangle-\int^{\infty}_{t_0} dt^{\prime}\chi^J_J(t-t^{\prime})\bi{A}_{\textrm{ext}}(t^{\prime})\nonumber\\
\end{eqnarray}
where $\delta\langle \hat{\bi{J}}(t)\rangle$ is the response of the current $\hat{\bi{J}}$, which can be computed from the the current-current response function
 \begin{eqnarray}\label{JJResponse}
     \chi^{J}_{J}(t-t^{\prime})=\frac{-\textrm{i}\Theta(t-t^{\prime})}{\hbar }\langle[\hat{\mathbf{J}}_{I}(t),\hat{\mathbf{J}}_{I}(t^{\prime})]\rangle.
 \end{eqnarray}
 Neglecting all contributions coming from $\bi{A}_{\textrm{ext}}(t)$, such that the current response $\delta\langle \hat{\bi{J}}\rangle$ stays in first order to $\bi{A}_{\textrm{ext}}$, we find for the commutator of Eq.~(\ref{JJResponse}) the following four terms
\begin{eqnarray}
    [\hat{\mathbf{J}}_{I}(t),\hat{\mathbf{J}}_{I}(t^{\prime})]&=&[\hat{\mathbf{J}}_{\textrm{p}, I}(t),\hat{\mathbf{J}}_{\textrm{p}, I}(t^{\prime})] +[\hat{\mathbf{J}}_{\textrm{d}, I}(t),\hat{\mathbf{J}}_{\textrm{p}, I}(t^{\prime})]\nonumber\\
    &+&[\hat{\mathbf{J}}_{\textrm{d}, I}(t),\hat{\mathbf{J}}_{\textrm{p}, I}(t^{\prime})] +[\hat{\mathbf{J}}_{\textrm{d}, I}(t),\hat{\mathbf{J}}_{\textrm{d}, I}(t^{\prime})]\nonumber\\
\end{eqnarray}
For the paramagnetic contribution using the self-adjointness of the paramagnetic current operator we have $\langle [\hat{\mathbf{J}}_{\textrm{p}, I}(t),\hat{\mathbf{J}}_{\textrm{p}, I}(t^{\prime})]\rangle=\langle\hat{\mathbf{J}}_{\textrm{p}, I}(t)\hat{\mathbf{J}}_{\textrm{p}, I}(t^{\prime})\rangle -\langle\hat{\mathbf{J}}_{\textrm{p}, I}(t)\hat{\mathbf{J}}_{\textrm{p}, I}(t^{\prime})\rangle^*$. Using the definition for the paramagnetic current operator in the interaction picture and the fact that the expectation value is computed in the ground state which has energy $E_{0,\bi{k}}$ we find $\langle \hat{\mathbf{J}}_{\textrm{p}, I}(t)\hat{\mathbf{J}}_{\textrm{p}, I}(t^{\prime})\rangle=e^{\textrm{i}E_{0,\bi{k}}(t-t^{\prime})/\hbar}\langle \hat{\mathbf{J}}_{p}e^{\textrm{i}\hat{H}(t^{\prime}-t)/\hbar}\hat{\mathbf{J}}_p\rangle$.
 Because the momentum operator commutes with the Hamiltonian $\hat{H}$, the ground state $|\Psi_{gs}\rangle=|\Phi_{0}\rangle\otimes|0,0\rangle_1|0,0\rangle_2$ is also an eigenstate of the paramagnetic current operator $\hat{\mathbf{J}}_{\textrm{p}}\sim \sum_j \nabla_j$. Acting with the paramagnetic current operator on the ground state we get the full paramagnetic current $\hat{\mathbf{J}}_{\textrm{p}}|\Psi_{gs}\rangle=\sum_j \mathbf{k}_j|\Psi_{gs}\rangle$, and because in the thermodynamic limit the ground state distribution of the momenta is the Fermi sphere, as we showed in section~\ref{Ground State}, the total paramagnetic current is zero and we have $\hat{\mathbf{J}}_{\textrm{p}}|\Psi_{gs}\rangle=0$. This means that all expectation values and correlators which involve $\hat{\bi{J}}_{\textrm{p}}$ are zero. This argument applies also to the mixed terms $[\hat{\mathbf{J}}_{\textrm{d}, I}(t),\hat{\mathbf{J}}_{\textrm{p}, I}(t^{\prime})]$ and $[\hat{\mathbf{J}}_{\textrm{p}, I}(t),\hat{\mathbf{J}}_{\textrm{d}, I}(t^{\prime})]$. Thus, the response function $\chi^J_{J}(t-t^{\prime})$ in Eq.~(\ref{JJResponse}) is given purely by the diamagnetic terms. Substituting the definition for the diamagnetic current $\hat{\bi{J}}_{\textrm{d}}$ of Eq.~(\ref{Current Operator}) we find the current-current response function $\chi^{J}_{J}(t-t^{\prime})$ to be proportional to the $\bi{A}$-field response function $\chi^A_A(t-t^{\prime})$
 \begin{eqnarray}
     \chi^{J}_{J}(t-t^{\prime})=\left(\frac{e^2 N}{m_{\textrm{e}}}\right)^2 \chi^A_A(t-t^{\prime}).
 \end{eqnarray}
 with $\chi^A_A(t-t^{\prime})$ given by Eq.~(\ref{A field response in time}). Since $\chi^J_J(t-t^{\prime})$ is proportional to $\chi^A_A(t-t^{\prime})$ the same will also hold in the frequency domain 
 \begin{eqnarray}\label{chiJJ to chiAA}
     \chi^J_J(w)=\left(\frac{e^2N}{m_{\textrm{e}}}\right)^2\chi^A_A(w)
 \end{eqnarray}
 where $\chi^A_A(w)$ is computed in appendix~\ref{appendix B}. Last, we need to compute the expectation value of the current $\langle \hat{\bi{J}}\rangle$ which is
\begin{eqnarray}
    \langle \hat{\bi{J}}\rangle=\langle \hat{\bi{J}}_{\textrm{p}}\rangle+\langle\hat{\bi{J}}_{\textrm{d}}\rangle-\frac{e^2N}{m_{\textrm{e}}}\langle\bi{A}_{\textrm{ext}}(t)\rangle.
\end{eqnarray}
As we already explained the contribution of the paramagnetic $\hat{\bi{J}}_{\textrm{p}}$ current is zero in the ground state $|\Psi_{gs}\rangle$. The diamagnetic part $\hat{\bi{J}}_{\textrm{d}}$ is proportional to the quantized field $\hat{\bi{J}}_{\textrm{d}}\sim \hat{\bi{A}}$. The quantized vector potential is the sum of an annihilation and a creation operator and the expectation values of these operators in the ground state is zero. Thus, we find that only the external field contributes to  $\langle \hat{\bi{J}}\rangle$
\begin{eqnarray}
    \langle \hat{\bi{J}}\rangle=-\frac{e^2N}{m_{\textrm{e}}}\bi{A}_{\textrm{ext}}(t).
\end{eqnarray}
The latter is the contribution of the the full background charge of the $N$ electrons in our system. From the equation for the full physical current in time $\langle\hat{\bi{J}}(t)\rangle$ given by Eq.~(\ref{current Expectation Value}) we can derive the relation between the current $\langle \hat{\bi{J}}(w)\rangle$ and the external vector potential $\bi{A}_{\textrm{ext}}(w)$ in the frequency domain by performing a Laplace transformation
\begin{eqnarray}\label{J n A }
    \langle \mathbf{J}(w)\rangle=\left(-\frac{e^2N}{m_{\textrm{e}}}-\chi^{J}_{J}(w)\right)\mathbf{A}_{\textrm{ext}}(w).
\end{eqnarray}
As we already explained, the vector potential and the electric field in the frequency domain are related via the relation $\bi{A}_{\textrm{ext}}(w)=\bi{E}_{\textrm{ext}}(w)/\textrm{i}(w+\textrm{i}\eta)$. Using this and dividing Eq.~(\ref{J n A }) by the volume $V$ in order to introduce the current density $\langle\bi{j}(w)\rangle=\langle\mathbf{J}(w)\rangle/V$ we can define the frequency dependent (or optical) conductivity $\sigma(w)$ as the ratio between the external electric field $\bi{E}(w)$ and the current density $\langle\bi{j}(w)\rangle$~\cite{Mermin,Vignale} 
\begin{eqnarray}
\langle\mathbf{j}(w)\rangle=\left(-\frac{e^2n_{\textrm{e}}}{m_{\textrm{e}}}-\frac{\chi^J_J(w)}{V}\right)\frac{\mathbf{E}_{\textrm{ext}}(w)}{\textrm{i}(w+\textrm{i}\eta)}=\sigma(w)\mathbf{E}_{\textrm{ext}}(w).\nonumber\\
\end{eqnarray}
The equation above is the \textit{Kubo formula} for the electrical conductivity~\cite{kubo, Vignale}. Using the result for the current-current response function $\chi^J_J(w)$ given by Eq.~(\ref{chiJJ to chiAA}), and introducing $\omega_p^2=e^2n_{\textrm{e}}/m_{\textrm{e}}\epsilon_0$ which is the plasma frequency in the cavity, we obtain the expression for the frequency dependent (or optical) conductivity $\sigma(w)$
\begin{widetext}
\begin{eqnarray}\label{Conductivity}
    \sigma(w)&=&\frac{\textrm{i}}{w+\textrm{i}\eta}\left(\frac{e^2n_{\textrm{e}}}{m_{\textrm{e}}}+\frac{\chi^J_J(w)}{V}\right)=\frac{\textrm{i}\epsilon_0\omega^2_p}{w+\textrm{i}\eta}- \frac{\textrm{i}\epsilon_0\omega^4_p}{(w+\textrm{i}\eta)2\widetilde{\omega}}\left[\frac{1}{w+\widetilde{\omega}+\textrm{i}\eta}-\frac{1}{w-\widetilde{\omega}+\textrm{i}\eta}\right] \;\;\textrm{with}\;\; \eta \rightarrow 0^{+}.
    \end{eqnarray}
\end{widetext}
The real $\Re[\sigma(w)]$ and imaginary $\Im[\sigma(w)]$ parts of the optical conductivity are given respectively by the expressions 
\begin{widetext}
\begin{eqnarray}\label{Real and Imaginary sigma}
\Re[\sigma(w)]&=&\frac{\epsilon_0\eta\omega^2_p}{w^2+\eta^2}-\frac{\eta\epsilon_0\omega^4_p}{2\widetilde{\omega}(w^2+\eta^2)}\left[\frac{2w+\widetilde{\omega}}{(w+\widetilde{\omega})^2+\eta^2}-\frac{2w-\widetilde{\omega}}{(w-\widetilde{\omega})^2+\eta^2}\right],\\
\Im [\sigma(w)]&=&\frac{\epsilon_0w\omega^2_p}{w^2+\eta^2}-\frac{\epsilon_0\omega^4_p}{2\widetilde{\omega}(w^2+\eta^2)}\left[\frac{w^2-\eta^2+w\widetilde{\omega}}{(w+\widetilde{\omega})^2+\eta^2}-\frac{w^2-\eta^2-w\widetilde{\omega}}{(w-\widetilde{\omega})^2+\eta^2}\right].\nonumber
\end{eqnarray}
\end{widetext}

\begin{figure}[h]
\includegraphics[height=5cm, width=\columnwidth]{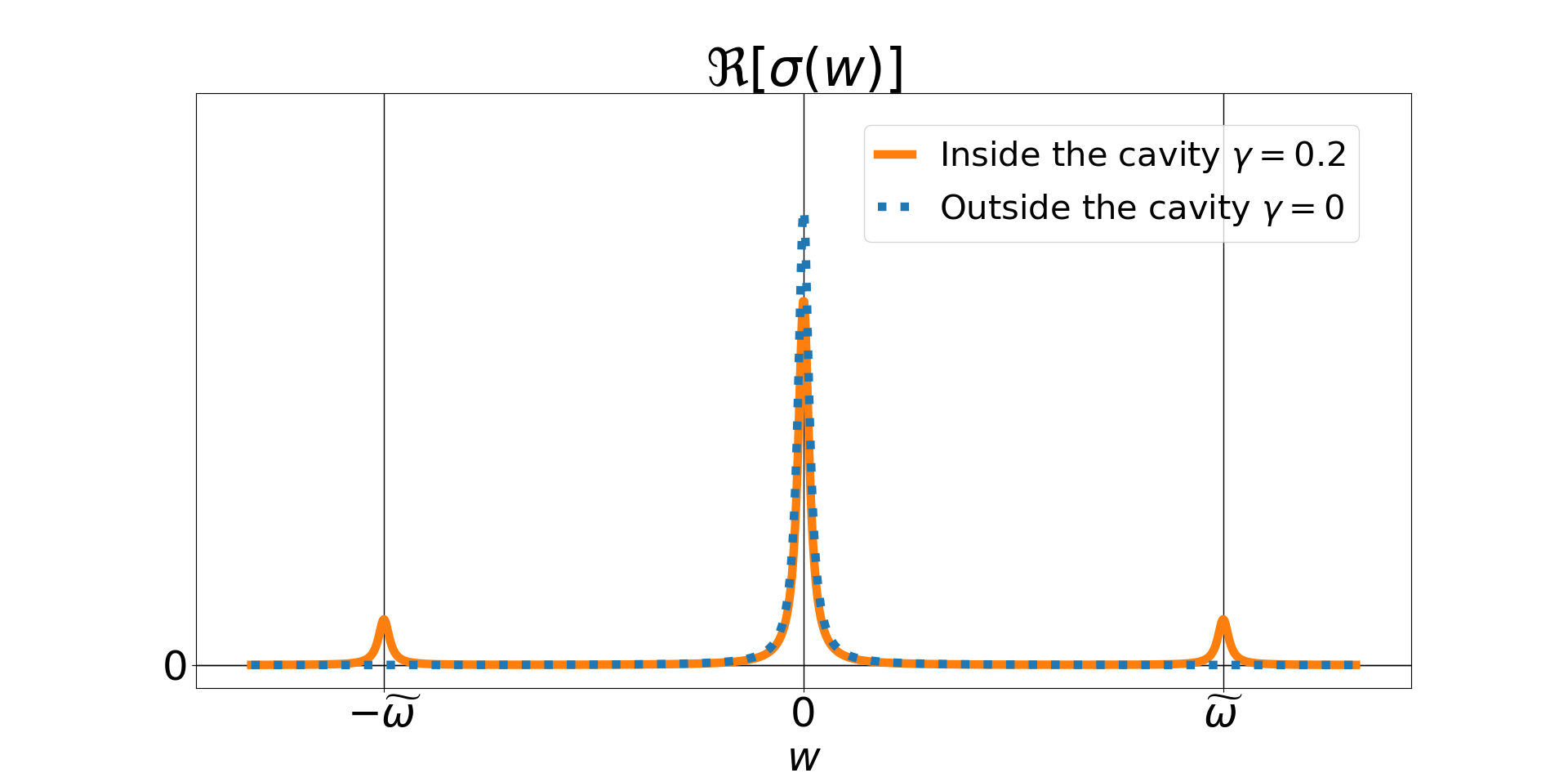}
\caption{\label{Real_Conductivity}Real part $\Re[\sigma(w)]$ of the conductivity $\sigma(w)$ of the 2DEG outside (blue dashed) and inside (orange solid line) the cavity for collective coupling $\gamma=0.2$. Inside the cavity the real part exhibits poles at the plasmon-polariton resonance $w=\pm\widetilde{\omega}$. At frequency $w=0$ the Drude peak of the 2DEG gets suppressed due to the cavity field.   }
\end{figure}

\begin{figure}[h]
\includegraphics[height=5cm, width=\columnwidth]{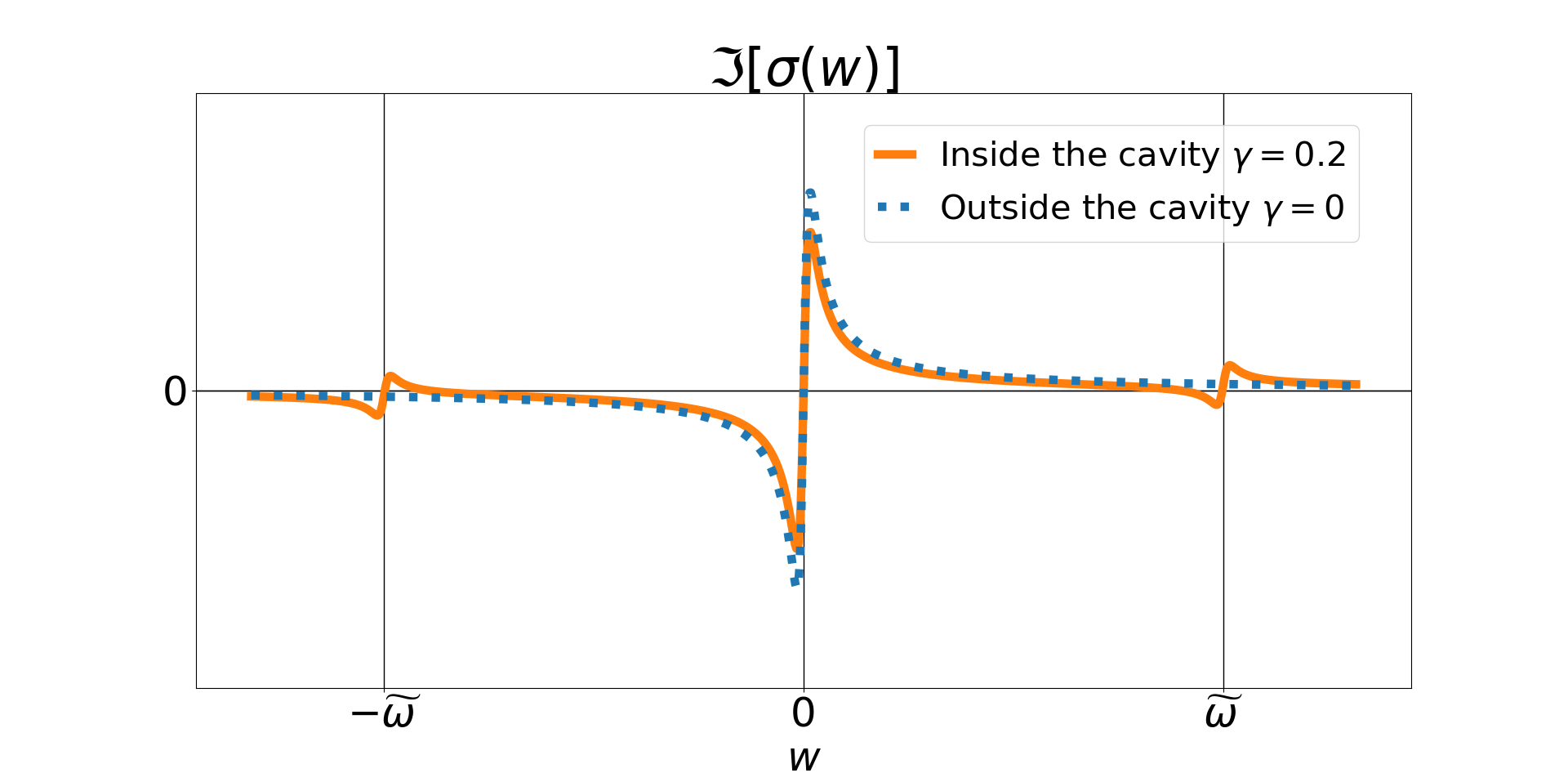}
\caption{\label{Imaginary_Conductivity}Imaginary part $\Im[\sigma(w)]$ of the conductivity $\sigma(w)$ of the 2DEG outside (blue dashed) and inside (orange solid line) the cavity for collective coupling $\gamma=0.2$. Inside the cavity the imaginary part has poles at the plasmon-polariton resonance $w=\pm\widetilde{\omega}$. The peak at $w=0$ gets suppressed by the cavity.  }
\end{figure}
In the optical conductivity $\sigma(w)$ there are two contributions. The first contribution comes from the full electron density $n_{\textrm{e}}$ via the plasma frequency $\omega^2_p=n_{\textrm{e}}e^2/m_{\textrm{e}}\epsilon_0$ and is of second order to $\omega_p$. This is the standard contribution of the free electron gas~\cite{Vignale}. The second contribution comes from the current-current response function $\chi^J_J(w)$. This one is purely due to the photon field in the cavity because $\chi^J_J(w)$ is proportional to the $\bi{A}$-field response function $\chi^A_A(w)$. The current-current response function is of fourth order in the plasma frequency $\omega_p$ and is a diamagnetic modification to the standard free electron gas conductivity. To be more specific, both the real and the imaginary part of the optical conductivity shown in Figs.~\ref{Real_Conductivity} and~\ref{Imaginary_Conductivity} respectively, exhibit resonances at the plasmon-polariton frequency $w=\pm\widetilde{\omega}$, which modify the optical conductivity of the 2DEG. In addition, in the real part of the conductivity we see that at $w=0$ the Drude peak~\cite{Basov, BasovHighTc} of the 2DEG is suppressed by the cavity field due to the higher-order diamagnetic contributions. As the Drude peak is very important for condensed matter systems and materials let us have a closer look at it.

\subsubsection{Cavity Suppression of the Drude Peak}

The Drude peak is defined as the $w\rightarrow 0$ limit of the real part of the optical conductivity and gives the (static) DC electrical conductivity of a material $\sigma_{\textrm{dc}}=\lim_{w\rightarrow 0}\Re[\sigma(w)]$~\cite{Mermin, Basov, Vignale}. In the case of an electron gas outside a cavity the DC conductivity is $\sigma^0_{dc}=\epsilon_0\omega^2_p/\eta$, which is the first term of $\Re[\sigma(w)]$ in Eq.~(\ref{Real and Imaginary sigma}) for $w\rightarrow 0$. However, for our system we have the extra diamagnetic contributions due to the electron-photon coupling and we find that the DC conductivity $\sigma_{\textrm{dc}}(\gamma)$ of the 2DEG in the cavity is a function of the collective coupling $\gamma$ (defined in Eq.~(\ref{collective coupling}))
\begin{eqnarray}
\sigma_{\textrm{dc}}(\gamma)=\sigma^0_{\textrm{dc}}\left(1-\frac{\gamma}{1+\eta^2/\widetilde{\omega}^2}\right) \;\; \textrm{with}\;\; \eta \rightarrow 0^+.
\end{eqnarray}
To zeroth order in the infinitesimal parameter $\eta$ we find that the DC conductivity in the cavity, i.e., the Drude peak, decreases linearly as function of the collective coupling constant $\gamma$
\begin{eqnarray}\label{DC Conductivity}
\sigma_{\textrm{dc}}(\gamma)=\sigma^0_{\textrm{dc}}\left(1-\gamma\right). 
\end{eqnarray}
This is a significant result because it shows that coupling materials to a cavity does not only modify the optical properties of the system, like the optical conductivity, but also the cavity can alter the static DC electrical conductivity. This phenomenon, of the decrease of the DC conductivity has also been reported for Landau polariton systems in~\cite{BartoloCiuti}. To be more specific, in the region of zero magnetic field (in which our theory is also applicable) an increase of the longitudinal resistivity was obtained, due to the cavity confinement~\cite{BartoloCiuti}. This implies that the DC conductivity due to the strong coupling to the cavity decreases, in accordance to our prediction. Most importantly, we would like to mention that this effect is also in agreement with magneto-transport measurements performed for such Landau polariton systems in~\cite{paravacini2019}. This is a firm confirmation of our work. We hope that further experimental measurements, focusing solely on the behavior of the Drude peak, under strong coupling to the photon field, will further explore this phenomenon and allow for a further quantitative test of our prediction about the modification of the Drude peak.

The fact that the photon field has the effect to decrease the conduction of electrons implies that the cavity field can be understood as viscous medium which slows down the motion of the charged particles. In such a picture the suppression of the Drude peak can be also understood as an increase in the effective mass of the electrons due to the coupling to the cavity field. From the expression for the DC conductivity in Eq.~(\ref{DC Conductivity}) we find that the effective (or renormalized) electron mass is $m_{\textrm{e}}(\gamma)=m_{\textrm{e}}/(1-\gamma)$. Such an increase of the effective electron mass we will also encounter later in section~\ref{Effective Field Theory} when we will couple the 2DEG to the full continuum of electromagnetic modes.

Lastly, we would like to mention that due to the fact that the collective coupling parameter has an upper bound $\gamma <1$ (see Eq.~\ref{collective coupling}) the Drude peak remains always larger than zero and the 2DEG is a conductor. However, if the coupling could reach the critical value $\gamma_c=1$ (which is forbidden) then the DC conductivity would be zero, which would imply that the cavity can turn the 2DEG into an insulator. For $\gamma >1$ the DC conductivity turns negative which implies that the system becomes unstable. This explains from a different point of view why the collective coupling must not exceed the upper bound 1.

 \subsection{Mixed Responses: Matter-Photon \& Photon-Matter}

In the beginning of this section we emphasized the fact that QED gives us the opportunity to access new mixed, cross-correlated responses. So let us now present how such mixed matter-photon and photon-matter response functions arise in QED and compute them.

\subsubsection{Matter-Photon Response}

The response of the current $\delta\langle \hat{\bi{J}}(t)\rangle$ is defined via Eq.~(\ref{response Observable}) and can be computed directly from the mixed response function $\chi^J_A(t-t^{\prime})$ which is proportional to the correlator $\langle[\hat{\bi{J}}_{I}(t),\hat{\bi{A}}_{I}(t^{\prime})]\rangle$ as we can deduce from Eq.~(\ref{def chi}). The full physical current $\hat{\bi{J}}$ given by Eq.~(\ref{Current Operator}), for $\bi{A}_{\textrm{ext}}=0$, includes two contributions. One from the paramagnetic current $\hat{\bi{J}}_{\textrm{p}}$ and one coming from the diamagnetic current $\hat{\bi{J}}_{\textrm{d}}$. The paramagnetic contribution as we explained in the previous subsection is zero because the ground state has zero paramagnetic current, and consequently only the diamagnetic current contributes. Substituting the definition for the diamagnetic current $\hat{\bi{J}}_{\textrm{d}}$ we find that the mixed response $\chi^{J}_{A}(t-t^{\prime})$ is proportional to the $\bi{A}$-field response function $\chi^{J}_{A}(t-t^{\prime})=\left(-e^2N/m_{\textrm{e}}\right)\chi^A_A(t-t^{\prime})$ where $\chi^A_A(t-t^{\prime})$ given by Eq.~(\ref{A field response in time}). The same relation between the two response functions also holds in the frequency domain 
\begin{eqnarray}\label{ChiJA and ChiAA}
  \chi^{J}_{A}(w)=\left(\frac{-e^2N}{m_{\textrm{e}}}\right)\chi^A_A(w).  
\end{eqnarray}
Lastly, we would like to emphasize that the mixed response function $\chi^J_A(w)$ is dimensionless and describes the ratio between the induced current $\delta\langle \hat{\bi{J}}(w)\rangle$ and the external current $\bi{J}_{\textrm{ext}}(w)$, $\delta\langle \hat{\bi{J}}(w)\rangle=\chi^J_A(w)\bi{J}_{\textrm{ext}}(w)$.

\subsubsection{Photon-Matter Response}

Having computed the matter-photon response function $\chi^J_A$ we want to compute also the photon-matter response function $\chi^A_J$ which corresponds to the inverse physical process with respect to $\chi^J_A$. Now we look into the response of the vector potential $\delta\langle \hat{\bi{A}}(t)\rangle$ given by the photon-matter response function $\chi^A_J(t-t^{\prime})$, which is proportional to the correlator $\langle[\hat{\mathbf{A}}_{I}(t),\hat{\mathbf{J}}_{I}(t^{\prime})]\rangle$ according to Eq.~(\ref{def chi}). To remain within linear response we neglect the contribution of $\bi{A}_{\textrm{ext}}(t)$ to the current operator $\hat{\bi{J}}$ which would result into higher order corrections.
 
 The paramagnetic contribution as we already explained is zero. Substituting the definition for the diamagnetic current $\hat{\bi{J}}_{\textrm{d}}$ we find that the mixed response function $\chi^{A}_{J}(t-t^{\prime})$ is proportional to the $\bi{A}$-field response function $\chi^{A}_{J}(t-t^{\prime})=\left(-e^2N/m_{\textrm{e}}\right)\chi^{A}_{A}(t-t^{\prime})$. Since this relation holds in time, it will also be true in the frequency domain,
\begin{eqnarray}\label{chiAJ to chiAA}
     \chi^{A}_{J}(w)=\left(\frac{-e^2N}{m_{\textrm{e}}}\right)\chi^{A}_{A}(w).
 \end{eqnarray}
From the result above we see that the response function $\chi^A_J(w)$ is the dimensionless ratio between the induced $\hat{\bi{A}}$-field and the external field $\bi{A}_{\textrm{ext}}$.

\subsection{Linear Response Equivalence Between the Electronic and the Photonic Sector}\label{Duality}

In this section we would like to compare the four fundamental response sectors we introduced and discussed above, and most importantly demonstrate how these sectors are connected and that actually are \textit{all equivalent} with respect to their pole structure. From all the response functions we computed in the different sectors we can can construct the following response table
\begin{eqnarray}\label{Response Table}
    \left(\begin{tabular}{c}
		$\delta\langle \hat{\bi{J}}(w)\rangle$ \\
	    $\delta\langle\hat{\bi{A}}(w)\rangle$  
	\end{tabular}\right)=\left(\begin{tabular}{ c c }
		$\chi^J_{J}(w)$ &$ \chi^J_{A}(w)$ \\
	    $\chi^A_{J}(w)$ &$ \chi^A_{A}(w)$ 
	\end{tabular}\right) \left(\begin{tabular}{c}
		$\bi{A}_{\textrm{ext}}(w)$ \\
	    $\bi{J}_{\textrm{ext}}(w)$  
	\end{tabular}\right)\nonumber\\
\end{eqnarray}
which summarizes all the different responses of the system. Looking back now into the Eqs.~(\ref{chiJJ to chiAA}),~(\ref{chiAJ to chiAA}) and~(\ref{ChiJA and ChiAA}) which give the response functions $\chi^J_J(w)$, $\chi^J_A(w)$ and $\chi^A_{J}(w)$ respectively, we see that all response functions are proportional to the $\bi{A}$-field response function $\chi^A_A(w)$. Thus, all elements of the response table can be written in terms of $\chi^A_A(w)$
\begin{eqnarray}\label{Response Table chiAA}
    \left(\begin{tabular}{c}
		$\delta\langle \hat{\bi{J}}(w)\rangle$ \\
	    $\delta\langle\hat{\bi{A}}(w)\rangle$  
	\end{tabular}\right)=\chi^A_A(w)\left(\begin{tabular}{ c c }
		$\left(\dfrac{e^2N}{m_{\textrm{e}}}\right)^2$ &\; $ \dfrac{-e^2N}{m_{\textrm{e}}}$ \\\\
	    $\dfrac{-e^2N}{m_{\textrm{e}}}$ &$1$ 
	\end{tabular}\right) \left(\begin{tabular}{c}
		$\bi{A}_{\textrm{ext}}(w)$ \\
	    $\bi{J}_{\textrm{ext}}(w)$  
	\end{tabular}\right)\nonumber\\
\end{eqnarray}
The fact that all response functions are proportional to $\bi{A}$-field response function $\chi^A_A(w)$ means that all response functions have exactly the same pole structure. This shows a deep and fundamental relation between the two sectors of the theory, namely that the photonic and the electronic sectors have exactly the same excitations and resonances. This implies that in an experiment, perturbing an interacting light-matter system with an external time dependent current, which couples to the photon field, and perturbing with an external electric field, which couples to the current, would give exactly the same information about the excitations of the system.

Furthermore, from the response table in Eq.~(\ref{Response Table chiAA}) we see that the current-current response function scales quadratically with the number of electrons $\chi^J_J(w)\sim N^2\chi^A_A(w)$, while the mixed response functions linearly $\chi^J_A(w)=\chi^A_J(w)\sim N \chi^A_A(w)$. The photon-photon response function $\chi^A_A(w)$ given by Eq.~(\ref{Re Im A-field}) also scales with respect to the area of the 2DEG as $1/S$. This implies that in the large $N,S$ limit only the responses involving matter ($\chi^J_J, \chi^J_A, \chi^A_J$) are finite, due to the dependence on $N$, while $\chi^A_A$ goes to zero. This is the same feature that appears also for the energy densities of the two sectors as we mentioned in section~\ref{Ground State}. Again, this hints towards the fact that in order to have a finite photon-photon response, we need to include a continuum of modes for the photon field because we are a considering a macroscopic 2D system. For a finite system such a problem would not arise and this shows another point in which coupling the photon field to a macroscopic system is different that to a finite system.

Moreover, the light-matter coupling $\gamma$ of Eq.~(\ref{collective coupling}) is proportional to the number of particles\footnote{This fact can be understood more easily from the coupling constant in the effective theory $g(\Lambda)$ in Eq.~(\ref{effectivecoupling}) but it is also true for the Dicke model~\cite{dicke1954}}. This implies that the strength of the responses actually depends on the coupling constant. This suggests that light and matter in quantum electrodynamics are not only equivalent with respect to their excitations and resonances, but also the strengths of the their respective responses are related through the light-matter coupling constant (or number of particles).

Lastly, we highlight that the response functions we computed throughout this section depend on the arbitrarily small yet finite auxiliary parameter $\eta$, which is standard to introduce in linear response, in order to have a well-defined Laplace transform~\cite{Vignale, flick2018light}. In the limit $\eta\rightarrow0$ the response functions go to zero (see for example Eq.~(\ref{Re Im A-field})) except of the frequencies $w=\pm\widetilde{\omega}$ where they diverge. This implies that $\eta$ works like a regulator which spreads the resonance over a finite range and describes the coupling of the system to an artificial \textit{environment} and how energy is dissipated to this environment~\cite{Vignale}. To remove this arbitrary broadening parameter $\eta$, one can treat the matter and the photon sectors on equal footing and perform the continuum-limit also for the photon field. This as we will see in the next section allows for the description of absorption and dissipation without the need of $\eta$. 

\section{Effective Quantum Field Theory in the Continuum}\label{Effective Field Theory}

Up to here we have investigated in full generality the behavior of the free electron gas in the large $N$ or thermodynamic limit for the electronic sector, coupled to a single quantized mode. The single mode approximation has been proven very fruitful and successful for quantum optics and cavity QED~\cite{faisal1987, cohen1997photons}, but as it is known from the early times of the quantum theory of radiation and the seminal work of Einstein~\cite{Einstein:1917zz} to describe even one of the most fundamental processes of light-matter interaction like spontaneous emission the full continuum of modes of the electromagnetic field have to be taken into account. Moreover, we should always keep in mind that in a cavity set-up of course a particular set of modes of the electromagnetic field are selected by the cavity, but it is never the case that only a single mode of the cavity contributes to the light-matter coupling. The single mode models like the Rabi, Jaynes-Cummings or Dicke model, describe effectively (with the use of an effective coupling) the exchange of energy between matter and the photon field as if there were only a single mode coupled to matter~\cite{harochekleppner}.  

 In our case the situation becomes even more severe because we consider a macroscopic system like the 2DEG, where the propagation of the in-plane modes becomes important. This implies that the 2D continuum of modes of the electromagnetic field has to be taken into account. Before we proceed with the construction of the theory for the photon field in the continuum let us give some more arguments why such a theory is needed and what particular observables and physical processes can only be described in such a theory.
\begin{center}
    \textit{Why a Quantum Field Theory}?
\end{center}
From the point of view of observables and physical processes the main reasons are: (i) As we saw in section~\ref{Ground State} the contribution of the single mode cavity field to the ground state energy density $E_{p}/S$ in the thermodynamic limit, where the number of electrons $N$ and the area $S$ become arbitrary large, becomes arbitrary small and tends to zero. This implies that in the single mode case no significant contribution to the ground state of the system comes from the photon field, because of the discrepancy between the amount of the electrons and the amount of modes. (ii) As we mentioned in the end of the previous section, absorption processes and dissipation can be described consistently and from first principles only when a continuum of modes is considered~\cite{Vignale}. (iii) Since the contribution of the cavity field to the energy density is zero, compared to the energy density of the electrons, no real contribution to the renormalized or effective mass of the electron can occur. This again is due to the fact that we consider a single mode of the photon field, and as it known from QED, mass renormalization shows up when electrons are coupled to the full continuum  of the electromagnetic field~\cite{Weinberg, Srednicki, Frohlich2010, CHEN20082555, Mandl}. (iv) Lastly, no macroscopic forces can appear between the cavity mirrors, like the well-known Casimir-Polder forces~\cite{casimir1948influence}, in the single mode limit. As it well known from the literature such forces show up only when the full continuum of modes is considered~\cite{buhmann2013dispersionI, buhmann2013dispersionII}. For all these reasons we proceed with the construction of the effective field theory for a continuum of modes.

\subsection{Effective Field Theory, Coupling and Cutoff}\label{Cutoff}

To promote the single mode theory to a field theory we need to perform the ``thermodynamic limit'' for the photon field (in analogy to the electrons), and integrate over all the in-plane modes of the electromagnetic field. Such a procedure can be performed for an arbitrary amount of photon modes, with the mode-mode interactions included (see appendix~\ref{Mode-Mode Interactions}). However, such a treatment would make the theory non-analytically solvable, and particularly in the thermodynamic limit. 

For the latter reason, we will follow an alternative approach. We will perform the integration in an \textit{effective} way, where we will neglect the mode-mode interactions and we will integrate the single mode spectrum of Eq.~(\ref{eigenspectrum}) over all the in-plane modes. In this way we will be able to construct an analytically solvable effective field theory, in the thermodynamic limit for both light and matter. Before we continue we would like to mention that the validity of the approximation to neglect the mode-mode interactions depends on the how large the diamagnetic shift $\omega_p$~\cite{faisal1987} is. We will investigate and test this approximation in more detail later in subsections~\ref{Mode-Mode Tests} and~\ref{Effective Mass}.

To construct this effective quantum field theory, first we need to introduce back the dependence to the momenta $\bm{\kappa}=(2\pi n_x/L,2\pi n_y/L,\pi n_z/L_z)$ of the all the parameters of the theory. The bare modes $\omega$ of the quantized electromagnetic field in terms of the momenta $\bm{\kappa}$ are $\omega(\bm{\kappa})=c|\bm{\kappa}|$. Furthermore, for the dressed frequency $\widetilde{\omega}=\sqrt{\omega^2+\omega^2_p}$  we also need to introduce the $\bm{\kappa}$-dependence by promoting it to $\widetilde{\omega}(\bm{\kappa})=\sqrt{\omega^2(\bm{\kappa})+\omega^2_p}$. As a consequence, also the single-mode (many-body) coupling constant $\gamma=\omega^2_p/\widetilde{\omega}^2$ becomes $\bm{\kappa}$-dependent $\gamma(\bm{\kappa})=\omega^2_p/\widetilde{\omega}^2(\bm{\kappa})$. With these substitutions and summing the eigenspectrum of Eq.~(\ref{eigenspectrum}) over all the momenta in the $(x,y)$ plane,  we find the expression for the ground state energy (where $n_{\lambda}=0$ for both $\lambda=1,2$) for the effective theory 
\begin{widetext}
\begin{eqnarray}\label{effective energy}
E_{\mathbf{k}}(\Lambda)=\frac{\hbar^2}{2m_{\textrm{e}}} \left[\sum\limits^{N}_{j=1}\mathbf{k}^2_j-\left(\sum^{\Lambda}_{\kappa_x,\kappa_y}\gamma(\bm{\kappa})\right)\frac{1}{N}\sum^2_{\lambda=1}\left(\bm{\varepsilon}_{\lambda}\cdot \mathbf{K}\right)^2\right]+\sum^{\Lambda}_{\kappa_{x},\kappa_{y}}\hbar\widetilde{\omega}(\bm{\kappa}).
\end{eqnarray}
\end{widetext}

\begin{figure}[h]
\includegraphics[height=5.5cm,width=6cm]{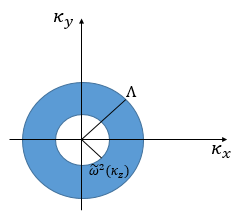}
\caption{\label{Photon Continuum}Representation of the frequency domain of the photon field in which the effective field theory is defined. The \textit{natural} lower cutoff of the theory is $\widetilde{\omega}^2(\kappa_z)$, while the highest allowed mode is $\Lambda$. }
\end{figure}

In the energy expression above we introduced the cutoff $\Lambda$ which defines the highest allowed frequency that we can consider in this effective field theory. Such a cutoff is necessary for effective field theories and it is standard to introduce it also for QED~\cite{greiner1996, spohn2004}. The sum over the single mode coupling constant $\gamma(\bm{\kappa})$ defines the effective coupling $g(\Lambda)$ of the effective field theory. For the effective coupling $g(\Lambda)$ we have
\begin{eqnarray}
    g(\Lambda)=\sum^{\Lambda}_{\kappa_x,\kappa_y}\gamma(\bm{\kappa})=\frac{e^2N}{\epsilon_0 m_{\textrm{e}}L_z}\frac{1}{S}\sum^{\Lambda}_{\kappa_x,\kappa_y}\frac{1}{\omega^2(\bm{\kappa})+\omega^2_p}.\nonumber\\
\end{eqnarray}
In the limit where the area of the cavity becomes macroscopic $S \rightarrow \infty$, the momenta $(\kappa_x,\kappa_y)$ of the photon field become continuous variables  and the sum gets replaced by an integral
\begin{eqnarray}\label{effectivecoupling}
    g(\Lambda)=\frac{e^2N}{\epsilon_0 m_{\textrm{e}}L_z}\frac{1}{4\pi^2}\iint\limits^{\Lambda}_{0}\frac{d \kappa_xd\kappa_y}{c^2\bm{\kappa}^2+\omega^2_p}=N\alpha\ln\left(\frac{\Lambda}{\widetilde{\omega}^2(\kappa_z)}\right)\nonumber\\
\end{eqnarray}
where we introduced the parameters
\begin{eqnarray}\label{alpha parameter}
    \alpha=\frac{e^2}{4\pi c^2\epsilon_0 m_{\textrm{e}}L_z} \;\; \textrm{and}\;\; \widetilde{\omega}^2(\kappa_z)=c^2\kappa^2_z+\omega^2_p,
\end{eqnarray}
and the momentum $\kappa_z=\pi/L_z$ (for $n_z=1$) depends on the distance between the cavity mirrors $L_z$ (see Fig.~\ref{HEG_Cavity}). Here comes a crucial point, the effective coupling $g(\Lambda)$ in Eq.~(\ref{effectivecoupling}) depends on the number of particles $N$. We would like to emphasize that the number of particles appears explicitly due to dipolar coupling, i.e. because in this effective field theory we couple \textit{all modes} to \textit{all particles} in the same way. However, in QED beyond the dipole approximation, each mode has a spatial profile which directly implies that the coupling is \textit{local}, in the sense that each mode couples to the local charge density and not to the full amount of electrons in the system. This is a second point in which the effectiveness of our field theory becomes manifest. This has implications because in the thermodynamic limit $N\rightarrow \infty$ the effective coupling $g(\Lambda)$ becomes arbitrarily large. Nevertheless, for the effective coupling $g(\Lambda)$ we can derive rigorously conditions under which the effective theory is stable and well defined.  

In section~\ref{Ground State} we found the ground state of the electron-photon system in the thermodynamic limit (with this limit performed only for the electrons) for all values of the single-mode coupling $\gamma$. Specifically we proved that if the coupling $\gamma$ exceeds the critical coupling $\gamma_c=1$ then the system is unstable and has no ground state. Now that we have promoted the single mode theory into an effective field theory we need to guarantee the stability of the theory by forbidding the effective coupling to exceed 1, $ 0 \leq g(\Lambda) \leq 1$. From this condition and given the definition of the effective coupling $g(\Lambda)$ in Eq.~(\ref{effectivecoupling}) we find the allowed range for the cutoff $\Lambda$ 
\begin{eqnarray}\label{Cutoffrange}
    \widetilde{\omega}^2(\kappa_z)\leq \Lambda \leq \widetilde{\omega}^2(\kappa_z)e^{1/N\alpha}.
\end{eqnarray}
From the expression above the highest allowed momentum for the photon field is $\widetilde{\omega}^2(\kappa_z)e^{1/N\alpha}$. Beyond this value the effective coupling $g(\Lambda)$ becomes larger than 1 and the system gets unstable and the energy diverges. In QED the finite momentum (or finite energy scale) for which the theory diverges is known as the \textit{Landau pole}~\cite{Srednicki}, and for that reason we will also refer here to the highest allowed momentum as the Landau pole 
\begin{eqnarray}\label{Landaupole}
    \Lambda_{\textrm{pole}}=\widetilde{\omega}^2(\kappa_z)e^{1/N\alpha}.
\end{eqnarray}
Moreover, from Eq.~(\ref{Cutoffrange}) it is clear that the cutoff $\Lambda$ is a multiple of the dressed frequency $\widetilde{\omega}^2(\kappa_z)$ which means that we can actually define $\Lambda$ in terms of a dimensionless parameter $\Lambda_0$ as
\begin{eqnarray}
    \Lambda=\widetilde{\omega}^2(\kappa_z)\Lambda_0\;\; \textrm{with}\;\; 1\leq \Lambda_0\leq e^{1/N\alpha}.
\end{eqnarray}
With this range chosen for $\Lambda_0$ the effective coupling is $0\leq g(\Lambda)\leq 1$ and the system is stable and has a ground state.

To complete this discussion on the construction of the effective field theory, we would like to see what is the infrared (IR) and the ultraviolet (UV) behavior of the field theory. From the expression for the effective coupling $g(\Lambda)$ in Eq.~(\ref{effectivecoupling}) it is clear that the effective coupling diverges if we allow the cutoff to go to infinity, $g(\Lambda)\rightarrow \infty$ for $\Lambda \rightarrow \infty$, which means that our theory has a UV divergence. This is the logarithmic divergence of QED which is known to exist for both relativistic and non-relativistic QED~\cite{Weinberg, Srednicki, greiner1996, spohn2004, Hiroshima}. On the other hand the effective coupling $g(\Lambda)$ of our theory has no IR divergence because for arbitrarily small momenta $\kappa_z=\pi/L_z$ the coupling goes to zero, $g(\Lambda)\rightarrow 0$ due to the parameter $\alpha$. The reason for which we have  an IR divergent-free theory is the appearance of the diamagnetic shift $\omega_p$ in Eq.~(\ref{effectivecoupling}) which defines the \textit{natural lower cutoff} of our theory~\cite{rokaj2019}. The diamagnetic shift appears due to the $\bi{A}^2$ term in the Pauli-Fierz Hamiltonian. Thus, we see that the diamagnetic term $\bi{A}^2$ makes non-relativistic QED IR divergent-free, while relativistic QED suffers from both UV and IR divergences. This is another fundamental reason for which the diamagnetic term $\bi{A}^2$ is of major importance.

\subsection{Mode-Mode Interactions}\label{Mode-Mode Tests}

For the sake of constructing an analytical effective field theory in the continuum, the mode-mode interactions in our treatment were neglected. The mode-mode interactions are an important element of QED because they are responsible for non-linear effects for the electromagnetic field beyond the classical regime. However, as we can understand from the extensive treatment presented in appendix~\ref{Mode-Mode Interactions}, the mode-mode interactions do not alter fundamentally the energy spectrum for the 2DEG coupled to the photon field. The mode-mode interactions shift the bare frequencies of the electromagnetic field and rotate the polarization vectors of the photon field (see also~\cite{faisal1987}). In a few-mode scenario these changes would be substantial modifications because the new normal modes would be at different points in the photonic frequency space and would probe different parts of the electronic spectrum.

\begin{figure}[h]
\includegraphics[height=5cm,width=6cm]{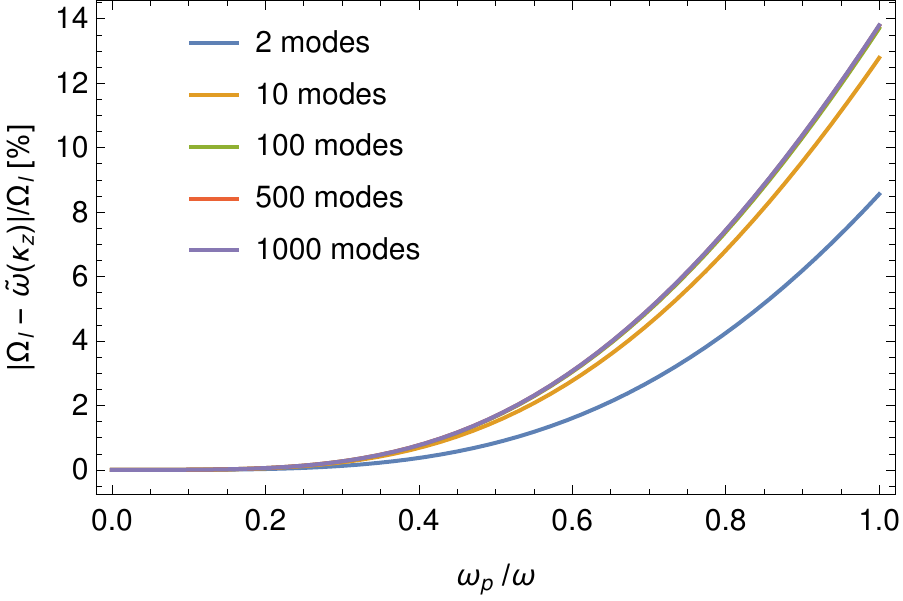}
\caption{\label{Lower Cutoff}Relative percentage difference of the effective lower cutoff $\widetilde{\omega}(\kappa_z)$ and the lowest normal mode $\Omega_l$ of the exact Hamiltonian (including the mode-mode interactions) plotted as function of the dimensionless ratio $\omega_p/\omega$. The computation is performed with different amount of modes. Above 100 modes the result is already converged. We assume all polarizations to be parallel to maximize the effect of the mode-mode interactions.}
\end{figure}

However, in the continuum these shifts in the frequencies are not of great importance, because upon integrating over all the photon frequencies a continuous domain of frequencies is spanned and all the modes within this domain are included (see Fig.\ref{Photon Continuum}). The only difference will be on how far the frequency domain extends. Consequently, the effect coming from the mode-mode interactions is to modify the lower and upper cutoff of the photon field. This means that the effective quantum field theory is not fundamentally different to a quantum field theory including the mode-mode interactions except of a re-definition of the cutoffs. In addition, the upper cutoff $\Lambda$ in the effective theory is left open and can be freely adopted depending on how far we aim to probe the photon field energetically. As a consequence the only approximation in the effective field theory is that the lower cutoff is assumed to be given by the expression $\widetilde{\omega}(\kappa_z)=\sqrt{c^2\kappa^2_z+\omega^2_p}$. To test quantitatively the validity of this approximation we compare the lower cutoff of the effective theory to the lowest normal mode $\Omega_l$ resulting from the exact diagonalization including the mode-mode interactions due to the $\bi{A}^2$ term, which is shown in appendix~\ref{Mode-Mode Interactions}. 

In Fig.~\ref{Lower Cutoff} we plot the relative percentage difference between the effective lower cutoff $\widetilde{\omega}(\kappa_z)$ and exact lowest frequency $\Omega_l$, as a function of the dimensionless ratio $\omega_p/\omega$. As it is shown the effective cutoff and the exact one differ by less than 10$\%$ from 0 until $\omega_p/\omega<0.9$. The relative difference exceeds 10$\%$ only for $\omega_p > 0.9\; \omega$. This result shows that the effective field theory is valid in the regime $0\leq \omega_p/\omega\leq 1$. Further, we would like to mention that the regime $0\leq \omega_p/\omega\leq 1$ is relevant for experiments performed even in the ultrastrong coupling regime~\cite{Keller2020, li2018}. We note that for $\omega_p > \omega$ the effective theory can be easily corrected by replacing the effective cutoff $\widetilde{\omega}(\kappa_z)$ with the exact lowest normal mode $\Omega_l$.

\textit{Running of the Coupling in 1D.}---To further test the validity of our effective field theory and of the approximation to neglect the mode-mode interactions we proceed by comparing the running of the coupling (in one dimension) in the effective theory, to the exact coupling as a function of the upper cutoff of the photon field. The effective coupling in one dimension is defined similarly to the effective coupling in the case of two dimensions~(\ref{effectivecoupling})
\begin{eqnarray}
g^{\textrm{1D}}(\Lambda)=\sum^{\Lambda}_{\kappa_z}\gamma(\kappa_z)
\end{eqnarray}
which upon introducing the fundamental cavity frequency $\omega=c\pi/L_z$ and summing over all the photon momenta from $0$ to $\Lambda$ in the thermodynamic limit, is found to be
\begin{eqnarray}
g^{\textrm{1D}}(\Lambda)=\frac{c\omega^2_p}{2\omega}\int^{\Lambda}_{0}\frac{d\kappa_z}{c^2\kappa^2_z+\omega^2_p}=\frac{\omega_p}{2\omega}\arctan\left(\frac{c\Lambda}{\omega_p}\right).\nonumber\\
\end{eqnarray}
From the above result we see that the effective coupling is an arctangent function of the upper cutoff $\Lambda$. Moreover, from the exact diagonalization performed in appendix~\ref{Mode-Mode Interactions} the exact coupling constant as a function of the amount of modes $M$ is
\begin{eqnarray}
g^{\textrm{1D}}_{\textrm{ex}}(M)=\sum^M_{\alpha=1}\frac{\omega^2_p(\widetilde{\varepsilon}_{\alpha})}{\Omega^2_{\alpha}},
\end{eqnarray}
where $\Omega_{\alpha}$ and $\widetilde{\varepsilon}_{\alpha}$ are the new normal modes and the new polarization vectors including also the mode-mode interactions. The exact coupling constant can be computed using the exact diagonalization presented in appendix~\ref{Mode-Mode Interactions}. In figure~\ref{Exact Coupling} we show the running of the exact coupling constant as a function of the amount of photon modes $M$.
\begin{figure}[h]
\includegraphics[height=6cm,width=7cm]{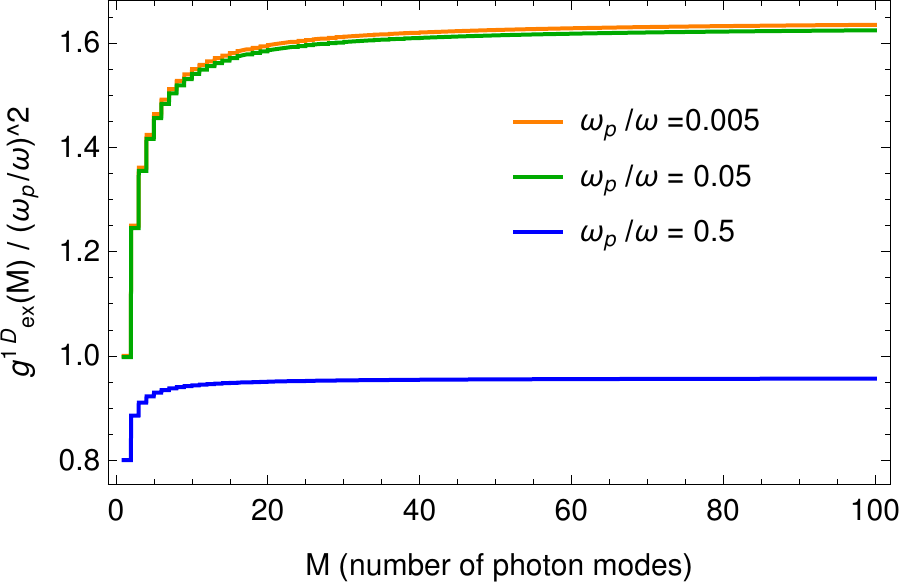}
\caption{\label{Exact Coupling}Exact many-mode coupling constant $g^{\textrm{1D}}_{\textrm{ex}}(M)$ normalized by $\left(\omega_p/\omega\right)^2$ in 1D, as a function of the number of photon modes $M$. The exact coupling is plotted for three different values of the ratio $\omega_p/\omega$ corresponding to different regimes of light-matter interaction. In all cases we see that the exact coupling has arctangent dependence on the amount of photon modes. The exact coupling is normalized by $\left(\omega_p/\omega\right)^2$ such that all graphs to be visible in one plot.}
\end{figure}

As it is depicted in Fig.~\ref{Exact Coupling} the exact many-mode coupling has an arctangent dependence on the amount of photon modes $M$. This implies that the exact coupling and the effective coupling exhibit the same running as a function of the upper cutoff of the photon field. This is very important because it demonstrates that the effective field theory that we constructed in the continuum not only describes accurately the lower and upper energetic (or frequency) cutoffs of the photon field but also captures the correct behavior of the coupling constant, which is of fundamental importance. This is a crucial benchmark for our effective field theory.

\subsection{Renormalized \& Effective Mass}\label{Effective Mass}

As it is known from relativistic QED when electrons interact with the full continuum of modes of the electromagnetic field the mass and charge of the electron get renormalized. Such renormalizations are known to lead to observable radiative corrections like vacuum polarization, the anomalous magnetic moment and the Lamb shift~\cite{Weinberg, Mandl}. In non-relativistic QED there is no need for charge renormalization, due to the elimination of positrons from the theory~\cite{HiroshimaSpohn}. However, mass renormalization effects show up. Here, we are interested in the renormalization of the electron mass due the interaction of the electron with the continuum of modes of the cavity.

\begin{figure}[h]
\includegraphics[height=6cm,width=7cm]{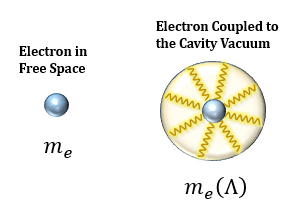}
\caption{\label{Penormalized Mass}Schematic depiction of an electron in free space with mass $m_{\textrm{e}}$ and an electron coupled to the vacuum of the electromagnetic field. The virtual photons of the cavity vacuum ``dress'' the electron and provide a radiative correction to the electron mass $m_{\textrm{e}}(\Lambda)$. }
\end{figure}

Generally computing the renormalized mass is a rather difficult task, in most cases performed perturbatively with methods ranging from dimensional regularization~\cite{Mandl}, renormalization group techniques~\cite{Srednicki, Wilson, Weinberg} or causal perturbation theory~\cite{EpsteinGlaser}. 

In non-relativistic QED the renormalized  electron mass for free electrons is defined via the energy dispersion of the electrons around $\bi{k}=0$ and is given by the formula~\cite{Frohlich2010, CHEN20082555}
\begin{eqnarray}\label{Def Renormalized Mass}
    m_{\textrm{e}}(\Lambda)=\left(\frac{1}{\hbar^2}\frac{\partial^2 E_{\bi{k}}(\Lambda)}{\partial \bi{k}^2_i}\right)^{-1}
\end{eqnarray}
 where $E_{\bi{k}}(\Lambda)$ is the energy dispersion of the electron-photon system, which depends on the momenta of the electrons and the cutoff of the theory.

\begin{figure}[h]
    \centering
    \includegraphics[width=\linewidth]{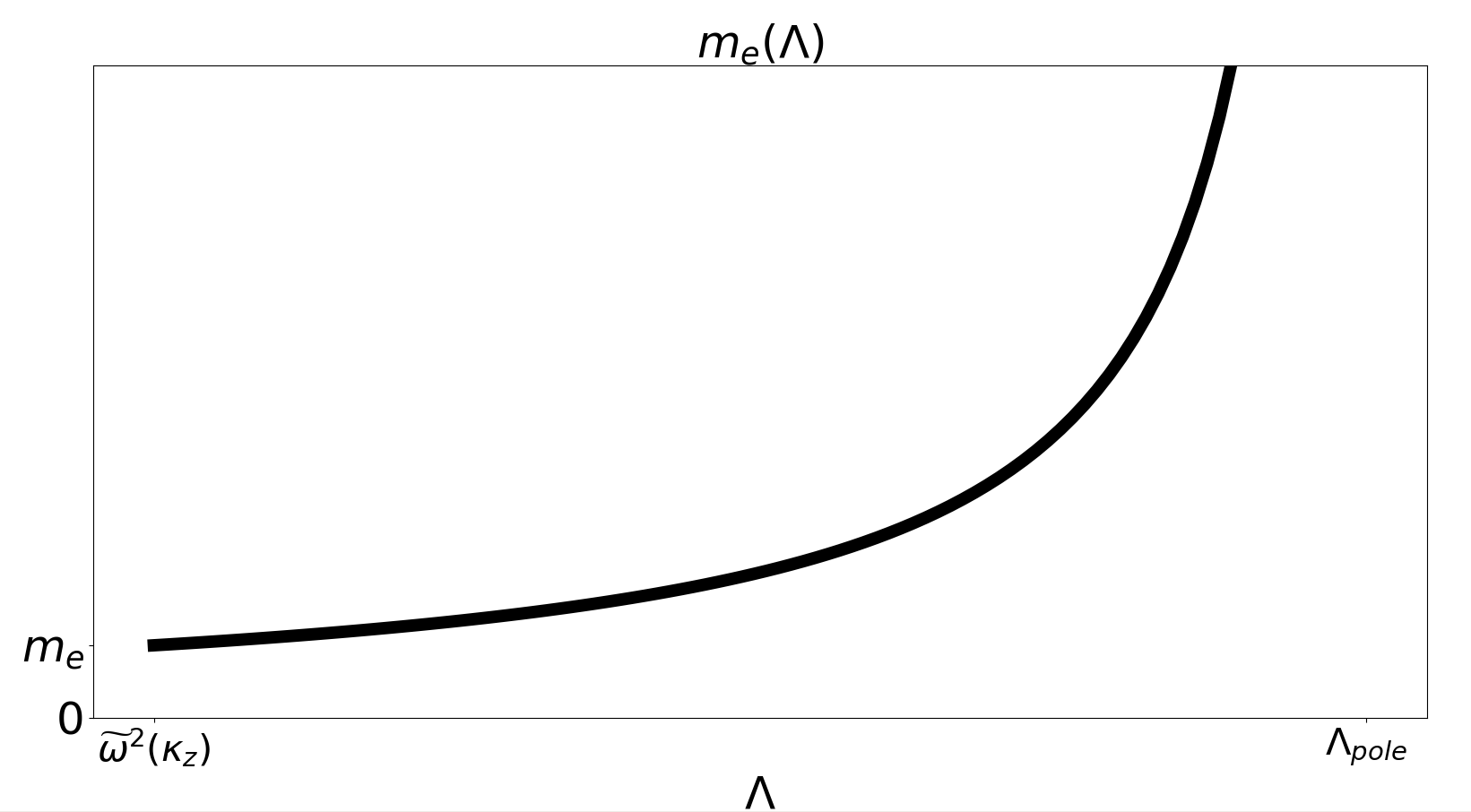}
    \caption{\label{mass_running}Renormalized mass $m_{\textrm{e}}(\Lambda)$ as a function of the cutoff $\Lambda$. For the cutoff $\Lambda$ being equal to the natural lower cutoff $\widetilde{\omega}^2(\kappa_z)$ the renormalized mass is equal to the free-space electron mass $m_{\textrm{e}}(\widetilde{\omega}^2(\kappa_z))=m_{\textrm{e}}$. As the cutoff increases the renormalized mass $m_{\textrm{e}}(\Lambda)$ gets larger than the free-space mass $m_{\textrm{e}}$ and eventually goes to infinity at the Landau pole $\Lambda_{\textrm{pole}}$. }
\end{figure}

In our case, we have diagonalized analytically the single mode Hamiltonian of Eq.~(\ref{single mode Hamiltonian}) and then we promoted the single mode energy spectrum given by Eq.~(\ref{eigenspectrum}) into the energy spectrum of Eq.~(\ref{effective energy}) which describes the effective field theory we constructed. Since we have an analytic expression for the energy spectrum $E_{\bi{k}}(\Lambda)$ of the effective theory given by Eq.~(\ref{effective energy}), for the computation of the renormalized mass we do not need to use any of the techniques we mentioned before, but we can straightforwardly use the definition for $m_{\textrm{e}}(\Lambda)$ of Eq.~(\ref{Def Renormalized Mass}). Thus, we find for the renormalized electron mass 
\begin{eqnarray}\label{Renormalized Mass}
    m_{\textrm{e}}(\Lambda)=m_{\textrm{e}}\left(1-\alpha\ln\left(\frac{\Lambda}{\widetilde{\omega}^2(\kappa_z)}\right)\right)^{-1}.
\end{eqnarray}
From the expression above we see that the renormalized electron mass $m_{\textrm{e}}(\Lambda)$ is larger than the electron mass in free space $m_{\textrm{e}}$ and increases as function of the coupling $\Lambda$. This behavior is in accordance with results coming from both relativistic and non-relativistic QED~\cite{Srednicki, Weinberg, Frohlich2010, CHEN20082555, HiroshimaSpohn}. Within the range of the cutoff $\Lambda$ given by Eq.~(\ref{Cutoffrange}) the renormalized mass is always positive and the effective theory is well-defined (see Fig.~\ref{mass_running}). If the cutoff though goes beyond the Landau pole $\Lambda_{\textrm{pole}}$ (which actually is a forbidden regime) the renormalized mass can become even negative and signifies that the theory becomes unstable, similarly to the single mode theory when the coupling coupling $\gamma$ goes beyond the critical coupling $\gamma_c$. In the limit where the cutoff $\Lambda$ takes its minimum value $\widetilde{\omega}^2(\kappa_z)$ the renormalized mass $m_{\textrm{e}}(\Lambda)$ is equal to $m_{\textrm{e}}$ (see Fig.~\ref{mass_running}). This explains also why in the single mode theory the electron mass does not get renormalized.

Moreover, from Eq.~(\ref{Renormalized Mass}) we see that the renormalized mass $m_{\textrm{e}}(\Lambda)$ depends also on the distance between the cavity mirrors $L_z$ (via $\alpha$) and most importantly on the full electron density in the cavity $n_{\textrm{e}}$ via the dressed frequency $\widetilde{\omega}(\kappa_z)$ given by Eq.~(\ref{alpha parameter}). The fact that $m_{\textrm{e}}(\Lambda)$ depends on the full electron density $n_{\textrm{e}}$ means that we can observe a \textit{many-body effect} in the renormalized mass $m_{\textrm{e}}(\Lambda)$. This many-body effect shows up because we consider here the many-body electron system of $N$ free electrons coupled to the electromagnetic field and our treatment is \textit{non-perturbative}. We emphasize that such a many-body mass renormalization effect does not show up for the usual single particle mass renormalization~\cite{Mandl, BetheRenorm} and is potentially very small for any finite system, but clearly not for extended systems. To the best of our knowledge such a many-body mass renormalization has not been demonstrated before. We note that the inclusion of the Coulomb interaction would result into further mass renormalization effects~\cite{Eicheffectivemass}.

\begin{figure}[h]
\includegraphics[width=\linewidth]{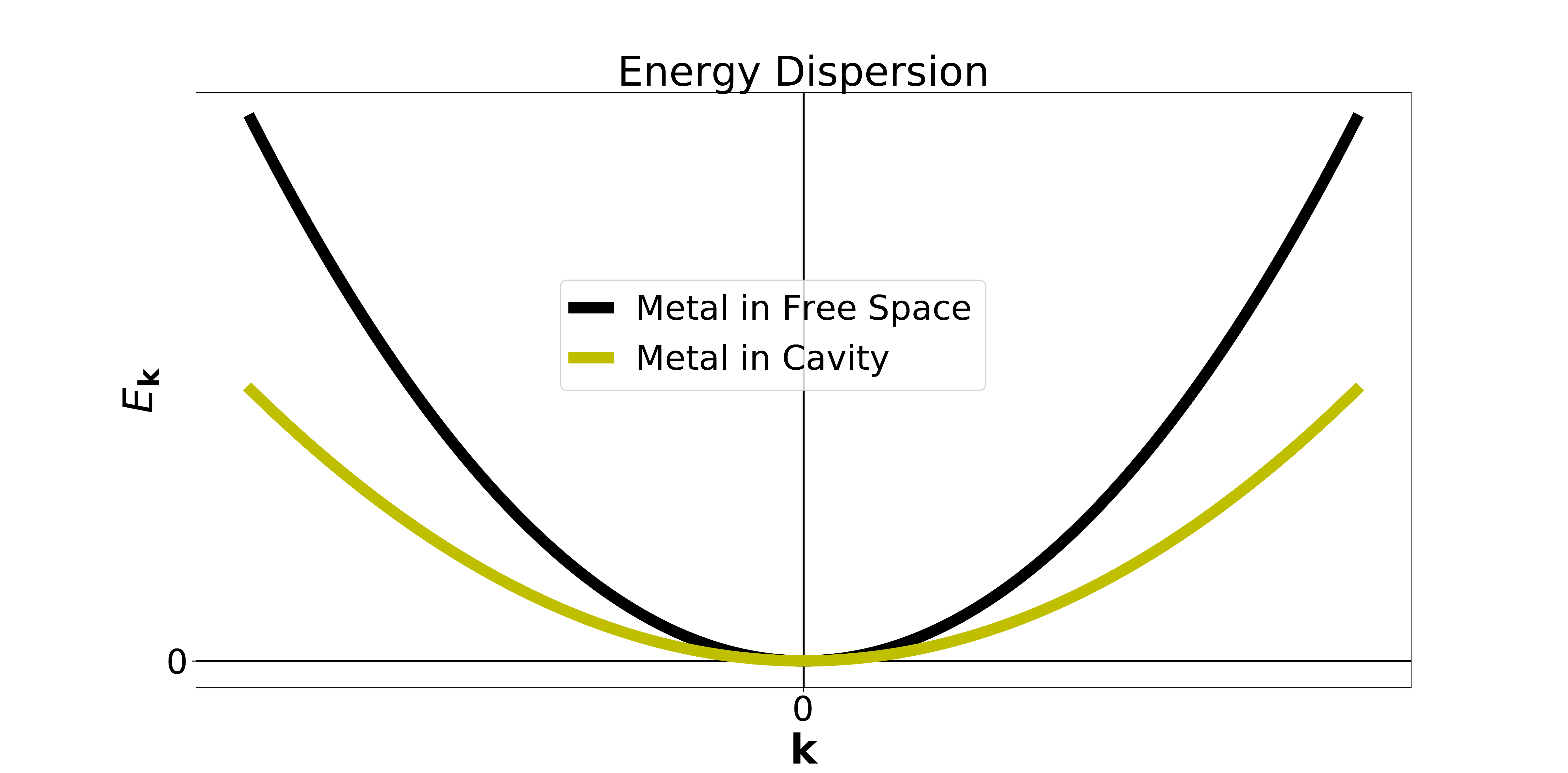}
\caption{\label{Parabola Renormalization} Energy dispersion for electrons in a metal outside a cavity (black parabola) and for electrons in a metal coupled to a cavity (yellow parabola). From the curvature of the parabolas the effective mass of the electrons in the respective environment can be obtained. The dispersion for the electrons in the cavity is less steep, because the electron mass is larger in the cavity due to the ``dressing'' by the cavity photons. }
\end{figure}

The renormalization of the electron mass due to the cavity field has experimental implications and can be measured experimentally by comparing the effective mass of the electrons \textit{outside} the cavity, to the effective mass \textit{inside} the cavity (see also Fig.~\ref{Parabola Renormalization}). The relation between the two is given by the formula which we derived for the renormalized mass $m_{\textrm{e}}(\Lambda)$ in Eq.~(\ref{Renormalized Mass}). Having obtained experimentally the ratio $m_{\textrm{e}}(\Lambda)/m_{\textrm{e}}$ the formula of Eq.~(\ref{Renormalized Mass}) allows us to deduce directly what is the highest momentum (the cutoff) $\Lambda$ to which the electrons are coupled to, and using Eq.~(\ref{effectivecoupling}) what is the coupling $g(\Lambda)$ to the cavity photons. 

We believe this provides a novel direct way to determine the light-matter coupling strength for extended systems in cavity QED and the effective volume of the cavity. We elaborate on this further in section~\ref{Summary}.  

\textit{Single-Particle Mass Renormalization in 3D.}---As an additional test of our effective quantum field theory, and our prediction for the many-body renormalization of the electron mass in the 2DEG, we will consider our effective theory in 3D and compute the singe-particle mass renormalization, for which several analytic results exist~\cite{BetheRenorm, HainzlMass, Mandl}.

The solution for the free electron gas coupled to a cavity mode is the same also in three dimensions. Consequently, the energy spectrum in the effective quantum field theory will have the same form as the one in 2D given by Eq.~(\ref{effective energy}). The only differences will be that the spectrum will depend on the 3D momenta of the electrons, and that the effective coupling $g^{\textrm{3D}}(\Lambda)$ will be given by the sum of the single-particle coupling constants $\gamma(\bm{\kappa})$ over all 3D photonic momenta $\bm{\kappa}=(\kappa_x,\kappa_y,\kappa_z)$ 
\begin{eqnarray}
g^{3\textrm{D}}(\Lambda)=\frac{2}{3}\sum_{\bm{\kappa}}\gamma(\bm{\kappa})=\frac{e^2N}{m_{\textrm{e}}\epsilon_0V}\sum_{\bm{\kappa}}\frac{1}{c^2\bm{\kappa}^2+\omega^2_p}.
\end{eqnarray}
We note that the prefactor $2/3$ is due to the fact that in 3D we have three spatial dimensions but only two possible polarizations for the photon field. In 2D, the respective prefactor is equal to 1. In the thermodynamic limit the sum above turns into an integral. Moreover, in the single particle case $N=1$ and the diamagnetic plasma frequency which depends on the electron density is negligible, $\omega_p \approx 0$. Thus, for the effective coupling we find
\begin{eqnarray}\label{effective g 3D}
g^{3\textrm{D}}(\Lambda)=\frac{2}{3}\frac{e^2}{m_{\textrm{e}}\epsilon_0(2\pi)^3}\iiint\limits^{\Lambda}_{0}\frac{d^3\bm{\kappa}}{c^2\bm{\kappa}^2}= \frac{4\alpha_{\textrm{fs}}}{3\pi}\frac{\hbar\Lambda}{m_{\textrm{e}}c}.\nonumber\\
\end{eqnarray}
To obtain the above result we also introduced the fine structure constant $\alpha_{\textrm{fs}}=e^2/4\pi\hbar c\epsilon_0$. Having the expression for the effective coupling, we can straightforwardly compute the renormalized electron mass according to the definition given in Eq.~(\ref{Def Renormalized Mass}), and we find
\begin{eqnarray}
m^{3\textrm{D}}_{\textrm{e}}(\Lambda)=\frac{m_{\textrm{e}}}{1-g^{3\textrm{D}}(\Lambda)}
\end{eqnarray}
which to the lowest order in the fine structure constant gives the following result for the renormalized mass 
\begin{eqnarray}
m^{\textrm{3\textrm{D}}}_{\textrm{e}}(\Lambda)=m_{\textrm{e}}+\frac{4\alpha_{\textrm{fs}}}{3\pi}\frac{\hbar\Lambda}{c}.
\end{eqnarray}
The above result reproduces exactly the single-particle mass renormalization for a free electron, which diverges linearly to the upper cutoff of the photon field, as discussed and obtained by several authors, for example Bethe~\cite{BetheRenorm}, Hainzl and Seiringer~\cite{HainzlMass} and Mandl and Shaw~\cite{Mandl}. This results provides another importation validation of our effective field theory.

Before we continue we would like to highlight another important point which emerges from our effective field theory in three dimensions. As we already discussed for our theory to be stable the effective coupling shall not exceed the value of one. Imposing this stability condition on $g^{\textrm{3D}}(\Lambda)$ given by Eq.~(\ref{effective g 3D}) we find that the maximum value for the upper cutoff $\Lambda^{\textrm{3D}}_{\textrm{pole}}$ is
\begin{eqnarray}\label{3D Pole}
\Lambda^{\textrm{3D}}_{\textrm{pole}}=\left(\frac{4\alpha_{\textrm{fs}}}{3\pi}\frac{\hbar}{m_{\textrm{e}}c}\right)^{-1}=0.84\times 10^{15}\textrm{m}^{-1}\sim \frac{1}{l_{\textrm{QCD}}}\nonumber\\
\end{eqnarray}
where $l_{\textrm{QCD}}=\hbar/m_pc=2.1\times 10^{-16}\textrm{m}$ is the length scale of quantum chromodynamics (QCD), defined with respect to the proton mass $m_p$, at which phenomena related to strong nuclear forces become important~\cite{Wilczek, Bettini}. Equation~(\ref{3D Pole}) shows that our effective quantum field theory is applicable for most regimes of light-matter interaction, and breaks down only at a scale inverse to the QCD length scale $l_{\textrm{QCD}}$. This means, that our effective quantum field theory is also able to predict the scale at which new phenomena take place, at the nucleus of the proton. Further, we would like to mention that since our theory is non-relativistic, we obtain a very much lower value for our pole than the relativistic value of the Landau pole~\cite{Weinberg, Srednicki}. This is a beautiful consistency check of our non-relativistic theory.

\subsection{Modified Fermi Liquid Quasiparticle Excitations}

Let us proceed by showing some further consequences of the effective field theory and more precisely implications for Fermi liquid theory.
In section~\ref{Ground State} we showed that the electronic ground state is one in which all single particle states with momenta less than the Fermi momentum $\bi{p}_{\textrm{F}}=\hbar \bi{k}_{\textrm{F}}$ are occupied. All other single particle states are empty. This is the starting assumption in Fermi liquid theory~\cite{Nozieres, LandauFermiLiquid}. The fundamental fermionic quasiparticle excitations of the Fermi liquid theory are generated by adding electrons with momentum greater than the Fermi momentum $\bi{p}_{\textrm{F}}$~\cite{Nozieres, Baym}. The energy of the quasiparticle at the Fermi surface is
\begin{eqnarray}
    \mu= E_{\bi{k}}(\Lambda, N+1)-E_{\bi{k}}(\Lambda, N),
\end{eqnarray}
where $E_{\bi{k}}(\Lambda, N)$ is the ground state energy of the system for $N$ electrons distributed on the Fermi sphere with their wavevectors in the region $0\leq\bi{k}<\bi{k}_{\textrm{F}}$ and $E_{\bi{k}}(\Lambda, N+1)$ is the energy of the system containing one more particle with $\bi{k}=\bi{k}_{\textrm{F}}$. In the ground state where the electrons are distributed on the Fermi sphere the collective momentum is zero, $\bi{K}=0$. Thus, in the energy $E_{\bi{k}}(\Lambda, N)$ the negative term appearing in the effective spectrum in Eq.~(\ref{effective energy}) does not contribute. However, in the $N+1$ state the last electron added on the Fermi surface with $\bi{k}=\bi{k}_{\textrm{F}}$ introduces a non-zero momentum to the system. Consequently, the photon-mediated negative term now gives a contribution. Thus, we find that the quasiparticle excitation energy at the Fermi surface is
\begin{eqnarray}
    \mu= \frac{\hbar^2}{2m_{\textrm{e}}}\left[\bi{k}^2_{\textrm{F}}-\alpha\ln\left(\frac{\Lambda}{\widetilde{\omega}^2(\kappa_z)}\right)\sum^2_{\lambda=1}\left(\bm{\varepsilon}_{\lambda}\cdot \bi{k}_{\textrm{F}}\right)^2\right].
\end{eqnarray} 
To obtain the above result we used that the effective coupling per particle is $g(\Lambda)/(N+1)=\alpha\ln\left(\Lambda/\widetilde{\omega}^2(\kappa_z)\right)$ as given by Eq.~(\ref{effectivecoupling}). Further, using that the polarization vectors are orthogonal and introducing the renormalized mass $m_{\textrm{e}}(\Lambda)$ the quasiparticle excitation at the Fermi surface takes the compact form
\begin{eqnarray}\label{chemical potential}
    \mu=\frac{\hbar^2\bi{k}^2_{\textrm{F}}}{2m_{\textrm{e}}(\Lambda)}.
\end{eqnarray}
The quasiparticle excitation at the Fermi surface $\mu$ is also known as the chemical potential. From Eq.~(\ref{chemical potential}) we see that the chemical potential depends on the renormalized (by the photon field) electron mass $m_{\textrm{e}}(\Lambda)$ given by Eq.~(\ref{Renormalized Mass}). This shows that in the effective field theory the photon field modifies the chemical potential. Moreover, in Fermi liquid theory the quasiparticle excitations in the neighborhood of the Fermi surface depend on the chemical potential~\cite{Baym}
\begin{eqnarray}
    \epsilon^x_{\bi{k}}=\mu+\hbar v_{\textrm{F}}(\bi{k}-\bi{k}_{\textrm{F}})=\frac{\hbar^2\bi{k}^2_{\textrm{F}}}{2m_{\textrm{e}}(\Lambda)}+\hbar v_{\textrm{F}}(\bi{k}-\bi{k}_{\textrm{F}}),\nonumber\\
\end{eqnarray}
where $v_{\textrm{F}}$ is the Fermi momentum at the Fermi surface. Thus, the quasiparticle excitations in the Fermi liquid get also modified and depend on the renormalized electron mass $m_{\textrm{e}}(\Lambda)$. This demonstrates that the effective field theory we constructed has direct implications for Fermi liquid theory. Lastly, we highlight that in the limit where the upper cutoff goes to the lower cutoff, $\Lambda \rightarrow \widetilde{\omega}^2(\kappa_z)$, the renormalized mass goes to the bare electron mass, $m_{\textrm{e}}(\Lambda) \rightarrow m_{\textrm{e}}$, and in this case no modification of the Fermi liquid shows up. This explains from another point of view why in the single-mode theory the Fermi liquid does not get modified and why a field theory in the continuum for the photon field is necessary to see a modification of the Fermi liquid.   

\subsection{Jellium Model \& Coulomb Interaction}

To further illustrate the applicability of the our effective field theory and discuss its implications for electronic systems interacting also via the Coulomb interaction we will consider the jellium model, which provides a first approximation to a metal or a plasma~\cite{Vignale, FetterWalecka, AltlandSimons}. The jellium model is an interacting electron gas placed in a uniformly distributed positive background representing the ions, chosen to ensure the neutrality of the full system. Due to the positive background, the Hamiltonian of the jellium model can then be written as the sum of the kinetic energy of all the electrons plus the regularized, by the positive background, Coulomb interaction. We note that this regularization is very important as it eliminates a particular divergent contribution of the Coulomb interaction~\cite{Vignale, FetterWalecka,AltlandSimons}. 

However, in the effective field theory that we constructed we do not have only the homogeneous background of the ions, but we also have the neutral homogeneous background of the photons in which the electrons are embedded. As we already saw in the previous subsections, the photonic background renormalizes the electrons and introduces fermionic quasi-particles with an effective electronic mass $m_{\textrm{e}}(\Lambda)$ given by Eq.~(\ref{Renormalized Mass}). This makes clear that the jellium model in our effective field theory shall be one consisting of the kinetic energy of these fermionic quasi-particles with effective mass $m_{\textrm{e}}(\Lambda)$, interacting via the regularized Coulomb interaction. Thus, the jellium model (in 2D) in our effective field theory is given by the following Hamiltonian 
\begin{eqnarray}\label{jellium}
\hat{H}_{\textrm{jell}}(\Lambda)=\underbrace{-\frac{\hbar^2}{2m_{\textrm{e}}(\Lambda)}\sum^N_{j=1}\nabla^2_j}_{\hat{T}(\Lambda)}+\underbrace{\frac{1}{8\pi \epsilon_0 S}\sum_{\bi{q}\neq 0}v_{\bi{q}}\left[\hat{n}_{-\bi{q}}\hat{n}_{\bi{q}}-\hat{N}\right]}_{\hat{W}_{\textrm{e-e}}}\nonumber\\
\end{eqnarray}
which depends parametrically on the upper cutoff $\Lambda$ of the photon field, via the effective mass. We note that in the jellium Hamiltonian $S=L^2$ is the area of the 2D electron gas, $v_{\bi{q}}=2\pi e^2/|\bi{q}|$ is the Fourier transform of the regularized Coulomb interaction in 2D, $\hat{n}_{\bi{q}}=\sum_je^{-\textrm{i}\bi{q}\cdot\bi{r}_j}$ is the Fourier transform of the electronic density operator and $\hat{N}=\hat{n}_0$ is the number operator~\cite{Vignale}.

The crucial step to treat the Coulomb interaction, comes from the observation that in this system there exists a natural length scale $r_sa_0$, where $a_0=4\pi \epsilon_0\hbar^2/m_{\textrm{e}}e^2$ is the Bohr radius, with respect to which the kinetic energy and the Coulomb energy scale differently. The parameter $r_s$ is defined with respect to the 2D electron density~\cite{Vignale}
\begin{eqnarray}
r_s=\frac{1}{a_0}\left(\frac{1}{\pi n_{\textrm{2D}}}\right)^{\frac{1}{2}}.
\end{eqnarray}
We note that the parameter $r_s$ is known as the Wigner-Seitz radius~\cite{Mermin} and that the length $r_sa_0$ is the radius enclosing on average one electron. With respect to the natural length $r_sa_0$ we can define the following scaled variables~\cite{Vignale, AltlandSimons, FetterWalecka}
\begin{eqnarray}
\widetilde{\bi{r}}=\frac{\bi{r}}{r_sa_0},\;\;\; \widetilde{\bi{q}}=r_sa_0\bi{q}\;\;\; \textrm{and}\;\;\; \widetilde{S}=\frac{S}{(r_sa_0)^2}.
\end{eqnarray}
With these definitions the jellium Hamiltonian takes the form
\begin{widetext}
\begin{eqnarray}
\hat{H}_{\textrm{jell}}(\Lambda)=\frac{e^2}{2a_0}\left(\frac{m_{\textrm{e}}}{m_{\textrm{e}}(\Lambda)}\frac{1}{r^2_s}\sum_j\widetilde{\nabla}^2_j+\frac{1}{r_s\widetilde{S}}\sum_{\widetilde{\bi{q}}\neq 0}\frac{2\pi}{|\widetilde{\bi{q}}|}\left[\hat{n}_{-\widetilde{\bi{q}}}\hat{n}_{\widetilde{\bi{q}}}-\hat{N}\right]\right).
\end{eqnarray}
\end{widetext}
We note that in the last step we introduced also the standard convention for the jellium model in which $4\pi \epsilon_0=1$ which implies that the energy is measured now in Rydberg units, $1\textrm{R}_{\textrm{y}}=e^2/2a_0$~\cite{FetterWalecka, Vignale}. From the expression of the jellium Hamiltonian in terms of the scaled variables it becomes evident that with respect to $r_s$ the kinetic energy and Coulomb energy scale differently. As a consequence, in the regime of small $r_s$, or equivalently large densities, the Coulomb interaction can be considered as perturbation to the kinetic contribution. Thus, in the large density regime the solution of the non-interacting electron gas is perfectly valid and we can use the many-body eigenfunctions we obtained in section~\ref{EG in cavity QED}, to compute the Coulomb contribution perturbatively. We note that in our photon-modified jellium model, the kinetic contribution depends on the ratio between the electron mass inside and outside the cavity $m_{e}/m_{\textrm{e}}(\Lambda)$. This ratio is what determines the electron-photon coupling in the effective field theory. 

First, we compute the kinetic energy per particle with respect to the ground state of our system given by~Eq.~(\ref{Thermodynamic GS}). Due to the fact that the electrons are distributed on the 2D Fermi sphere we find~\cite{Vignale}
\begin{eqnarray}
\tau(r_s,\Lambda)=\frac{\langle \Psi_{gs}|\hat{T}(\Lambda)|\Psi_{gs}\rangle}{N}= \frac{e^2}{2a_0}\frac{m_{\textrm{e}}}{m_{\textrm{e}}(\Lambda)} \frac{1}{r^2_s}.
\end{eqnarray}
As we see the kinetic energy per particle is modified by the photons because it depends on the ratio $m_{\textrm{e}}/m_{\textrm{e}}(\Lambda)$. Then, to first order in perturbation theory the contribution of the Coulomb energy per particle is~\cite{Vignale}
\begin{eqnarray}
\epsilon_x(r_s)=\frac{\langle \Psi_{gs}|\hat{W}_{\textrm{e-e}}|\Psi_{gs}\rangle}{N}=-\frac{8\sqrt{2}}{3\pi}\frac{e^2}{2a_0}\frac{1}{r_s}.
\end{eqnarray}
The above contribution is also known as the exchange energy and as we see it does not depend on any photonic parameter. The total energy per particles is of course the sum of these two contributions $\mathcal{E}(r_s,\Lambda)=\tau(r_s,\Lambda)+\epsilon_x(r_s)$ and is a function of the Wigner-Seitz radius $r_s$. Minimizing the energy with respect to $r_s$, we find that the minimum Wigner-Seitz radius for the interacting electron gas is
\begin{eqnarray}
(r_s)_{\textrm{min}}(\Lambda)=\frac{3\pi}{4\sqrt{2}}\frac{m_{\textrm{e}}}{m_{\textrm{e}}(\Lambda)}
\end{eqnarray}
which is a function of the upper cutoff $\Lambda$. Since $m_{\textrm{e}}(\Lambda)$ is larger than $m_{\textrm{e}}$ (see Eq.~(\ref{Renormalized Mass}) or Fig.~\ref{mass_running}) the Wigner-Seitz radius for the interacting electrons coupled to the photon field is smaller than the uncoupled. This implies that when the interacting electrons are coupled to the photon field the radius containing one electron on average becomes smaller. This means that the photon field localizes the electrons, in the sense that (on average) an electron occupies a smaller volume in space. This is a significant result because most electronic-structure properties depend on the Wigner-Seitz radius and the average electron density. We note that such localization effects on the electronic density have also been reported for atomic and molecular systems under strong coupling to a cavity, with first-principle calculations~\cite{flick2015, flick2017}.  

We would like to mention that for the upper cutoff being equal to the lower cutoff, $\Lambda=\widetilde{\omega}^2(\kappa_z)$, $m_{\textrm{e}}(\Lambda)$ becomes equal to $m_{\textrm{e}}$ and the photon modified jellium goes to the standard jellium Hamiltonian and we recover all the respective known results of the jellium model~\cite{Vignale}. 

\textit{Beyond the First Order Coulomb Contribution}.---It is important to emphasize that in second or higher order contributions, the excited states of the electron-photon system will contribute to the Coulomb energy. The excited states of our system, as can be easily seen from Eq.~(\ref{eigenstates}), are correlated states between the electrons and the photons and as a consequence the photonic states will contribute to the higher orders and modify the correlation energy. This computation of the correlation energy is non-trivial and needs a separate treatment. 

\textit{Beyond the Large Density Regime}.---It is worth to mention that the perturbative treatment we performed here has certain limitations and it is not straightforwardly applicable in the intermediate and the low density regimes~\cite{AltlandSimons}. The intermediate density regime is the realm of Landau's Fermi liquid theory~\cite{LandauFermiLiquid, Baym} which is based on the concept of adiabatic continuity. Under this assumption, the non-interacting ground state evolves smoothly into the interacting one, without ruining the existence of a well-defined Fermi surface and well-defined quasi-particle excitations. The interaction between these quasi-particles is then usually treated with perturbative Green's functions methods. In the low density regime, the many-body ground state is no longer the one of the non-interacting (or weakly interacting) electrons, as the system it is believed to undergo a phase transition in which a Wigner crystal is formed~\cite{Vignale, AltlandSimons, FetterWalecka}. This is the regime in which the free electron gas does not provide a good starting point for the many-body ground state.

\subsection{Repulsive Casimir Force for a Non-Empty Cavity}\label{Repulsive Casimir Forces}
  
 Having defined and constructed the effective field theory for the continuum of modes, we want to proceed by computing the zero point energy of the electromagnetic field. The zero point energy of the electromagnetic field it is known to be responsible for forces like the interatomic van der Waals forces, the Casimir-Polder forces between an atom and a body~\footnote{By body here we mean a macroscopic object.} ~\cite{casimir1948influence, buhmann2013dispersionI}, and the Casimir force between parallel conducting plates~\cite{Casimir:1948dh}. Since we consider a 2D material in a cavity we fall in the third category and the macroscopic forces in the system should be Casimir forces. To find the Casimir force between the mirrors of the cavity we need to compute the zero point energy of the electromagnetic field $E_p $ per area $S$ of the cavity mirrors. From the energy expression of the effective theory in Eq.~(\ref{effective energy}) we deduce that the ground state energy ($n_{\lambda}=0$) per area is
  \begin{eqnarray}
     \frac{E_p}{S}=\frac{1}{S}\sum_{\kappa_x,\kappa_y}\hbar\widetilde{\omega}(\bm{\kappa})=\frac{\hbar}{4\pi^2}\iint\limits^{\Lambda}_{0}d\kappa_xd\kappa_y\sqrt{c^2\bm{\kappa}^2+\omega^2_p},\nonumber\\
 \end{eqnarray}
 where we also took the limit $S\rightarrow\infty$ in which the sum gets promoted into an integral. Going now to polar coordinates and performing the integral we obtain the result for the photon energy per area
 \begin{eqnarray}
     \frac{E_p}{S}=\frac{\hbar(\Lambda^{3/2}_0-1)}{6\pi c^2}\widetilde{\omega}^3(\kappa_z).
 \end{eqnarray}
 Using the expression for $\widetilde{\omega}(\kappa_z)$ given by Eq.~(\ref{alpha parameter}) and taking the derivative of the photon energy per area $E_p/S$ with respect to the distance of the cavity mirrors $L_z$ we find the force per area (the pressure) 
 \begin{widetext}
 \begin{equation}\label{Casimir force}
     \frac{F_{\textrm{c}}}{S}=-\frac{\partial(E_p/S)}{\partial L_z}=\frac{\hbar(\Lambda^{3/2}_0-1)}{4\pi c^2}\left(\frac{2\pi^2 c^2}{L^3_z}+\frac{e^2n_{\textrm{2D}}}{m_{\textrm{e}}\epsilon_0 L^2_z}\right)\sqrt{\frac{\pi^2 c^2}{L^2_z}+\frac{e^2n_{\textrm{2D}}}{m_{\textrm{e}}\epsilon_0 L_z}}.
 \end{equation}
 \end{widetext}
We note that to obtain the above result we also took into account the dependence of $\omega_p$ to the distance between the cavity mirrors $L_z$, as given by Eq.~(\ref{plasma frequency}). The force (or pressure) above describes the force that the parallel plates of the cavity feel due to the zero point energy of the photon field of the interacting hybrid system in the cavity. The force given by Eq.~(\ref{Casimir force}) is positive because $\Lambda^{3/2}_0\geq 1$. This indicates that the Casimir force is \textit{repulsive}. The possibility of repulsive Casimir forces has been discussed in many different settings~\cite{HoyeBrevik, Boyer1974, Milonni, Butcher_2012, Henkel} and has even been experimentally observed for interacting materials immersed in a fluid~\cite{Munday2009}. In our case we do not have a fluid between the cavity mirrors but a 2DEG which interacts with the cavity field.

\subsection{Absorption and Dissipation in the Effective Field Theory}\label{Linear Response EFT}
 
 In section~\ref{Cavity Responses} we performed linear response for the electronic and the photonic sectors of the theory, in the single mode case. Our goal now is to study the linear response behavior of the effective theory we constructed, and to see how the response functions get modified by the infinite amount of in-plane modes. Here, we focus on the linear response in the photonic sector, which as we showed in subsection~\ref{Duality} is adequate for the description of all the resonances of the system.  
 
To perturb the photon field we apply an external time-dependent current $\bi{J}_{\textrm{ext}}(t)$ which couples to the quantized cavity field, as shown in Fig.~\ref{Cavity_Induction}. Thus, we consider the external perturbation $\hat{H}_{\textrm{ext}}(t)=-\mathbf{J}_{\textrm{ext}}(t)\cdot\hat{\mathbf{A}}$ as we did in section~\ref{Photonic Response}. The external current is chosen to be in the $x$-direction $\bi{J}_{\textrm{ext}}(t)=\bi{e}_xJ_{\textrm{ext}}(t)$. The vector potential in the effective theory is the sum over all the in-plane modes 
\begin{eqnarray}\label{manymodeA}
	\hat{\mathbf{A}}=\left(\frac{\hbar}{\epsilon_0 V}\right)^{\frac{1}{2}}\sum_{\kappa_x,\kappa_y}\frac{\bi{e}_x}{\sqrt{2\omega(\bm{\kappa}})}\left(\hat{a}_{\bm{\kappa}}+\hat{a}^{\dagger}_{\bm{\kappa}}\right),
\end{eqnarray}
where we only kept the polarization in the $x$-direction, because it is the only one that couples to the external perturbation. To perform linear response we need to introduce and define the Hamiltonian of the effective theory $\hat{H}_{\textrm{eff}}$. For the effective Hamiltonian it is not necessary to give a particular expression in terms of electronic and photonic operators. The effective Hamiltonian can be defined also by giving a definition of the ground state of $\hat{H}_{\textrm{eff}}$ and its excited states. The ground state of the effective Hamiltonian $\hat{H}_{\textrm{eff}}$ we define it as
\begin{eqnarray}
    |\Psi_{gs}\rangle=|\Phi_{0}\rangle \otimes \prod_{\kappa_x,\kappa_y}|0,0\rangle_{\kappa_x,\kappa_y}
\end{eqnarray}
where $|\Phi_0\rangle$ is ground state of the electronic sector given by the Slater determinant in Eq.~(\ref{Slater determinant}), with the electrons distributed on the 2D Fermi sphere (see Fig.~\ref{Fermi Sphere}), which consequently have zero total momentum $\bi{K}=\sum_{j}\bi{k}_j=0$. Furthermore, the set of states $|0,0\rangle_{\kappa_x,\kappa_y}$ get annihilated by the operator $\hat{c}_{\bm{\kappa}}$, $\hat{c}_{\bm{\kappa}} |0,0\rangle_{\kappa_x,\kappa_y} =0, \;\; \forall\;\; \bm{\kappa}$. Having the ground state we can define the excited states of the system by applying the creation operators $\hat{c}^{\dagger}_{\bm{\kappa}}$ on it. Thus, we find that the excited states of $\hat{H}_{\textrm{eff}}$ satisfy the equation 
\begin{eqnarray}
    \hat{H}_{\textrm{eff}}\frac{(\hat{c}^{\dagger}_{\bm{\kappa}})^m}{\sqrt{m!}}|\Psi_{gs}\rangle=\left(E_{\mathbf{k}}+\hbar\widetilde{\omega}(\bm{\kappa})\left(m+\frac{1}{2}\right)\right)\frac{(\hat{c}^{\dagger}_{\bm{\kappa}})^m}{\sqrt{m!}}|\Psi_{gs}\rangle\nonumber\\
\end{eqnarray}
where $E_{\bi{k}}=\sum_j\hbar^2\bi{k}^2_j/2m_{\textrm{e}}$ is the kinetic energy of the electrons. We also note that the operators $\{\hat{c}_{\bm{\kappa}},\hat{c}^{\dagger}_{\bm{\kappa}^{\prime}}\}$ satisfy bosonic commutation operators $[\hat{c}_{\bm{\kappa}},\hat{c}^{\dagger}_{\bm{\kappa}^{\prime}}]=\delta_{\bm{\kappa}\bm{\kappa}^{\prime}}$ $\forall \;\; \bm{\kappa},\bm{\kappa}^{\prime}$. With the definition of the effective Hamiltonian the full time dependent Hamiltonian under the external perturbation is $\hat{H}(t)=\hat{H}_{\textrm{eff}}-\bi{J}_{\textrm{ext}}(t)\cdot\hat{\bi{A}}$. The vector potential in terms of the renormalized annihilation and creation operators of Eq.~(\ref{c operators}) is
\begin{eqnarray}\label{manymodeAinC}
	\hat{\mathbf{A}}=\left(\frac{\hbar}{\epsilon_0 V}\right)^{\frac{1}{2}}\sum_{\kappa_x,\kappa_y}\frac{\bi{e}_x}{\sqrt{2\widetilde{\omega}(\bm{\kappa}})}\left(\hat{c}_{\bm{\kappa}}+\hat{c}^{\dagger}_{\bm{\kappa}}\right).
\end{eqnarray}
With these definitions we can define all operators of the theory in the interaction picture as $\hat{\mathcal{O}}_{I}(t)=e^{\textrm{i}t\hat{H}_{\textrm{eff}}/\hbar}\hat{\mathcal{O}}e^{-\textrm{i}t\hat{H}_{\textrm{eff}}/\hbar}$ and the wavefunctions respectively as $\Psi_{I}(t)=e^{\textrm{i}t\hat{H}_{\textrm{eff}}/\hbar}\Psi(t)$. Here, we are interested in the $\bi{A}$-field response function $\chi^A_A(t-t^{\prime})$ which is defined through Eq.~(\ref{def chi}). Substituting the expression for $\hat{\bi{A}}$ given by Eq.~(\ref{manymodeAinC}) and using the fact that the effective Hamiltonian is a sum of non-interacting modes and that the ground state $|\Psi_{gs}\rangle$ of the effective theory in the thermodynamic limit is a tensor product of the photonic states of all the modes, we find the response function to be the sum of all the single mode response functions given by Eq.~(\ref{A field response in time})
\begin{eqnarray}
    \chi^{A}_{A}(t-t^{\prime})=-\sum_{\kappa_x,\kappa_y}\frac{\Theta(t-t^{\prime})\sin(\widetilde{\omega}(\bm{\kappa})(t-t^{\prime}))}{\epsilon_0V\widetilde{\omega}(\bm{\kappa})}.
\end{eqnarray}
Since the response function in time is the sum of the single mode responses, also the response function in the frequency domain $\chi^A_A(w)$ is the sum of all the single mode response functions given in appendix~\ref{appendix B}
\begin{widetext}
\begin{eqnarray}\label{EFT response frequency}
    \chi^A_A(w)=\sum_{\kappa_x,\kappa_y}\frac{-1}{2\epsilon_0\widetilde{\omega}(\bm{\kappa})V}\lim_{\eta \to 0}\left[\frac{1}{w+\widetilde{\omega}(\bm{\kappa})+\textrm{i}\eta}-\frac{1}{w-\widetilde{\omega}(\bm{\kappa})+\textrm{i}\eta}\right].
\end{eqnarray}
\end{widetext}
In the thermodynamic limit the above sum turns into an integral and following the derivation shown in appendix~\ref{Integrals for EFT} we find the analytic expressions for the real and the imaginary parts of the $\bi{A}$-field response function  $\chi^A_A(w)$ for the effective field theory
\begin{widetext}
\begin{eqnarray}\label{Re and Im of EFT Response}
    \Re[\chi^A_A(w)]&=&\frac{1}{8\pi c^2\epsilon_0L_z}\left[\ln\left(\frac{(w-\widetilde{\omega}(\kappa_z))^2+\eta^2}{(w-\sqrt{\Lambda})^2+\eta^2}\right)+\ln\left(\frac{(w+\widetilde{\omega}(\kappa_z))^2+\eta^2}{(w+\sqrt{\Lambda})^2+\eta^2}\right)\right]\;\; \textrm{and}\\
    \Im[\chi^A_A(w)]&=&\frac{1}{4\pi c^2\epsilon_0L_z}\left[\tan^{-1}\left(\frac{\sqrt{\Lambda}+w}{\eta}\right)-\tan^{-1}\left(\frac{\widetilde{\omega}(\kappa_z)+w}{\eta}\right)+\tan^{-1}\left(\frac{\widetilde{\omega}(\kappa_z)-w}{\eta}\right)-\tan^{-1}\left(\frac{\sqrt{\Lambda}-w}{\eta}\right)\right].\nonumber
\end{eqnarray}
\end{widetext}
If we take now the limit $\eta \rightarrow 0$  for the artificial broadening $\eta$, we find for the imaginary part
\begin{eqnarray}\label{Im EFT}
    \Im[\chi^A_A(w)]&=&
	\begin{cases}
		\dfrac{1}{4 c^2\epsilon_0L_z}\;,\;\; \textrm{for} \;\; -\sqrt{\Lambda}<w<-\widetilde{\omega}(\kappa_z)  \\\\
		-\dfrac{1}{4 c^2\epsilon_0L_z}\;,\;\; \textrm{for} \;\; \widetilde{\omega}(\kappa_z)<w<\sqrt{\Lambda} \\
		0\;\;, \;\;\;\;\;\;\; \textrm{elsewhere}. 
	\end{cases}
\end{eqnarray}

\begin{figure}[ht]
\includegraphics[height=6cm, width=\linewidth]{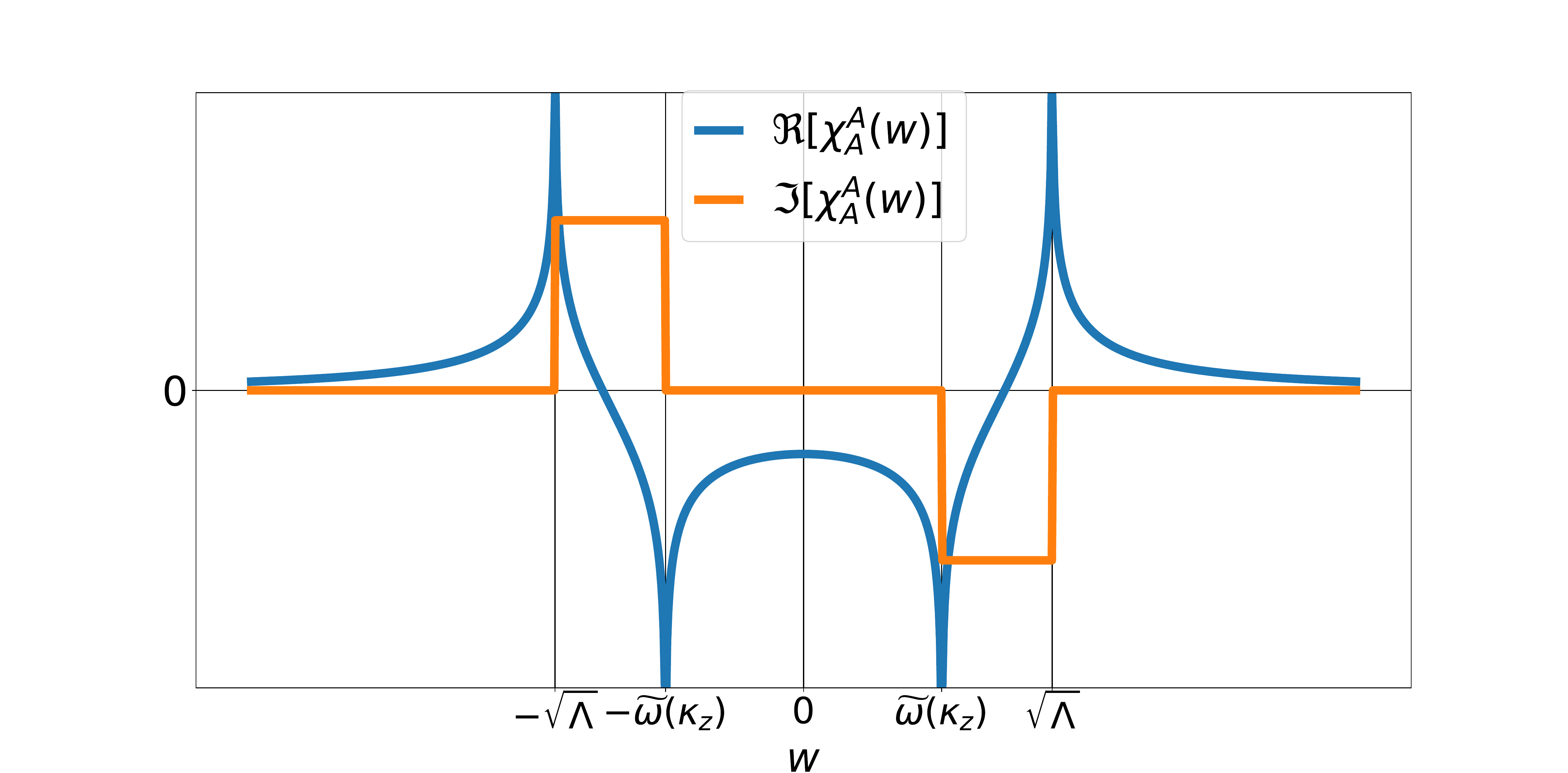}
\caption{\label{EFT Response} Real $\Re[\chi^A_A(w)]$ and imaginary $\Im[\chi^A_A(w)]$ parts of the $\bi{A}$-field response function $\chi^A_A(w)$ in the effective field theory with $\eta=0$. The imaginary part has a finite value within the frequency window $\widetilde{\omega}(\kappa_z)<|w|<\sqrt{\Lambda}$ which indicates that in this frequency range the system can continuously absorb energy. The real part though diverges at the natural lower cutoff $w=\pm \widetilde{\omega}(\kappa_z)$ and the upper cutoff $w=\pm \sqrt{\Lambda}$ and shows that the system in the effective field theory has two scales.}
\end{figure}

From the expression above we see that the imaginary part $\Im[\chi^A_A(w)]$ is well-defined in the limit $\eta \rightarrow 0$ for all $w$ without any divergences appearing. This is in contrast to $\Im[\chi^A_A(w)]$ of Eq.~(\ref{Re Im A-field}) in the single-mode theory  which was divergent for $w=\pm \widetilde{\omega}$. Further, the imaginary part in Eq.~(\ref{Im EFT}) takes a constant value in the region $\widetilde{\omega}(\kappa_z)<|w|<\sqrt{\Lambda}$ and is zero everywhere else, as shown also in Fig.~\ref{EFT Response}. This means that our system can absorb energy continuously with the same strength in the frequency window $\widetilde{\omega}(\kappa_z)<|w|<\sqrt{\Lambda}$. This is because it is exactly this frequency range in which the effective field theory is defined, see Fig.~\ref{Photon Continuum}, and all modes are excited by the external current with the same strength.

The fact that the imaginary part is well-defined and does not diverge means that absorption can be consistently described in the effective field theory and the absorption rate $W$ of Eq.~(\ref{Absorption Rate}) is well-defined and can be computed properly. This proves our claim that by constructing a theory of infinitely many modes in the continuum we can indeed describe absorption processes and dissipation from first-principles, without the need of the artificial broadening $\eta$, and without having to introduce some kind of environment for our system. This demonstrates that a system with its photon field works like its own heat bath~\cite{Vignale}, and more precisely in our case the continuum of modes describes the full \textit{photon bath}.

The real part $\Re[\chi^A_A(w)]$ though for $\eta=0$ diverges at the frequencies $w=\pm\widetilde{\omega}(\kappa_z)$ and $w=\pm\sqrt{\Lambda}$ and gives us information about the resonances of the system. In the single mode case, in section~\ref{Photonic Response}, there was only one resonance appearing at frequency $w=\pm \widetilde{\omega}$, while now we have two resonances at the frequency of the plasmon-polariton $\widetilde{\omega}(\kappa_z)$ and the cutoff $\sqrt{\Lambda}$. This indicates that there are now two scales in the system, the natural lower cutoff $\widetilde{\omega}(\kappa_z)$ and upper cutoff of the effective field theory as also shown in Fig.~\ref{Photon Continuum}.

Lastly, we would like to highlight that in the large $N,S$ limit the imaginary and the real parts of the response function $\chi^A_A(w)$ of Eq.~(\ref{Re and Im of EFT Response}) have a well-defined finite value and do not vanish. This is in contrast to the single mode response function given by Eq.~(\ref{Re Im A-field}). This shows again that by going to the continuum of modes and constructing this effective field theory, the photon observables become well-defined and have a substantial contribution to the macroscopic 2DEG in the cavity.

\section{Summary, Experimental Implications \& Future Directions}\label{Summary}

\textit{Summary}.---In this article we investigated Sommerfeld's theory of the free electron gas~\cite{Sommerfeld1928} in the framework of cavity QED. In the long-wavelength limit (or dipole approximation) and in case where the quantized cavity field consists of a single mode we showed that the system is analytically solvable. This allowed us to perform the thermodynamic limit for the electrons, in which the ground state $\bi{k}$-space distribution of the electrons is the 2D Fermi sphere. This means that our system is a Fermi liquid. In addition, we showed that the hybrid electron-photon ground state contains virtual photons. Moreover, we provided the full phase diagram of the interacting electron-photon system for all possible couplings, and we found that when the coupling reaches its maximum value (critical coupling) the ground state becomes \textit{infinitely degenerate}. Such an infinite degeneracy appears also for Landau levels~\cite{Landau} in the integer quantum Hall effect~\cite{Klitzing}. This fact hints towards a novel connection between QED and the semiclassical theory. 

Beyond the critical coupling the system has no ground state and equilibrium is not well-defined. The non-existence of a ground state also occurs if the diamagnetic $\bi{A}^2$ term is neglected. This is in stark contrast to the finite-system models of quantum optics, like the Rabi and the Dicke model, in which a ground state always exists. This demonstrates that extended systems indeed behave very much differently than finite systems and that the well-known (finite-system) models of quantum optics might not be straightforwardly applicable for materials in cavity QED. We believe this result clarifies further the ongoing discussion about whether the diamagnetic $\bi{A}^2$ term can be neglected or not~\cite{vukics2014, schaeferquadratic, DiStefano2019, bernadrdis2018breakdown}. The elimination of the diamagnetic $\bi{A}^2$ term it is known to be responsible for the notorious superradiant phase transition~\cite{Lieb, Wang, Birula}.

Then, we performed linear response for the 2DEG in the cavity and we introduced the four fundamental response sectors: the matter-matter, photon-photon, matter-photon and photon-matter. In addition, we demonstrated that all response sectors are equivalent with respect to their pole structure and that their strengths are related via the electron-photon coupling (or the number of particles). All responses show \textit{plasmon-polariton} resonances which modify the conductive and radiation properties of the 2DEG.

To bridge the discrepancy between the electronic sector, in which the energy density is finite, and the photonic sector, in which the single mode energy density is zero, we promoted the single mode theory into an effective field theory in the continuum, by integrating over all the in-plane modes of the photon field. In this way the energy density of the photon field becomes macroscopic and induces a radiative correction~\cite{Weinberg, Srednicki, Mandl} to the electron mass and renormalizes it~\cite{CHEN20082555, Frohlich2010, HiroshimaSpohn}. The renormalized mass depends on the full electron density $n_{\textrm{e}}$ in the cavity. To the best of our knowledge such a many-body effect has not been reported so far. This is a special feature of the 2DEG due to its macroscopicity. Further, the renormalized electron mass modifies the chemical potential of the system and the fermionic quasiparticle excitations of the Fermi liquid. The concept of the fermionic quasi-particles of renormalized mass allowed us to introduce within our effective quantum field theory a jellium model for these quasi-particles and to include perturbatively the Coulomb interaction. In this model the photon field shrinks the Wigner-Seitz radius which implies a localization effect on the electrons. Moreover, the energy density of the photon field makes itself manifest by producing a Casimir force~\cite{Casimir:1948dh, casimir1948influence, buhmann2013dispersionI, buhmann2013dispersionII} between the mirrors of the cavity, which is repulsive due to the light-matter coupling. Then, we performed linear response in the effective field theory and we showed that due to the continuum of photon modes we are able to describe dissipation and absorption processes without the need of any artificial damping parameter or having to introduce an environment for the system.  

\textit{Experimental Implications}.---To a large extent this work is motivated by the great experimental progress in cavity QED materials~\cite{paravacini2019, li2018, Keller2020, AtacPRL, AtacPRX} and polaritonic chemistry~\cite{ebbesen2016, george2016, hutchison2013, hutchison2012, orgiu2015}. We believe that several of the results presented throughout the article are measurable and have experimental implications. So let us elaborate a bit further on the main ones. 

\begin{itemize}
    \item \textit{Cavity Modified Conductivity}.---In section~\ref{Electronic Response} we computed the optical conductivity $\sigma(w)$ of the 2DEG in the cavity, given by Eq.~(\ref{Conductivity}). The standard optical conductivity of the free electron gas gets modified by the appearance of plasmon-polariton resonances 
    which show up in both the real and the imaginary part of $\sigma(w)$ (see Figs.~\ref{Real_Conductivity} and \ref{Imaginary_Conductivity}). Since $\sigma(w)$ gets modified by the cavity field this implies that also the dielectric function $\epsilon(w)$ will be modified as well (as $\epsilon(w)= \epsilon_0+\textrm{i}\sigma(w)/w$~\cite{Mermin}). These modifications can be observed with optical transmission measurements. In addition, we showed that in the static limit the DC conductivity and the Drude peak get suppressed due the coupling to the cavity field (see Eq.~(\ref{DC Conductivity})). We would like to mention that such modifications of the conduction properties of 2D materials confined in cavities have already been observed~\cite{paravacini2019, A.Thomas2019}. We believe our work can provide further insights into these experiments and motivate new directions to be explored.
    
    \item{ \textit{Measurement of the Effective Mass, Coupling and Cutoff}.---Metals in solid state theory are described in most cases using the free electron model~\cite{Mermin} in which the energy dispersion of the electrons is described by a parabola $E_{\bi{k}}=\hbar^2\bi{k}^2/m^*_{\textrm{e}}$ with some effective mass $m^*_{\textrm{e}}$ for the electrons. Measuring the dispersion of the electrons by angle-resolved photoemission spectroscopy (ARPES)~\cite{Damascelli_2004} one can obtain the effective mass $m^*_{\textrm{e}}$. The effective mass appears because in a metal electrons are not completely non-interacting, but there are Coulomb interactions and the potential of the ions which modify the free electron behavior~\cite{Callaway, Dresselhaus, Eicheffectivemass}. This pictures indicates that the mass of the electron is not purely inherent but depends on its \textit{environment}.  
     
     In our case the 2DEG is coupled to a cavity. Thus, the environment of the electrons includes also the interactions with the photon field. Consequently, if one measures the energy dispersion of the electrons in a metal confined in a cavity, one should find a parabolic dispersion but with a different effective mass (see also Fig.~\ref{Parabola Renormalization}). The contribution of the cavity photons to the effective mass of the electrons in the metal is given by the expression of the renormalized electron mass $m_{\textrm{e}}(\Lambda)$ of Eq.~(\ref{Renormalized Mass}). 
     
     We propose that the renormalized electron mass due to the cavity photons can be measured by comparing the effective masses \textit{outside and inside} the cavity. Furthermore, from such an experimental measurement of the effective electron mass and the formula in Eq.~(\ref{Renormalized Mass}) one can deduce directly the cutoff $\Lambda$. The cutoff $\Lambda$ gives the highest frequency (or momentum) with which the electrons interact inside the cavity. Having the expression for the effective coupling $g(\Lambda)$ given by Eq.~(\ref{effectivecoupling}) we can obtain also the coupling strength between the electrons and the photons in the cavity. In most cases for finite systems the light-matter coupling strength is defined via the Rabi (or Dicke) model~\cite{shore1993} and the corresponding Rabi split. Our theory provides a \textit{novel way} to measure the electron-photon coupling for extended systems in cavity QED, which goes beyond standard quantum optics models, via the effective mass. }
    
    \item \textit{Modified Casimir Forces}.---As it is known since the seminal works of Casimir and Polder~\cite{Casimir:1948dh, casimir1948influence} macroscopic vacuum forces can emerge due to the vaccuum energy of the electromagnetic field between perfectly conducting plates, like for example two cavity mirrors. In section~\ref{Repulsive Casimir Forces} we computed the Casimir force due to the vaccuum energy of the 2DEG in the cavity, and we found that due to the light-matter coupling the Casimir force gets modified by the electron density $n_{\textrm{e}}$ and turns out to be \textit{repulsive}. Such repulsive vacuum forces have been reported in the case of a cavity immersed in a fluid~\cite{Munday2009}. Our theory provides an example of a such a repulsive force and a first-principles explanation on how such repulsive vacuum forces can emerge, due to strong light-matter interaction, and opens new pathways for manipulating and engineering Casimir forces in cavity QED.

\end{itemize}

\textit{Future Directions}.---The presented theory has many implications. Besides the ones we have pointed out so far, we would like to mention a few further research directions which are potentially interesting and to our opinion worthwhile to pursue. 

\begin{itemize}
    \item \textit{Coulomb Interaction and Fermi Liquid Theory in QED}.---In this work the Coulomb interaction was only treated perturbatively, as a first-order perturbation within the jellium model, and a particular modification of the jellium model due to the photon field was presented. The perturbative approach is valid in the large density regime and can be further extended to include the correlation energy of the electron-photon system in which the photonic states will introduce non-trivial effects. Moreover, for the regime of intermediate densities the paradigmatic theory is the Fermi liquid theory~\cite{LandauFermiLiquid} in which the interacting electronic system is described using fermionic quasi-particles, and the Coulomb interaction is treated using Green's function techniques~\cite{Nozieres}. What defines a Fermi liquid is the fact that electrons in $\bi{k}$-space are distributed on the Fermi sphere. In section~\ref{Ground State} we showed that this is the case also for the 2DEG coupled to a cavity. Further, we showed how the photon modifies the quasiparticle excitations of the Fermi liquid. Thus, we believe that a theory for materials in cavity QED can be constructed along the lines of the Fermi liquid theory.  
    
    \item \textit{LDA functional in QEDFT}.---The local density approximation within DFT~\cite{HohenbergKohn} is one of the most successful methods for the computation of properties of materials~\cite{RubioReview}. Recently, a generalization of DFT in the framework of QED has been introduced~\cite{TokatlyPRL, ruggenthaler2014} and has already been applied~\cite{CamillaPRL, flick2018light}. The original LDA was constructed from the analytic solution of the free electron gas. In this work we solved exactly the free electron gas in cavity QED and this gives the opportunity to construct an LDA functional in the framework of QEDFT. If such an LDA-type functional in QEDFT shares the success of the original LDA, then we would have a really powerful new tool for the description of materials in the emerging field of cavity QED. Our result on the shrinking of the Wigner-Seitz radius in the jellium model, can be potentially helpful in developing local-density-type approximations for the light-matter interactions.

    \item \textit{Superradiance}.---Superradiance as predicted by Dicke~\cite{dicke1954} is the enhancement of spontaneous emission due to the collective coupling of emitters. In addition, an equilibrium superradiant phase transition was also predicted for the Dicke model~\cite{Lieb, Wang}, which since then has triggered an ongoing debate~\cite{Birula, CiutiSuperradiance, MarquardtNoGO, MazzaSuperradiance, AndolinaNoGo}. In both cases these phenomena emerge due to the collective coupling of many particles or dipoles to the quantized photon field. The theory we presented here involves a large number of electrons coupled to the quantized field of a cavity and we believe will serve as a new playground for the investigation of superradiance and other collective phenomena in cavity QED. 
    
\end{itemize}

\begin{acknowledgements}
We would like to thank M.~A.~Sentef, D.~Welakuh, M.~Penz, I.~Theophilou, J.~Faist, M.~Altarelli and S.~Y.~Buhmann for useful discussions. This work was supported by the European Research Council (ERC-2015-AdG694097), the Cluster of Excellence ``Advanced Imaging of Matter'' (AIM), Grupos Consolidados (IT1249-19) and SFB925 ``Light induced dynamics and control of correlated quantum systems''. The Flatiron Institute is a division of the Simons Foundation.
\end{acknowledgements}

 \appendix

 \section{Linear Response in the Photonic Sector}\label{appendix A}
 
 The aim of this appendix is to give the details of the linear response computations, in the photonic sector, of the subsection~\ref{Photonic Response}. In subsection~\ref{Photonic Response} we perturbed our system with external time-dependent current by adding to the Hamiltonian of Eq.~(\ref{single mode Hamiltonian}) the perturbation $\hat{H}_{\textrm{ext}}=-\bi{J}_{\textrm{ext}}\cdot \hat{\bi{A}}$. With this perturbation the full time-dependent Hamiltonian is given by Eq.~(\ref{Current Perturbation}). The external current is chosen to be in the $x$-direction, $\bi{J}_{\textrm{ext}}(t)=\bi{e}_xJ_{\textrm{ext}}(t)$, and in this case only the $x$-component of the vector potential $\hat{\bi{A}}$ is of interest. Following the standard linear response formalism~\cite{flick2018light, kubo}, which we introduced in the subsection~\ref{Response Formalism}, the response of any observable $\hat{\mathcal{O}}$ due to this perturbation is 
\begin{eqnarray}
    \delta\langle \hat{\mathcal{O}}(t)\rangle=-\frac{\textrm{i}}{\hbar }\int^{t}_{0}\langle [\hat{\mathcal{O}}_{I}(t),\hat{\bi{A}}_{I}(t^{\prime})]\rangle (-\bi{J}_{\textrm{ext}}(t^{\prime})).
\end{eqnarray}
where $\hat{\bi{A}}_{I}(t^{\prime})$ is the quantized vector potential in the interaction picture and the correlator $\langle [\hat{\mathcal{O}}_{I}(t),\hat{\bi{A}}_{I}(t^{\prime})]\rangle$ is defined with respect to the ground state $|\Psi_{gs}\rangle$ of the unperturbed Hamiltonian. From the previous expression the respective response function is
\begin{eqnarray}\label{Response of O}
    \chi^{\mathcal{O}}_{A}(t-t^{\prime})=-\frac{\textrm{i} \Theta(t-t^{\prime})}{\hbar }\langle[\hat{\mathcal{O}}_{I}(t),\hat{\bi{A}}_{I}(t^{\prime})]\rangle.
\end{eqnarray}
In section~\ref{Ground State} we found the ground state $|\Psi_{gs}\rangle$ of the unpurturbed Hamiltonian $\hat{H}$ in the thermodynamic limit to be given by Eq.~(\ref{Thermodynamic GS}). Having $|\Psi_{gs}\rangle$ we can compute the response function for any observable $\hat{\mathcal{O}}$. To obtain the response function $\chi^{\mathcal{O}}_{A}(t-t^{\prime})$ of Eq.~(\ref{Response of O}) we need to compute the commutator
\begin{eqnarray}\label{Commutator}
    \langle [\hat{\mathcal{O}}_{I}(t),\hat{\bi{A}}_{I}(t^{\prime})]\rangle&=& \langle \hat{\mathcal{O}}_{I}(t)\hat{\bi{ A}}_{I}(t^{\prime})\rangle-\langle\hat{\mathcal{O}}_{I}(t)\hat{\bi{ A}}_{I}(t^{\prime})\rangle^*.\nonumber\\
\end{eqnarray}
where we used also the hermiticity of the operator $\hat{\mathcal{O}}_{I}(t)\hat{\bi{ A}}_{I}(t^{\prime})$ which implies that $\langle \hat{ \bi{A}}_{I}(t^{\prime}) \hat{\mathcal{O}}_{I}(t)\rangle=\langle\hat{\mathcal{O}}_{I}(t)\hat{\bi{ A}}_{I}(t^{\prime})\rangle^*$. Thus, we only need to compute the correlator $\langle\hat{\mathcal{O}}_{I}(t)\hat{\bi{ A}}_{I}(t^{\prime})\rangle$. Using the definition for the operators in the interaction picture and the fact that $|\Psi_{gs}\rangle$ is the ground state of the Hamiltonian $\hat{H}$, which means that $e^{-\textrm{i}\hat{H} t^{\prime}/\hbar}|\Psi_{gs}\rangle=e^{-\textrm{i}E_{0,\bi{k}}t^{\prime}/\hbar}|\Psi_{gs}\rangle$, we find
\begin{eqnarray}
    \langle\hat{\mathcal{O}}_{I}(t)\hat{\bi{ A}}_{I}(t^{\prime})\rangle=e^{\frac{\textrm{i}E_{0,\bi{k}}(t-t^{\prime})}{\hbar}}\langle\hat{\mathcal{O}}e^{\frac{-\textrm{i}\hat{H}(t-t^{\prime})}{\hbar}} \hat{\bi{A}}\rangle.
\end{eqnarray}
Where $E_{0,\bi{k}}=E_{\bi{k}}+\hbar\widetilde{\omega}$ is the ground state energy given by Eq.~(\ref{eigenspectrum}), with $n_{\lambda}=0$ for both $\lambda=1,2$. To continue we need to apply the vector potential $\hat{\bi{A}}$ to the ground state $|\Psi_{gs}\rangle$. For that we need the expression of $\hat{\bi{A}}$ in terms of the annihilation and creation operators $\{\hat{c}_{\lambda},\hat{c}^{\dagger}_{\lambda}\}$. From Eqs.~(\ref{Ainb}) and~(\ref{c operators}), and for $\mathbf{K}=0$ (which is true in the ground state)  we find for the quantized vector potential 
\begin{eqnarray}\label{AinC}
    \hat{\bi{A}}=\left(\frac{\hbar}{2\epsilon_0\tilde{\omega}V}\right)^{\frac{1}{2}}\left(\hat{c}_1+\hat{c}^{\dagger}_1\right)\bi{e}_x.
\end{eqnarray}
Applying now $\hat{\bi{A}}$ to the ground state $|\Psi_{gs}\rangle$ we have
\begin{eqnarray}
    &&\langle\hat{\mathcal{O}}_{I}(t)\hat{\bi{ A}}_{I}(t^{\prime})\rangle=\left(\frac{\hbar}{2\epsilon_0\tilde{\omega}V}\right)^{\frac{1}{2}}e^{\frac{\textrm{i}E_{0,\bi{k}}(t-t^{\prime})}{\hbar}} \times \nonumber\\
    &&\times\langle \Psi_{gs}|\hat{\mathcal{O}}e^{\frac{-\textrm{i}\hat{H}(t-t^{\prime})}{\hbar}}|\Phi_{0}\rangle\otimes |1,0\rangle_{1}|0,0\rangle_2.
\end{eqnarray}
From the expression above we see that quantized field $\hat{\bi{A}}$ gets the ground state to the first excited state for $n_1=1$. The state $|\Phi_{0}\rangle\otimes |1,0\rangle_{1}|0,0\rangle_2$ is the first excited state of $\hat{H}$ with eigenenergy $E_{1,\bi{k}}=E_{\bi{k}}+2\hbar\widetilde{\omega}$. Using this we find
\begin{eqnarray}
    &&\langle \hat{\mathcal{O}}_{I}(t)\hat{\bi{ A}}_{I}(t^{\prime})\rangle=\left(\frac{\hbar}{2\epsilon_0\tilde{\omega}V}\right)^{\frac{1}{2}}e^{-\textrm{i}\widetilde{\omega}(t-t^{\prime})} \times \nonumber\\
    &&\times\langle \Psi_{gs}|\hat{\mathcal{O}}|\Phi_{0}\rangle\otimes |1,0\rangle_{1}|0,0\rangle_2.
\end{eqnarray}
From the previous result we obtain the following expression for the commutator of Eq.~(\ref{Commutator}) 
\begin{widetext}
\begin{eqnarray}\label{O commutator}
    \langle [\hat{\mathcal{O}}_{I}(t),\hat{\bi{A}}_{I}(t^{\prime})]\rangle=\left(\frac{\hbar}{2\epsilon_0\tilde{\omega}V}\right)^{\frac{1}{2}}\left[e^{-\textrm{i}\widetilde{\omega}(t-t^{\prime})} \langle \Psi_{gs}|\hat{\mathcal{O}}|\Phi_{0}\rangle\otimes |1,0\rangle_{1}|0,0\rangle_2-e^{\textrm{i}\widetilde{\omega}(t-t^{\prime})} \left(\langle \Psi_{gs}|\hat{\mathcal{O}}|\Phi_{0}\rangle\otimes |1,0\rangle_{1}|0,0\rangle_2\right)^*\right].\nonumber\\
\end{eqnarray}
\end{widetext}
The formula above is very important because it applies to any observable $\hat{\mathcal{O}}$ and we will use it for the computation of several different response functions.

\section{$\bi{A}$-Field Response Function}\label{appendix B}
 
 Having derived Eq.~(\ref{O commutator}) we will use this formula to compute the response function $\chi^{A}_{A}(t-t^{\prime})$ for the quantized vector potential $\hat{\bi{A}}$. From Eq.~(\ref{O commutator}) it is clear that all we have to compute is $\langle \Psi_{gs}|\hat{\bi{A}}|\Phi_{0}\rangle\otimes |1,0\rangle_{1}|0,0\rangle_2$. Using Eq.~(\ref{AinC}) which gives the $\hat{\bi{A}}$-field in terms of the operators $\{\hat{c}_1,\hat{c}^{\dagger}_1\}$ we find 
\begin{eqnarray}
\langle \Psi_{gs}|\hat{\bi{A}}|\Phi_{0}\rangle\otimes |1,0\rangle_{1}|0,0\rangle_2=\left(\frac{\hbar}{2\epsilon_0\widetilde{\omega}V}\right)^{\frac{1}{2}}.
\end{eqnarray}
Combining the result above with Eqs.~(\ref{O commutator}) and Eq.~(\ref{Response of O}) we find the response function in time $\chi^A_A(t-t^{\prime})$
\begin{eqnarray}\label{A field response}
\chi^A_A(t-t^{\prime})=-\frac{\Theta(t-t^{\prime})\sin(\widetilde{\omega}(t-t^{\prime}))}{\epsilon_0\widetilde{\omega}V} .
\end{eqnarray}
The response function above is also the propagator of the $\bi{A}$-field. Making use of the integral form of the $\Theta$-function and performing a Laplace transform for $\chi^A_A(t-t^{\prime})$ we find the response of the $\bi{A}$-field in the frequency domain $\chi^A_A(w)$
\begin{eqnarray}\label{A frequencyresponse}
\chi^A_A(w)=\frac{-1}{2\epsilon_0\widetilde{\omega}V}\lim_{\eta \to 0^+}\left[\frac{1}{w+\widetilde{\omega}+\textrm{i}\eta}-\frac{1}{w-\widetilde{\omega}+\textrm{i}\eta}\right].\nonumber\\
\end{eqnarray}

\section{$\bi{E}$-Field Response Function}\label{appendix C}
 
Now we would also like to compute the response of the electric field due to the external time-dependent current $\bi{J}_{\textrm{ext}}(t)$. The electric field in dipole approximation, in the $x$-direction, is~\cite{rokaj2017}
\begin{equation}\label{Electric field appendix}
	\hat{\mathbf{E}}=\textrm{i}\left(\frac{\hbar\omega }{2\epsilon_0 V}\right)^{\frac{1}{2}}\left(\hat{a}_{1}-\hat{a}^{\dagger}_{1}\right) \bi{e}_x.
\end{equation}
To make use of Eq.~(\ref{O commutator}) we need to write the electric field in terms of the operators $\{\hat{c}_{1},\hat{c}^{\dagger}_{1}\}$. Using Eqs.~(\ref{boperators}) and (\ref{c operators}) we find for the electric field
\begin{equation}
	\hat{\mathbf{E}}=\textrm{i}\left(\frac{\hbar\widetilde{\omega} }{2\epsilon_0 V}\right)^{\frac{1}{2}}\left(\hat{c}_{1}-\hat{c}^{\dagger}_{1}\right) \bi{e}_x.
\end{equation}
Substituting the expression for the electric field operator into Eq.~(\ref{O commutator}) and then using the definition of the response function in time given by Eq.~(\ref{Response of O}) we find the response function $\chi^E_{A}(t-t^{\prime})$
\begin{eqnarray}\label{E field response}
    \chi^E_{A}(t-t^{\prime})=\frac{\Theta(t-t^{\prime})\cos(\widetilde{\omega}(t-t^{\prime}))}{\epsilon_0 V}.
\end{eqnarray}
From the response function in time by performing a Laplace transform we can obtain the response function in the frequency domain
    \begin{eqnarray}\label{E response frequency}
    \chi^{E}_{A}(w)=\frac{\textrm{i}}{2\epsilon_0V}\lim_{\eta \to 0^+}\left[\frac{1}{w+\widetilde{\omega}+\textrm{i}\eta}+\frac{1}{w-\widetilde{\omega}+\textrm{i}\eta}\right].
\end{eqnarray}
Moreover, we can also deduce the real and the imaginary parts of $ \chi^{E}_{A}(w)$
\begin{eqnarray}
    \Re[\chi^E_A(w)]=\frac{\eta}{2\epsilon_0V}\left[\frac{1}{(w+\widetilde{\omega})^2+\eta^2}-\frac{1}{(w-\widetilde{\omega})^2+\eta^2}\right],\nonumber\\
    \Im[\chi^E_A(w)]=\frac{1}{2\epsilon_0V}\left[\frac{w+\widetilde{\omega}}{(w+\widetilde{\omega})^2+\eta^2}-\frac{w-\widetilde{\omega}}{(w-\widetilde{\omega})^2+\eta^2}\right]\nonumber\\
\end{eqnarray}

\section{Computation of the Response Functions in the Effective Field Theory} \label{Integrals for EFT}\label{appendix D}

Here we would like to give the details of the computation of the real and the imaginary part of the $\bi{A}$-field response function $\chi^A_A(w)$ in the effective field theory. In subsection~\ref{Linear Response EFT} we showed that the response of the $\bi{A}$-field in the frequency domain, for the effective theory,  is given by the expression (see Eq.~(\ref{EFT response frequency})) 
\begin{widetext}
\begin{eqnarray}
    \chi^A_A(w)=\sum_{\kappa_x,\kappa_y}\frac{-1}{2\epsilon_0\widetilde{\omega}(\bm{\kappa})V}\lim_{\eta \to 0^+}\left[\frac{1}{w+\widetilde{\omega}(\bm{\kappa})+\textrm{i}\eta}-\frac{1}{w-\widetilde{\omega}(\bm{\kappa})+\textrm{i}\eta}\right].
\end{eqnarray}
\end{widetext}
In the thermodynamic limit the above sum turns into an integral and we find the following expression for the real and the imaginary part of the response function $\chi^A_A(w)$
\begin{widetext}
\begin{eqnarray}\label{Re and Im Integrals}
    \Re\left[\chi^A_A(w)\right]&=&\frac{1}{8\pi^2\epsilon_0L_z}\iint\limits^{\Lambda}_0\left[\frac{w-\widetilde{\omega}(\bm{\kappa})}{\widetilde{\omega}(\bm{\kappa})[(w-\widetilde{\omega}(\bm{\kappa}))^2+\eta^2]}-\frac{w+\widetilde{\omega}(\bm{\kappa})}{\widetilde{\omega}(\bm{\kappa})[(w+\widetilde{\omega}(\bm{\kappa}))^2+\eta^2]}\right]d\kappa_xd\kappa_y\;\; \textrm{and}\nonumber\\
    \Im\left[\chi^A_A(w)\right]&=&\frac{\eta}{8\pi^2\epsilon_0 L_z}\iint\limits^{\Lambda}_0\left[\frac{1}{\widetilde{\omega}(\bm{\kappa})[(w+\widetilde{\omega}(\bm{\kappa}))^2+\eta^2]}-\frac{1}{\widetilde{\omega}(\bm{\kappa})[(w-\widetilde{\omega}(\bm{\kappa}))^2+\eta^2]}\right]d\kappa_xd\kappa_y.
\end{eqnarray}
\end{widetext}
In the definition of the real and the imaginary parts $\Re[\chi^A_A(w)]$ and $\Im[\chi^A_A(w)]$ appear the following four integrals
\begin{widetext}
\begin{eqnarray}\label{Integrals}
    \mathcal{A}&=&\iint\limits^\Lambda_0\frac{d\kappa_xd\kappa_y}{\widetilde{\omega}(\bm{\kappa})[(w-\widetilde{\omega}(\bm{\kappa}))^2+\eta^2]},\;\;\;
    \mathcal{B}=\iint\limits^\Lambda_0\frac{d\kappa_xd\kappa_y}{(w-\widetilde{\omega}(\bm{\kappa}))^2+\eta^2},\\
    \mathcal{C}&=&\iint\limits^\Lambda_0\frac{d\kappa_xd\kappa_y}{\widetilde{\omega}(\bm{\kappa})[(w+\widetilde{\omega}(\bm{\kappa}))^2+\eta^2]},\;\;\;
    \mathcal{D}=\iint\limits^\Lambda_0\frac{d\kappa_xd\kappa_y}{(w+\widetilde{\omega}(\bm{\kappa}))^2+\eta^2}.\nonumber
    \end{eqnarray}
\end{widetext}
To simplify these integrals we go to polar coordinates $(\kappa_x,\kappa_y) \rightarrow (\kappa_r,\kappa_{\theta})$. In polar coordinates the integration measure is $d\kappa_xd\kappa_y=\kappa_rd\kappa_r d\kappa_{\theta}$, and the dressed frequency is $\widetilde{\omega}(\bm{\kappa})=\sqrt{c^2\kappa^2_r+c^2\kappa^2_z+\omega^2_p}$. Furthermore, we define the frequency $\widetilde{\omega}^2(\kappa_z)=c^2\kappa^2_z+\omega^2_p$ and we make the variable substitution $u=c^2\kappa^2_r+\widetilde{\omega}^2(\kappa_z)$ to all the integrals in Eq.~(\ref{Integrals}) and we obtain
\begin{widetext}
\begin{eqnarray}\label{Integrals in u}
    \mathcal{A}&=&\frac{\pi}{c^2}\int\limits^\Lambda_{\widetilde{\omega}^2(\kappa_z)}\frac{du}{\sqrt{u}[(w-\sqrt{u})^2+\eta^2]},\;\;\;
    \mathcal{B}=\frac{\pi}{c^2}\int^{\Lambda}_{\widetilde{\omega}^2(\kappa_z)}\frac{du}{(w-\sqrt{u})^2+\eta^2},\\
    \mathcal{C}&=&\frac{\pi}{c^2}\int\limits^\Lambda_{\widetilde{\omega}^2(\kappa_z)}\frac{du}{\sqrt{u}[(w+\sqrt{u})^2+\eta^2]},\;\;\;
    \mathcal{D}=\frac{\pi}{c^2}\int^{\Lambda}_{\widetilde{\omega}^2(\kappa_z)}\frac{du}{(w+\sqrt{u})^2+\eta^2}.\nonumber
    \end{eqnarray}
\end{widetext}
\marginpar

Performing the integration now over the variable $u$ we obtain the following expressions for the previous integrals 
    \begin{widetext}
    \begin{eqnarray}\label{Integrals Result}
    \mathcal{A}&=&\frac{2\pi}{c^2\eta}\left[\tan^{-1}\left(\frac{\sqrt{u}-w}{\eta}\right)\right]^{\Lambda}_{\widetilde{\omega}^2(\kappa_z)},\;\;\;
    \mathcal{B}=\frac{\pi}{c^2}\left[\frac{2w}{\eta}\tan^{-1}\left(\frac{\sqrt{u}-w}{\eta}\right)+\ln\left((w-\sqrt{u})^2+\eta^2\right)\right]^{\Lambda}_{\widetilde{\omega}^2(\kappa_z)},\\
    \mathcal{C}&=&\frac{2\pi}{c^2\eta}\left[\tan^{-1}\left(\frac{\sqrt{u}+w}{\eta}\right)\right]^{\Lambda}_{\widetilde{\omega}^2(\kappa_z)},\;\;\;
    \mathcal{D}=\frac{\pi}{c^2}\left[\ln\left((w+\sqrt{u})^2+\eta^2\right)-\frac{2w}{\eta}\tan^{-1}\left(\frac{\sqrt{u}+w}{\eta}\right)\right]^{\Lambda}_{\widetilde{\omega}^2(\kappa_z)}.\nonumber
    \end{eqnarray}
    \end{widetext}
The real part $\Re[\chi^A_A(w)]$ and the imaginary part $\Im[\chi^A_A(w)]$ of the response function $\chi^A_A(w)$, given by Eq.~(\ref{Re and Im Integrals}), in terms of the integrals $\mathcal{A}, \mathcal{B}, \mathcal{C}$ and $\mathcal{D}$ are 
\begin{eqnarray}
    \Re[\chi^A_A(w)]&=&\frac{1}{8\pi^2\epsilon_0L_z}\left(w\mathcal{A}-\mathcal{B}-w\mathcal{C}-\mathcal{D}\right) \;\; \textrm{and}\nonumber\\
    \Im[\chi^A_A(w)]&=&\frac{\eta}{8\pi^2\epsilon_0L_z}\left(\mathcal{C}-\mathcal{A}\right).
\end{eqnarray}
 Using the expressions we found for $\mathcal{A}, \mathcal{B}, \mathcal{C}$ and $\mathcal{D}$ in Eq.~(\ref{Integrals Result}) we obtain the real part $\Re[\chi^A_A(w)]$ and the imaginary part $\Im[\chi^A_A(w)]$ of $\chi^A_A(w)$ given in Eq.~(\ref{Re and Im of EFT Response}).

 \section{Exact Diagonalization with Mode-Mode Interactions}\label{Mode-Mode Interactions}
 
 The aim of this appendix is to show that within the presented framework the interactions between the different modes of the electromagnetic field can be treated exactly without introducing any fundamental changes with respect to the effective field theory in section~\ref{Effective Field Theory}. 
 
 The Pauli-Fierz Hamiltonian for $N$ free electrons coupled to the full photon field is
\begin{eqnarray}
\hat{H}&=&\frac{1}{2m_{\textrm{e}}}\sum\limits^{N}_{j=1}\left(\textrm{i}\hbar \mathbf{\nabla}_{j}+e \hat{\mathbf{A}}\right)^2 +\sum\limits_{\bm{\kappa},\lambda}\hbar\omega(\bm{\kappa})\left[\hat{a}^{\dagger}_{\bm{\kappa},\lambda}\hat{a}_{\bm{\kappa},\lambda}+\frac{1}{2}\right]\nonumber\\
\end{eqnarray}
where the quantized $\hat{\bi{A}}$-field is in dipole approximation. To treat the many-mode case and the mode-mode interaction it is convenient to introduce for the description of the annihilation and creation operators the displacement coordinates $q_{\bm{\kappa},\lambda}$ and the conjugate momenta $\partial/\partial q_{\bm{\kappa},\lambda}$~\cite{rokaj2017}
\begin{eqnarray}
q_{\bm{\kappa},\lambda}=\frac{1}{\sqrt{2}}\left(\hat{a}_{\bm{\kappa},\lambda}+\hat{a}^{\dagger}_{\bm{\kappa},\lambda}\right)\;\; \& \;\;\frac{\partial}{\partial q_{\bm{\kappa},\lambda}}=\frac{1}{\sqrt{2}}\left(\hat{a}_{\bm{k},\lambda}-\hat{a}^{\dagger}_{\bm{\kappa},\lambda}\right).\nonumber\\
\end{eqnarray}
The vector potential in terms of the displacement coordinates is~\cite{rokaj2017}
\begin{eqnarray}
\hat{\bi{A}}=\sqrt{\frac{\hbar}{\epsilon_0V}}\sum_{\bm{\kappa},\lambda}\frac{\bm{\varepsilon}_{\lambda}(\bm{\kappa}) }{\sqrt{\omega(\bm{\kappa}})}q_{\bm{\kappa},\lambda},
\end{eqnarray}
and the Hamiltonian upon expanding the covariant kinetic term and writing the diamagnetic $\hat{\bi{A}}^2$ explicitly takes the form
\begin{widetext}
\begin{eqnarray}
\hat{H}=\sum\limits^{N}_{j=1}\left[-\frac{\hbar^2}{2m_{\textrm{e}}}\nabla^2_j +\frac{\textrm{i}e\hbar}{m_{\textrm{e}}} \hat{\mathbf{A}}\cdot\nabla_j\right]+\underbrace{\frac{\hbar \omega^2_p}{2}\sum_{\bm{\kappa},\bm{\kappa}^{\prime},\lambda,\lambda^{\prime}}\frac{\bm{\varepsilon}_{\lambda}(\bm{\kappa})\cdot \bm{\varepsilon}_{\lambda^{\prime}}(\bm{\kappa}^{\prime})}{\sqrt{\omega(\bm{\kappa})\omega(\bm{\kappa}^{\prime})}} q_{\bm{\kappa},\lambda}q_{\bm{\kappa}^{\prime},\lambda^{\prime}}}_{\frac{Ne^2}{2m_{\textrm{e}}}\hat{\bi{A}}^2}+\sum_{\bm{\kappa},\lambda}\frac{\hbar\omega(\bm{\kappa})}{2}\left(-\frac{\partial^2}{\partial q^2_{\bm{\kappa},\lambda}} +q^2_{\bm{\kappa},\lambda}\right).
\end{eqnarray}
\end{widetext}
The purely photonic part can be separated into a part being quadratic in the displacement coordinates $q^2_{\bm{\kappa},\lambda}$ and a part being bilinear $q_{\bm{\kappa},\lambda}q_{\bm{\kappa}^{\prime},\lambda^{\prime}}$ 
\begin{widetext}
\begin{eqnarray}
\hat{H}=\sum\limits^{N}_{j=1}\left[-\frac{\hbar^2}{2m_{\textrm{e}}}\nabla^2_j +\frac{\textrm{i}e\hbar}{m_{\textrm{e}}} \hat{\mathbf{A}}\cdot\nabla_j\right]+\frac{\hbar \omega^2_p}{2}\sum_{\bm{\kappa}\neq \bm{\kappa}^{\prime},\lambda,\lambda^{\prime}}\frac{\bm{\varepsilon}_{\lambda}(\bm{\kappa})\cdot \bm{\varepsilon}_{\lambda^{\prime}}(\bm{\kappa}^{\prime})}{\sqrt{\omega(\bm{\kappa})\omega(\bm{\kappa}^{\prime})}} q_{\bm{\kappa},\lambda}q_{\bm{\kappa}^{\prime},\lambda^{\prime}}+\sum_{\bm{\kappa},\lambda}\frac{\hbar\omega(\bm{\kappa})}{2}\left(-\frac{\partial^2}{\partial q^2_{\bm{\kappa},\lambda}} +q^2_{\bm{\kappa},\lambda}\left(1+\frac{\omega^2_p}{\omega^2(\bm{\kappa})}\right)\right).\nonumber\\
\end{eqnarray}
\end{widetext}
Further, we introduce a new  scaled set of coordinates $u_{\bm{\kappa},\lambda}=q_{\bm{\kappa},\lambda}\sqrt{\hbar/\omega(\bm{\kappa})}$ and the dressed frequencies $\widetilde{\omega}^2(\bm{\kappa})=\omega^2(\bm{\kappa})+\omega^2_p$. Then, the Hamiltonian takes the form
\begin{widetext}
\begin{eqnarray}
\hat{H}=\sum\limits^{N}_{j=1}\left[-\frac{\hbar^2}{2m_{\textrm{e}}}\nabla^2_j +\frac{\textrm{i}e\hbar}{m_{\textrm{e}}} \hat{\mathbf{A}}\cdot\nabla_j\right]+\sum_{\bm{\kappa},\lambda}\left(-\frac{\hbar^2}{2}\frac{\partial^2}{\partial u^2_{\bm{\kappa},\lambda}} +\frac{\widetilde{\omega}^2(\bm{\kappa})}{2}u^2_{\bm{\kappa},\lambda}\right) +\frac{\omega^2_p}{2}\sum_{\bm{\kappa}\neq \bm{\kappa}^{\prime},\lambda,\lambda^{\prime}} \bm{\varepsilon}_{\lambda}(\bm{\kappa})\cdot\bm{\varepsilon}_{\lambda^{\prime}}(\bm{\kappa}^{\prime}) u_{\bm{\kappa},\lambda} u_{\bm{\kappa}^{\prime},\lambda^{\prime}}.\nonumber\\
\end{eqnarray}
\end{widetext}
The vector potential in terms of the scaled coordinates is $\hat{\mathbf{A}}=\sqrt{1/\epsilon_0 V}\sum_{\bm{\kappa},\lambda}\bm{\varepsilon}_{\lambda}(\bm{\kappa})u_{\bm{\kappa},\lambda}$. For simplicity and convenience we introduce the enlarged ``4-tuple'' variable $\alpha\equiv (\bm{\kappa},\lambda)=(\kappa_x,\kappa_y,\kappa_z,\lambda)$ and everything can be written in a more compact form
 \begin{widetext}
\begin{eqnarray}
\hat{H}=\sum\limits^{N}_{j=1}\left[-\frac{\hbar^2}{2m_{\textrm{e}}}\nabla^2_j +\frac{\textrm{i}e\hbar}{m_{\textrm{e}}} \hat{\mathbf{A}}\cdot\nabla_j\right]-\frac{\hbar^2}{2}\sum^{M}_{\alpha=1}\frac{\partial^2}{\partial u^2_{\alpha}}+\frac{1}{2}\sum^{M}_{\alpha, \beta=1}W_{\alpha\beta}u_{\alpha}u_{\beta}, \;\;\;\textrm{where}\;\;\; W_{\alpha\beta}=\widetilde{\omega}^2_{\alpha}\delta_{\alpha \beta}+\omega^2_p\mathcal{E}_{\alpha \beta},
\end{eqnarray}
\end{widetext}
with the matrix $\mathcal{E}_{\alpha \beta}$ being zero for $\alpha=\beta$, $\mathcal{E}_{\alpha \alpha}=0$, while for $\alpha \neq \beta$ this matrix is defined as the inner product of the polarization vectors $\mathcal{E}_{\alpha \beta}= \bm{\varepsilon}_{\alpha}\cdot \bm{\varepsilon}_{\beta}$. The matrix $W$ is real and symmetric and as a consequence can be diagonalized by an orthogonal matrix $U$. This means that it can be brought into a diagonal form
\begin{eqnarray}\label{diagform}
\sum_{\gamma,\delta} U^{-1}_{\alpha\gamma}W_{\gamma\delta}U_{\delta\beta}= \Omega^2_{\alpha}\delta_{\alpha\beta}
\end{eqnarray}
where $\Omega^2_{\alpha}$ are the eigen-values of the matrix $W_{\alpha\beta}$. Further, because the matrix $U$ is an orthogonal matrix it is also invertible and for its inverse $U^{-1}$ holds that is equal to its transpose $U^{T}$. Using the orthogonal matrix $U$ we can define the normal coordinates $z_{\gamma}$ and the canonical momenta $\partial/\partial z_{\gamma}$~\cite{faisal1987} 
\begin{eqnarray}
z_{\gamma}=\sum_{\alpha} U_{\alpha\gamma} u_{\alpha} \;\;\; \&\;\;\;\; \frac{\partial}{\partial z_{\gamma}}=\sum_{\alpha}U_{\alpha\gamma} \frac{\partial}{\partial u_{\alpha}} .
\end{eqnarray}
In terms of these coordinates and momenta the Hamiltonian takes the form~\cite{faisal1987}
 \begin{widetext}
\begin{eqnarray}
\hat{H}=\sum\limits^{N}_{j=1}\left[-\frac{\hbar^2}{2m_{\textrm{e}}}\nabla^2_j +\frac{\textrm{i}e\hbar}{m_{\textrm{e}}} \hat{\mathbf{A}}\cdot\nabla_j\right]+\sum^M_{\gamma=1}\left(-\frac{\hbar^2}{2}\frac{\partial^2}{\partial z^2_{\gamma}}+\frac{\Omega^2_{\gamma}}{2}z^2_{\gamma}\right)\;\; \textrm{with}\;\; \hat{\mathbf{A}}=\sqrt{\frac{1}{\epsilon_0 V}}\sum_{\gamma=1}\widetilde{\bm{\varepsilon}}_{\gamma} z_{\gamma}.
\end{eqnarray}
\end{widetext}
The new polarization vectors $\widetilde{\bm{\varepsilon}}_{\gamma}$ are defined as $\widetilde{\bm{\varepsilon}}_{\gamma}=\sum_{\alpha=1}\bm{\varepsilon}_{\alpha}U_{\alpha\gamma}$. The Hamiltonian is translationally invariant and thus the wavefunctions of the electronic part are given by the Slater determinant $\Phi_{\bi{K}}$ defined in Eq.~(\ref{Slater determinant}). Applying $\hat{H}$ on $\Phi_{\bi{K}}$  we obtain
\begin{widetext}
\begin{eqnarray}
\hat{H}\Phi_{\bi{K}}=\left[\frac{\hbar^2}{2m_{\textrm{e}}}\sum\limits^{N}_{j=1}\mathbf{k}^2_j+\sum^M_{\gamma=1}\left(-\frac{\hbar^2}{2}\frac{\partial^2}{\partial z^2_{\gamma}}+\frac{\Omega^2_{\gamma}}{2}z^2_{\gamma}-\widetilde{g}z_{\gamma} \widetilde{\bm{\varepsilon}}_{\gamma}\cdot\mathbf{K}  \right)\right]\Phi_{\bi{K}}, \;\; \textrm{where}\;\; \widetilde{g}=\frac{e\hbar}{m_{\textrm{e}}\sqrt{\epsilon_0 V}}.
\end{eqnarray}
\end{widetext}
We perform a square completion and the part of the Hamiltonian depending on $z_{\gamma}$ can be written as a sum of displaced harmonic oscillators with frequencies $\Omega_{\gamma}$ 
\begin{widetext}
\begin{eqnarray}
\hat{H}\Phi_{\bi{K}}=\left[\frac{\hbar^2}{2m_{\textrm{e}}}\sum\limits^{N}_{j=1}\mathbf{k}^2_j+\sum^M_{\gamma=1}\left(-\frac{\hbar^2}{2}\frac{\partial^2}{\partial z^2_{\gamma}}+\frac{\Omega^2_{\gamma}}{2}\left(z_{\gamma}-\frac{\widetilde{g}\widetilde{\bm{\varepsilon}}_{\gamma}\cdot\mathbf{K} }{\Omega^2_{\gamma}}\right)^2-\frac{\left(\widetilde{g}\widetilde{\bm{\varepsilon}}_{\gamma}\cdot\mathbf{K}\right)^2}{2\Omega^2_{\gamma}} \right)\right] \Phi_{\bi{K}}.
\end{eqnarray}
\end{widetext}
The eigenspectrum of each displaced harmonic oscillator is $E_{n_{\gamma}}=\hbar\Omega_{\gamma}(n_{\gamma}+1/2)$ and thus we find that the energy spectrum of the free electron gas coupled to an arbitrary amount of photon modes with the mode-mode interactions included is.
\begin{widetext}
\begin{eqnarray}
E_{\bi{k}}=\frac{\hbar^2}{2m_{\textrm{e}}}\left[\sum\limits^{N}_{j=1}\mathbf{k}^2_j -\frac{\omega^2_p}{N}\sum^M_{\gamma=1}\frac{\left(\widetilde{\bm{\varepsilon}}_{\gamma}\cdot\mathbf{K}\right)^2}{\Omega^2_{\gamma}}\right]+\sum^M_{\gamma=1}\hbar\Omega_{\gamma}\left(n_{\gamma}+\frac{1}{2}\right)
\end{eqnarray}
\end{widetext}
Thus, we see that the structure of the energy spectrum even with the inclusion of the mode-mode interactions is the same with the one in the effective quantum field theory in Eq.~(\ref{effective energy}), and that the mode-mode interactions do not introduce any fundamental modification. To obtain the expression for the eigenspectrum we substituted the parameter $\widetilde{g}$. Also we would like mention that in connection to the effective energy spectrum defined in Eq.~(\ref{effective energy}) and the coupling constant in the effective field theory, the exact coupling as a function of the number of the photon modes on each particular direction is 
\begin{eqnarray}
g^{i}_{\textrm{ex}}(M)=\sum^{M}_{\gamma=1}\frac{\omega^2_p\left(\widetilde{\varepsilon}^{i}_{\gamma}\right)^2}{\Omega^2_\gamma}\;\;\; \textrm{where}\;\;\; i=x,y,z.
\end{eqnarray}

\bibliography{cavity_QED}

\end{document}